\documentclass[twocolumn]{aastex631}

\usepackage{multirow}
\usepackage{amsmath}

\shorttitle{DAVOS: COSMOS}
\shortauthors{Burke et al.}

\begin{document}

\title{DAVOS: Dwarf Active Galactic Nuclei from Variability for the Origins of Seeds: \\ Properties of Variability-Selected Active Galactic Nuclei in the COSMOS Field and Expectations for the Rubin Observatory}

\correspondingauthor{Colin J. Burke}
\email{colin.j.burke@yale.edu}

\author[0000-0001-9947-6911]{Colin J. Burke}
\affiliation{Department of Astronomy, Yale University, 266 Whitney Avenue, New Haven, CT 06511, USA}

\author[0000-0003-4247-0169]{Yichen Liu}
\affiliation{Department of Astronomy, University of Illinois at Urbana-Champaign, 1002 W. Green Street, Urbana, IL 61801, USA}
\affiliation{National Center for Supercomputing Applications, University of Illinois at Urbana-Champaign, 605 East Springfield Avenue, Champaign, IL 61820, USA}


\author[0000-0002-4557-6682]{Charlotte A. Ward}
\affiliation{Department of Astrophysical Sciences, Princeton University, Princeton, NJ 08544, USA}

\author[0000-0003-0049-5210]{Xin Liu}
\affiliation{Department of Astronomy, University of Illinois at Urbana-Champaign, 1002 W. Green Street, Urbana, IL 61801, USA}
\affiliation{National Center for Supercomputing Applications, University of Illinois at Urbana-Champaign, 605 East Springfield Avenue, Champaign, IL 61820, USA}
\affiliation{Center for Artificial Intelligence Innovation, University of Illinois at Urbana-Champaign, 1205 West Clark Street, Urbana, IL 61801, USA}

\author[0000-0002-5554-8896]{Priyamvada Natarajan}
\affiliation{Department of Astronomy, Yale University, 266 Whitney Avenue, New Haven, CT 06511, USA}
\affiliation{Department of Physics, Yale University, 217 Prospect Street, New Haven, CT 06520, USA}
\affiliation{Black Hole Initiative, Harvard University, 20 Garden Street, Cambridge, MA 02138, USA}

\author[0000-0002-5612-3427]{Jenny E. Greene}
\affiliation{Department of Astrophysical Sciences, Princeton University, Princeton, NJ 08544, USA}

\DeclareRobustCommand{\yichen}[1]{{\sethlcolor{lime}\hl{(YL:) #1}}}



\begin{abstract}

We study the black hole mass $-$ host galaxy stellar mass relation, $M_{\rm{BH}}-M_{\ast}$, of a sample of $z<4$ optically-variable AGNs in the COSMOS field. The parent sample of 491 COSMOS AGNs were identified by optical variability from the Hyper Suprime-Cam Subaru Strategic Program (HSC-SSP) program. Using publicly-available catalogs and spectra, we consolidate their spectroscopic redshifts and estimate virial black hole masses using broad line widths and luminosities. We show that variability searches with deep, high precision photometry like the HSC-SSP can identity AGNs in low mass galaxies up to $z\sim1$. However, their black holes are more massive given their host galaxy stellar masses than predicted by the local relation for active galaxies. We report that $z\sim 0.5-4$ variability-selected AGNs are meanwhile more consistent with the $M_{\rm{BH}}-M_{\ast}$ relation for local inactive early-type galaxies. This result is in agreement with most previous studies of the $M_{\rm{BH}}-M_{\ast}$ relation at similar redshifts and indicates that AGNs selected from variability are not intrinsically different from the broad-line Type 1 AGN population at similar luminosities. Our results demonstrate the need for robust black hole and stellar mass estimates for intermediate-mass black hole candidates in low-mass galaxies at similar redshifts to anchor this scaling relation. Assuming that these results do not reflect a selection bias, they appear to be consistent with self-regulated feedback models wherein the central black hole and stars in galaxies grow in tandem.
\end{abstract}



\section{Introduction} \label{sec:intro}

The observed local scaling relations between supermassive black hole (SMBH) mass $M_{\rm{BH}}$ and host galaxy properties (total galaxy stellar mass, bulge stellar mass, and bulge stellar velocity dispersion: $M_{\rm{BH}}-M_{\ast}$, $M_{\rm{BH}}-M_{\ast,\rm{bulge}}$, $M_{\rm{BH}}-\sigma_{\ast}$, and bulge luminosity) in both active and inactive galaxies anchor our understanding of SMBH-host galaxy co-evolution (e.g., \citealt{Magorrian1998,Haehnelt1998,Kormendy2013,Reines2015}). These relations are usually interpreted as evidence for some form of self-regulated feedback in active galactic nuclei (AGNs). However, given that there is no consensus model of AGN feedback, observational studies of these host scaling relations across redshift and extending to lower luminosities are critical for placing new constraints on such models (e.g., \citealt{Ricarte2018}). In addition to understanding feedback, the low-mass end of these scaling relations could also be sensitive to currently not well constrained initial mass distribution of SMBH seeds at high redshift and accretion physics \citep{Volonteri2009,Natarajan2011}.

Actively accreting AGNs and inactive local black hole populations appear to differ in the slope and amplitude of the $M_{\rm{BH}}-M_{\ast}$ scaling relation on which they lie \citep{Reines2015}. Scaling relations have been studied for accreting SMBH populations selected across wavelengths and redshifts. Moving beyond $z\sim0$, \citet{Merloni2010} studied the $M_{\rm{BH}}-M_{\ast}$ relation at $z\sim0.5-2.5$ for broad-line AGNs from the zCOSMOS survey \citep{Lilly2007}, finding evidence that the black hole to host galaxy stellar mass ratio, $M_{\rm{BH}}/M_{\ast}$, increases with redshift. \citet{Suh2020} and \citet{Zhuang2023} studied the $z\sim0.5-2.5$ relation for X-ray selected AGNs selected from the Chandra-COSMOS Legacy Survey \citep{Civano2016}. Similarly, \citet{Ding2020} studied the $M_{\rm{BH}}-M_{\ast}$ relation for broad-line X-ray selected AGNs in deep fields. \citet{Li2021} measured the $M_{\rm{BH}}-M_{\ast}$ relation for SDSS quasars with black hole masses estimated from reverberation mapping. \citet{Mezcua2023} show the $M_{\rm{BH}}-M_{\ast}$ for seven $z\sim0.4-0.9$ AGNs in dwarf galaxies from the VIPERS survey. In a more recent paper, focusing on galaxies at cosmic noon,$z \sim 1 -3$, \cite{Mezcua+2024} report these sources also appear to host over-massive black holes compared to the local $M_{\rm{BH}}-M_{\ast}$ relation. With the notable exception of \citet{Suh2020}, these results generally suggest that these \textbf{intermediate and high redshift AGNs have over-massive black holes compared to the local ($z<0.055$) AGN relation of $M_{\rm{BH}}/M_{\ast} \sim 0.025\%$ \citep{Reines2015}. Instead, they more closely follow the relations for local \emph{inactive} early-type galaxies.}

Recently, exploring a higher redshift population, \citet{Pacucci2023} report that $z \sim 4-7$ quasars discovered by the \emph{James Webb Space Telescope} (\emph{JWST}; \citealt{Harikane2023,Maiolino2023,Kocevski2023,Ubler2023}) have black holes masses $\sim 10 - 100$ times more massive compared to local AGNs with comparable stellar mass hosts. \citet{Kokorev2023} report a $z=8.5$ AGN with a $M_{\rm{BH}}/M_{\ast}$ ratio of at least $\sim 30$ percent discovered with \emph{JWST}. Leveraging {\emph{JWST}} and the \emph{Chandra X-ray Observatory},  \citet{Bogdan2023,Natarajan2023,Goulding2023} discovered a $z\approx10.1$ quasar UHZ1 with a $M_{\rm{BH}}/M_{\ast} \sim 1$. The ratio is also found to be skewed for the $z=10.6$ source GN-z11 \citep{Maiolino2023}. Meanwhile with detailed statistical modeling \cite{Li+2024} make the case that for the $z > 6$ {\emph{JWST}} AGN a combination of selection biases and measurement uncertainties might be skewing $M_{\rm{BH}}-M_{\ast}$ scalings. Our understanding of scaling relations at high redshifts is rapidly evolving. Departure from local scaling relations at these extremely high redshifts $9<z<12$ is predicted to be a signature of the formation of heavy black hole seeds in the early Universe \citep{Natarajan+2017}.

These observations of currently small samples of individual sources are strongly affected by selection biases, whereby AGNs selected only by luminosity, can produce a false evolution in host galaxy scaling relations \citep{Lauer2007}. Meanwhile, uncertainties from single-epoch virial black hole mass measurements can lead to the systematic over-estimation of black hole masses, especially at the high mass end of these relations \citep{Shen2010}. Therefore, it is imperative to select AGNs with lower luminosities across redshifts in order to mitigate these selection biases (e.g., \citealt{Izumi2019,Izumi2021,Suh2020}).

Optical variability is becoming an established and potentially powerful approach to identify AGNs in low luminosity sources and/or with low mass black holes in dwarf galaxies \citep{Baldassare2018,Baldassare2020,Halevi2019,Guo2020,Burke2020tess,Burke2022des,Burke2023,Ward2022}. Besides, as noted earlier, the low mass end of these scaling relations may potentially encode additional information on black hole formation. 

In this paper, we obtain black hole masses, redshifts, and reasonably robust stellar masses for variability-selected low luminosity AGNs selected by \citet{Kimura2020} from the Hyper Suprime-Cam (HSC) UltraDeep survey within the Cosmic Evolution Survey (COSMOS) field \citep{Scoville2007}. About 90 percent of the sources are detected in the X-ray \citep{Kimura2020}. Using these data, we measure the $M_{\rm{BH}}-M_{\ast}$ relation at $z \sim 0.5 - 3$. Our sample has a bolometric luminosity range of $L_{\rm{bol}} \sim 10^{44-47}$ erg s$^{-1}$, comparable to the COSMOS X-ray selected sample of \citet{Suh2020}. We find that this sample of $z \sim 0.5 - 3$ variability-selected AGNs have over-massive black holes compared to the local AGN relation of \citet{Reines2015}, broadly consistent with most previous studies of $z \gtrsim 0.5$ AGNs selected with other techniques \citep{Merloni2010,Mezcua2023,Li2021,Zhang2023,Zhuang2023,Stone2023,Tanaka2024}. 

Our paper is organized as follows. In \S\ref{sec:data}, we describe our procedure for obtaining the archival spectra and photometry and construction of a spectroscopic database for our variable AGN sample. In \S\ref{sec:sed}, we describe our procedure for fitting the spectral energy distribution (SED) to broad-band photometry and the reliability of our stellar mass estimates in the presence of an AGN. In \S\ref{sec:spec}, we describe our spectral fitting approach and resulting broad-line black hole mass estimates. In \S\ref{sec:relation}, we place the sources on the $M_{\rm{BH}}-M_{\ast}$ relation and compare with previous work. We discuss our results \S\ref{sec:discussion} and conclude in \S\ref{sec:conclusion}.

\subsection{Hyper Suprime-Cam Subaru Strategic Program}

In this work, we use the variability-selected AGNs from the UltraDeep Hyper Suprime-Cam Subaru Strategic Program (HSC-SSP) program \citep{Aihara2022} within the $\sim1.5$ deg$^2$ COSMOS field \citep{Kimura2020}. The UltraDeep HSC-SSP program has a deep single-epoch limiting magnitude of $r\sim26$, and has therefore been used to study high redshift galaxies, supernova, and active galactic nuclei (AGNs; \citealt{Kimura2020}). Also see \citet{Zhong2022}, who studied the morphologies of the hosts. The single-epoch photometric precision is significantly better than previous surveys like the Dark Energy Survey (DES) supernova program ($r\sim24.5$; \citealt{Kessler2015,Burke2022des}), and even better than what is planned for LSST Rubin without co-adding multiple exposures ($r \sim 24.5$\footnote{\url{https://pstn-054.lsst.io}}; \citealt{Ivezic2019}) albeit with much a smaller survey volume. The very deep single-epoch photometry enabled identification of low luminosity variability from Type 1 AGNs up to $z\sim4$ \citep{Kimura2020}, providing a more complete sample for detailed follow-up studies of the black hole mass -- host galaxy relationships.

\section{Data Analysis} \label{sec:data}

\subsection{Spectra Database} 

\begin{table}
\scriptsize
\centering
\caption{Sources of optical/NIR spectroscopic redshifts for the HSC-SSP variable AGNs. \label{tab:redshifts}}
\begin{tabular}{lll}
\hline\hline
Count & Reference & Name \\
\hline
177 & \citet{Trump2009} & Magellan XMM AGN \\
134 & \citet{Lilly2009} & zCOSMOS 10k-bright $^a$ \\
78 & \citet{Hasinger2018} & DEIMOS 10K     $^a$ \\
64 & \citet{Cool2013} & PRIMUS         $^a$ \\
45 & \citet{Hasinger2018} & DEIMOS 10K \\
45 & \citet{Ahumada2020} & SDSS DR16      $^a$ \\
35 & \citet{Silverman2015} & FMOS COSMOS DR2$^a$ \\
31 & \citet{Lilly2007} & zCOSMOS DR3 \\
17 & \citet{Marchesi2016} & Chandra COSMOS legacy \\
14 & \citet{Momcheva2016} & 3D-HST v4.1.5  $^a$ \\
11 & \citet{vanderWel2021} & LEGA-C DR3 \\
6 & \citet{Silverman2015} & FMOS-COSMOS \\
6 & \citet{Straatman2018} & LEGA-C DR2     $^a$ \\
5 & \citet{Schulze2018} & FMOS-COSMOS AGN NIR \\
4 & \citet{Damjanov2018} & hCOSMOS \\
4 & \citet{Alam2015} & SDSS DR12 \\
4 & \citet{Paris2014} & SDSS DR10 quasar cat \\
3 & \citet{DESI2023} & DESI EDR \\
3 & \citet{Balogh2014} & GEEC2 \\
3 & \citet{Kartaltepe2015} & FMOS-COSMOS NIR \\
3 & \citet{Monzon2020} & CLAMATO \\
2 & \citet{Paris2018} & SDSS DR14 quasar cat \\
2 & \citet{Boutsia2018} & IMACS faint AGN \\
2 & \citet{LeFevre2013} & VVDS DRFinal   $^a$ \\
1 & \citet{Ahn2012} & SDSS DR9 \\
1 & \citet{Knobel2012} & zCOSMOS 20k Group \\
1 & \citet{Brusa2009} & z$>$3 X-ray QSOs \\
1 & \citet{Allevato2012} & z$<$1 X-ray AGN \\
1 & \citet{Jamal2018} & VVDS reprocessed \\
1 & \citet{Onodera2016} & Star-forming Galaxies NIR \\
1 & \citet{Harrison2016} & KASHz \\
1 & \citet{Koprowski2016} & SCUBA-2 \\
1 & \citet{Harrison2017} & KROSS \\
1 & \citet{Straatman2018} & LEGA-C DR2 \\
1 & \citet{Ono2018} & GOLDRUSH \\
1 & \citet{Masters2019} & C3R2 DR2 \\
1 & \citet{Rosani2020} & SMUVS Lya emitters \\
1 & \citet{Mukae2020} & HETDEX LAEs/eBOSS QSOs \\
1 & \citet{Lyke2020} & SDSS DR16 QSOs \\
1 & \citet{Stanford2021} & C3R2 DR3 \\
1 & \citet{Masters2017} & C3R2 DR1       $^a$ \\
\hline
\end{tabular}
{\sc Note.} --- ($a$) Data taken from the HSC PDR3 catalog of spectroscopic redshifts. 
\end{table}

\begin{figure}
\centering
\includegraphics[width=0.49\textwidth]{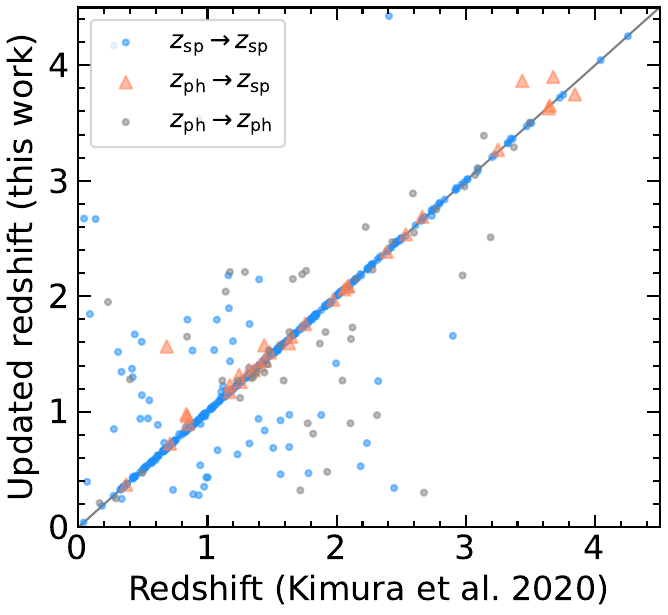}
\caption{Redshift from \citet{Kimura2020} versus our updated redshifts from the HSC PDR3 redshift catalog or SIMBAD (blue circle symbols and orange triangle symbols). Photometric redshifts are from the COSMOS2020 catalog (gray square symbols; \citealt{Weaver2022}).} \label{fig:zupdated}
\end{figure}

\begin{figure}
\centering
\includegraphics[width=0.49\textwidth]{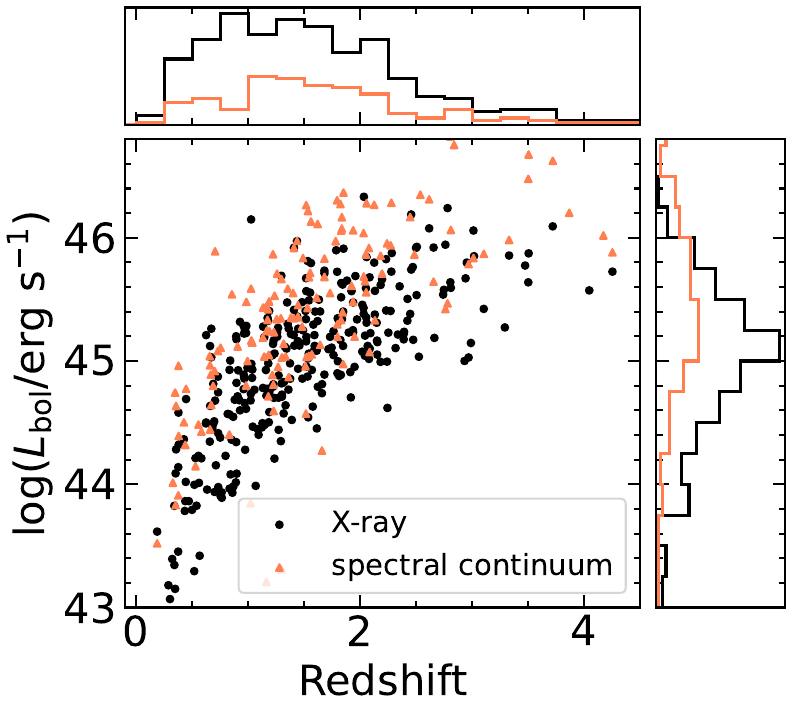}
\caption{Bolometric AGN luminosity (calculated from 10 $\times$ the 2$-$10 keV X-ray luminosity; e.g. \citealt{Duras2020}) versus our updated redshifts. The corresponding redshift and bolometric luminosity distributions are shown.} \label{fig:Lbolz}
\end{figure}

\begin{figure*}
\centering
\includegraphics[width=0.98\textwidth]{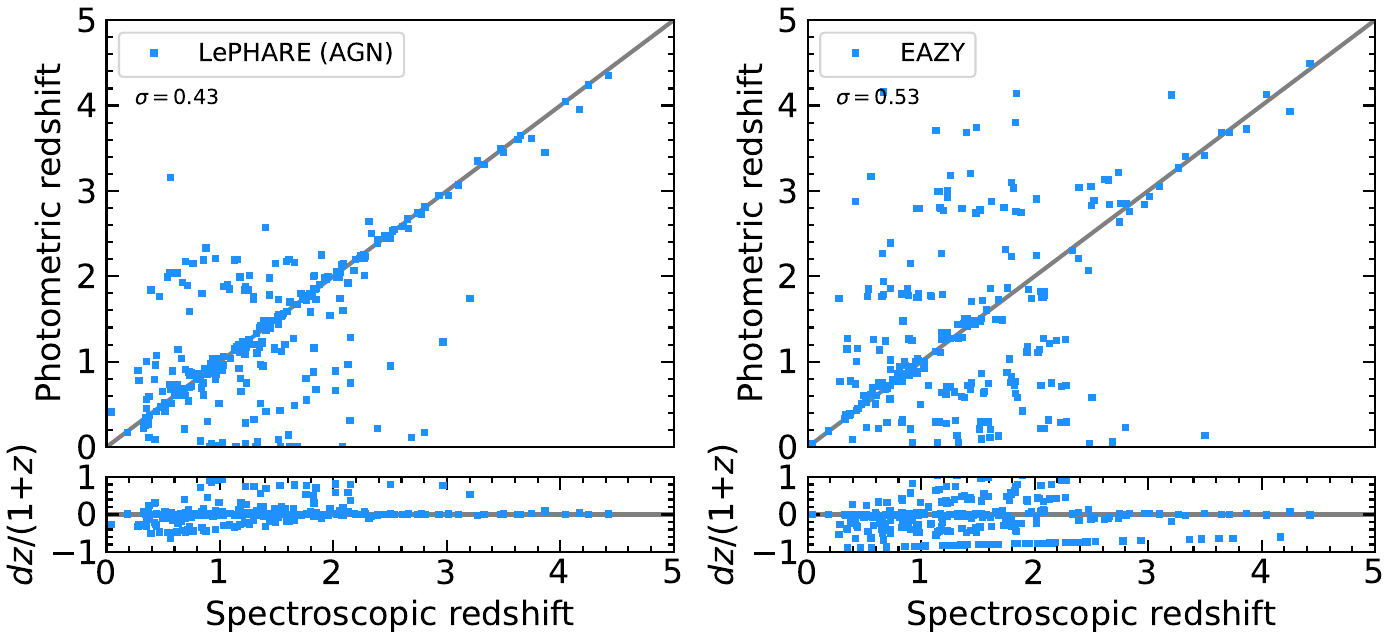}
\caption{Comparison of COSMOS2020 \citep{Weaver2022} photometric redshifts with LePHARE (\emph{left}) and EAZY (\emph{right}) against the spectroscopic redshifts in this work. The LePHARE photometric redshifts includes AGN SED templates, and are therefore more appropriate for the HSC-SSP variable AGNs. The root mean square error (RMSE) values are shown in the upper left hand corner of both figure panels. The gray $y=x$ line is shown for comparison. \label{fig:photoz}}
\end{figure*}


\citet{Kimura2020} have compiled spectroscopic redshifts for their HSC-SSP variable AGNs from the HSC ancillary data products. Specifically, they matched their variable AGNs to the HSC public data release 2 (PDR2) catalog of spectroscopic redshifts\footnote{\url{https://hsc-release.mtk.nao.ac.jp/doc/index.php/specz-2/}}, and spectroscopic redshifts from the Chandra-COSMOS Legacy Survey and DEIMOS catalogs \citep{Hasinger2018}. This database includes spectroscopic redshifts overlapping with the HSC-SSP COSMOS field from zCOSMOS \citep{Lilly2009}, 3D-HST \citep{Skelton2014,Momcheva2016}, FMOS-COSMOS \citep{Silverman2015,Kashino2019}, VUDS \citep{LeFevre2013}, SDSS DR12 \citep{Alam2015}, PRIMUS DR1 \citep{Coil2011,Cool2013}. \citet{Kimura2020} also matched to the DEIMOS 10k Spectroscopic Survey Catalog \citep{Hasinger2018}, which was not included in the HSC PDR2 catalog of spectroscopic redshifts. 

We have instead matched to the HSC PDR3 version of the same catalog, which includes updated SDSS redshifts from SDSS DR15 and additional spectroscopic redshifts overlapping with the HSC-SSP COSMOS field from the DEIMOS 10k sample \citep{Hasinger2018} and LEGA-C DR2 \citep{Straatman2018}. Using an \textsc{astroquery} \citep{Ginsburg2019} SIMBAD search, we identified additional spectroscopic redshifts from the COSMOS XMM-Newton AGN spectroscopic survey \citep{Trump2009} and several additional surveys in the literature that have been indexed by SIMBAD (see references in Table~\ref{tab:redshifts}). We restrict the SIMBAD search to spectroscopic redshifts derived from reliable optical or near-infrared (NIR) spectra using the flags ${\rm RVZ\_WAVELENGTH} == {\rm 'O'}$ or ${\rm RVZ\_WAVELENGTH} == {\rm 'N'}$ and ${\rm rvz\_qual} != {\rm'E'}$. The first constraint restricts the search to optical or NIR spectra. We did not find any ALMA sub-mm redshifts for our sources that have been indexed into the spectroscopic redshift database of SIMBAD. The second constraint excludes photometric redshifts. Finally, we download the publicly-available spectra from these sources (see Appendix~\ref{sec:availibility}). We also obtained spectroscopic redshifts from the Dark Energy Spectroscopic Instrument (DESI) Early Data Release \citep{DESI2023}, which have not yet been indexed by SIMBAD at the time of writing. We choose our fiducial spectroscopic redshifts according to the following priority: SIMBAD (316 sources), HSC PDR3 catalog of spectroscopic redshifts (93 sources), DESI (3 sources), and Chandra-COSMOS Legacy Survey spectroscopic redshifts (17 sources). But significantly inconsistent redshifts are resolved by hand after careful visual inspection (described in \S\ref{sec:inconsistentspecz}). If no spectroscopic redshift was found, we use the photometric redshifts from the COSMOS2020 catalog (50 sources; \S\ref{sec:photoz}). There are also 12 sources with no COSMOS2020 match, and therefore with no redshift. Our updated redshifts are shown in Figure~\ref{fig:zupdated}. The bolometric luminosity and redshift distributions are shown in Figure~\ref{fig:Lbolz}. A table showing the sources of the public spectroscopic redshifts for the HSC-SSP AGNs is shown in Table~\ref{tab:redshifts}, and our updated redshifts are given in Table~\ref{tab:mass}. To include cases where a single source has more than one available spectrum from different programs, we always repeat the matching between the HSC-SSP AGNs and the spectroscopic sample when downloading the spectra from publicly-available sources.

\subsection{Inconsistent spectroscopic redshifts} \label{sec:inconsistentspecz}

Using $\Delta z>0.1$ as our criterion, we found 36 sources with inconsistent spectroscopic redshifts between SIMBAD, HSC PDR3, DESI, and Chandra-COSMOS Legacy Survey catalogs and 20 inconsistent optical spectra between Magellan, zCOSMOS, DEIMOS, DESI, and LEGA-C spectral files, from which we will use to estimate BH masses. For each source with inconsistent spectroscopic redshifts, we plotted their publicly available spectra against a Type 1 quasar template spectrum (e.g., \citealt{VandenBerk2001,Richards2006}) and attempted to identify which spectroscopic redshift is correct by visually matching the spectrum with the strong emission lines from the template. At least two of the authors have visually inspected and agreed to these corrected redshifts. In most cases the correct spectroscopic redshift can be obviously identified. For 7 of these sources, we were unable to determine the correct spectroscopic redshift due to poor spectral quality or lack of obvious emission lines or features in the spectrum. We adopt the COSMOS2020 photometric redshifts for these 7 sources. 28 of the clearly incorrect spectroscopic redshifts originate from the PRIMUS catalog. The PRIMUS spectra are not publicly available, preventing us from visually inspecting them. We suspect these incorrect PRIMUS redshifts are due to a misidentified emission line given the relatively low resolution of the prism spectra of $R = \lambda/\Delta\lambda \sim 40$ \citep{Coil2011}. 

\subsection{Photometric redshifts} \label{sec:photoz}

For sources without a reliable spectroscopic redshift, we use the \textsc{LePHARE} \citep{Arnouts1999,Ilbert2006} photometric redshifts from the COSMOS2020 catalog \citep{Weaver2022}. The \textsc{LePHARE} redshifts are derived by fitting galaxy and AGN templates to the photometry. When compared to \textsc{EAZY} \citep{Brammer2008} photometric redshifts without AGN templates, we find that the \textsc{LePHARE} redshifts are generally much more reliable. Using the spectroscopic redshifts as our ground truth, we estimate the scatter in the photometric redshifts as $\sigma = 1.4826$ MAD, where MAD is the median absolute deviation, which is robust to outliers (e.g., \citealt{Burke2022des}). We find $\sigma = 0.43$ for the \textsc{LePHARE} photometric redshifts and $\sigma = 0.53$ for the \textsc{EAZY} photometric redshifts. A comparison between the two photometric redshifts for the HSC-SSP variable AGNs are shown in Figure~\ref{fig:photoz}. 

\section{SED fitting}  \label{sec:sed}

\begin{table*}
\scriptsize
\centering
\caption{\textsc{X-CIGALE} SED fitting parameters.}
\label{tab:cigale}
\begin{tabular}{lll} \hline\hline
Module & Parameter & Values \\
\hline
\multirow{2}{*}{\shortstack[l]{Star formation history:\\ 
                              delayed model, $\mathrm{SFR}\propto t \exp(-t/\tau)$ }} 
& $e$-folding time, $\tau$ (Gyr) & 0.1, 0.5, 1, 5 \\
& Stellar age, $t$ (Gyr) & 0.5, 1, 3, 5, 7 \\ 
\hline
\multirow{2}{*}{\shortstack[l]{Simple stellar population:\\ \cite{Bruzual2003}}} 
& Initial mass function & \cite{Chabrier2003} \\
& Metallicity, $Z$ & 0.02 \\
\hline
\multirow{2}{*}{\shortstack[l]{Galactic dust attenuation:\\ \cite{Calzetti2000} \& \cite{Leitherer2002} }} 
& $E(B-V)$ of starlight for the young population  
& 0.0, 0.5, 1.0, 1.5 \\
& $E(B-V)$ ratio between the old and young populations 
& 0.44 \\
\hline
\multirow{1}{*}{\shortstack[l]{Galactic dust emission: \cite{Dale2014}}} 
& $\alpha$ slope in $dM_{\rm dust} \propto U^{-\alpha} dU$
& 1.5, 2.0, 2.5 \\
\hline
\multirow{11}{*}{\shortstack[l]{AGN (UV--IR): \\ SKIRTOR}} 
& Torus optical depth at 9.7 microns $\tau_{9.7}$ & 7.0 \\
& Torus density radial parameter $p$ 
    ($\rho \propto r^{-p} \mathrm{e}^{-q|\cos (\theta)|}$) & 1.0 \\
& Torus density angular parameter $q$ 
    ($\rho \propto r^{-p} \mathrm{e}^{-q|\cos (\theta)|}$) & 1.0 \\
& Angle between the equatorial plan and edge of the torus & 40$^\circ$ \\
& Ratio of the maximum to minimum radii of the torus & 20 \\
& Viewing angle $\theta$ (face on: $\theta=0^\circ$, edge on: 
                          $\theta=90^\circ$) & 30$^\circ$ (type~1) \\
& Power-law index of the disk UV/optical slope $\delta$ & $-1.0$, $-0.5$, $-0.36$, 0.0, 0.5, 1.0 \\
& UV AGN fraction $0.1-0.3\ \mu$m  $f_{\rm{AGN}}$ & $0-0.9$ (step 0.1), 0.95, 0.9999 \\
& Extinction law of polar dust & SMC \\
& $E(B-V)$ of polar dust & 0.0, 0.1, 0.3, 0.5 \\
& Temperature of polar dust (K) & 100 \\
& Emissivity of polar dust & 1.6 \\
\hline
\multirow{2}{*}{\shortstack[l]{X-ray: }} 
& AGN photon index $\Gamma$ & 1.8 \\ 
& Maximum deviation from the $\alpha_{\rm{ox}}$-$L_{\rm{2500} \AA}$ relation & 0.2 \\
\hline
\end{tabular}
\begin{flushleft}
{\sc Note.} --- See \citet{Yang2020,Yang2022} for details of \textsc{x-cigale} SED parameters and models.
\end{flushleft}
\label{tab:parsed}
\end{table*}


We obtained deep, broad-band photometry by matching to version 2.2 of the COSMOS2020 catalog \citep{Weaver2022}. We identified 455 matches with the COSMOS2020 catalog. We use the photometry derived from \textsc{The Farmer} pipeline \citep{Weaver2023}, which is based on source modeling using \textsc{The Tractor} software \citep{Lang2016}. We use included deblended photometry extracted from HSC-SSP PDR2 optical imaging, UltraVISTA near-infrared imaging \citep{McCracken2012,Moneti2023}, Spitzer/IRAC mid-infrared images from the Cosmic Dawn Survey \citep{Euclid2022}, and near and far ultraviolet imaging from the COSMOS GALEX catalog \citep{Zamojski2007}. The X-ray 0.5$-$10 keV photometry is from the \emph{Chandra} COSMOS Legacy Survey \citep{Civano2016,Marchesi2016}. Inclusion of the optical to mid-infrared photometry is essential to constrain the star-formation and reprocessed dust emission. Additionally, including the X-ray photometry or even upper limits can help constrain the AGN emission \citep{Yang2020}, helping to eliminate \emph{some} degeneracies between star-formation and AGN emission which can lead to spurious stellar mass estimates. 

We use version 2022.1 of the \textsc{cigale} code \citep{Burgarella2005,Noll2009,Boquien2019,Yang2020,Yang2022} to perform SED fitting. This version of the code includes X-ray fitting modules from \textsc{x-cigale}, with detailed AGN emission models that have been extensively tested with both galaxy and AGNs \citep{Yang2020,Yang2022}. These codes work by imposing a self-consistent energy balance constraint between different emission and absorption mechanisms across the EM spectrum. A large grid of models is computed and fitted to the data, allowing for an estimation of the star formation rate (SFR), stellar mass, and AGN contribution via a Bayesian-like analysis of the likelihood distribution.

We use a delayed exponential star formation history and vary the $e$-folding time and age of the stellar population assuming solar metallicity. Nevertheless, different choices of the initial mass function, stellar population models, and star formation histories can introduce systematic uncertainties of $\sim 0.3$ dex \citep{Conroy2013}. \citet{Zou2022} found that different \emph{parametric} star formation histories results in systematic differences in stellar mass of $\sim 0.1$ dex for a sample of $z = 0 - 6$ AGN. We adopt the commonly-used \citet{Chabrier2003} initial stellar mass function with the stellar population models of \citet{Bruzual2003} and adopt the nebular emission template of \citet{Inoue2011}. We use the \citet{Leitherer2002} extension of the \citet{Calzetti2000} model for reddening due to dust extinction, and the \citet{Draine2014} updates to the \citet{Draine2007} model for dust emission. Finally, we adopt the SKIRTOR clumpy two-phase torus AGN emission model \citep{Stalevski2012,Stalevski2016} allowing for additional polar extinction. We do not force the UV part of the SED to be completely dominated by the AGN component, because this can result in over-estimation of the stellar masses. We assume Type-1-like inclination angle of $30$ deg for the HSC-SSP variable AGNs. Our choice of considering only a single viewing angle close to the average values for Type 1 AGNs is justified by previous studies, which find that different viewing angles were largely degenerate with the average values of 30 and 70 degrees for Type 1 and 2 AGNs, respectively \citep[e.g.,][]{Mountrichas2021,Padilla2022}. 

Table~\ref{tab:cigale} shows our \textsc{cigale} input parameters, with example SED fitting results shown in Figure~\ref{fig:seds}. Our stellar masses range from $M_{\ast} \sim 10^{9-11.5}\ M_{\odot}$. Inclusion of far-infrared photometry could further improve the results by constraining the dust emission component and reducing degeneracies in the modeling, given \textsc{cigale}'s energy conservation principle. Unfortunately, there are difficulties with spatially associating sources given the large PSF sizes in the far-infrared. Our stellar masses could also be refined using a future COSMOS2020 super-deblended catalog with far-infrared photometry (Jin et al. in preparation).

\subsection{Reliability of stellar mass estimates}
\label{sec:Mstarrecovery}

\begin{figure*}
\centering
\includegraphics[width=0.48\textwidth]{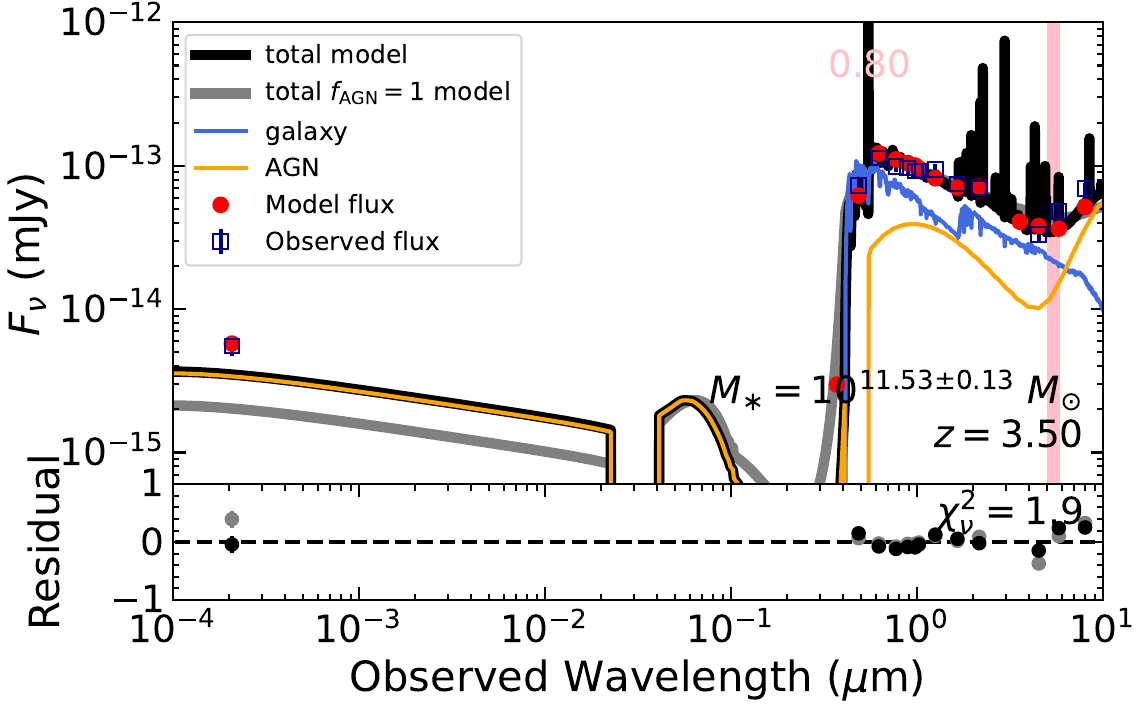}
\includegraphics[width=0.48\textwidth]{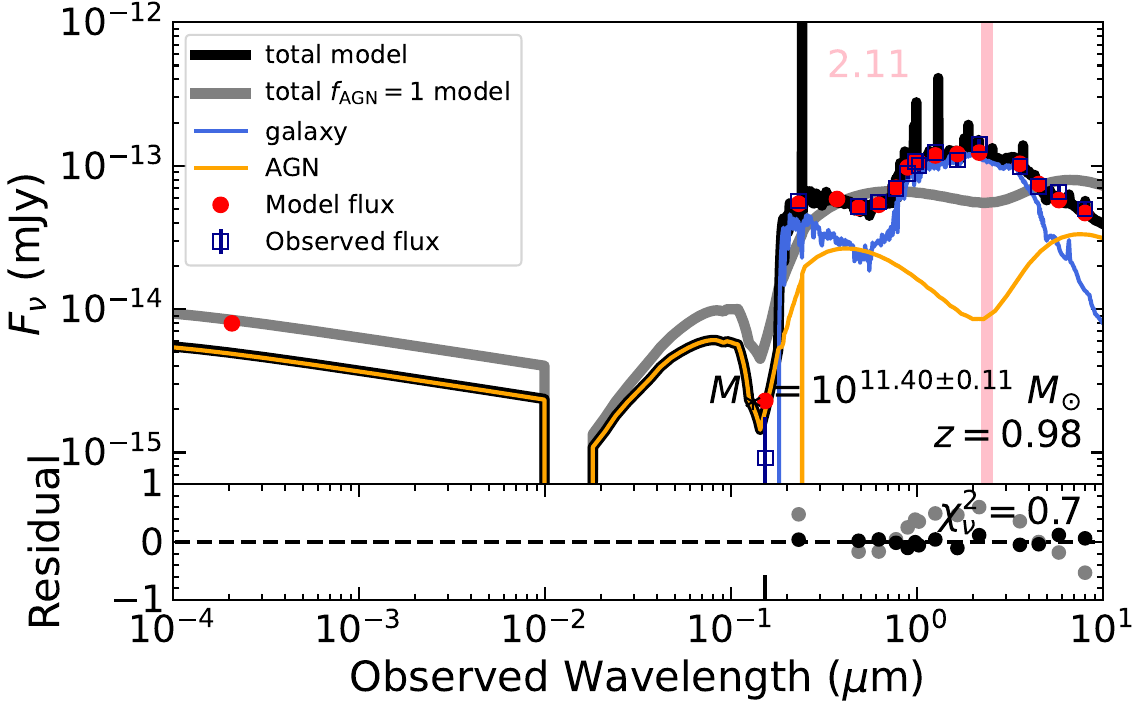}
\caption{\emph{Left panel:} Example SED fitting result for an AGN-dominated source (\citet{Kimura2020} ID = 2). Including a stellar emission component does not significantly improve the fit in this source. Therefore, the stellar emission and associated parameters (e.g., stellar mass, star formation history) cannot be reliably constrained. \emph{Right panel:} Example SED fitting result for a non AGN-dominated source (\citet{Kimura2020} ID = 3). The inclusion of a stellar emission component improves the best-fit $\chi_\nu^2$, and the resulting best-fit stellar emission is a significant excess of the AGN continuum emission near $\sim 1.2$ $\mu$m, enabling a reliable stellar mass estimate. The uncertainties on the stellar mass here are estimated by \textsc{cigale} and do not include uncertainties beyond the parameter choices in Table~\ref{tab:cigale}. The vertical red line is the region near $1.2$ $\mu$m, where the AGN emission is at a minimum, and the red number is the ratio of SF emission (blue) from the SF$+$AGN model to total emission (black) of an AGN-dominated model (gray). Both redshifts are spectroscopic. \label{fig:seds}}
\end{figure*}

\begin{figure}
\centering
\includegraphics[width=0.5\textwidth]{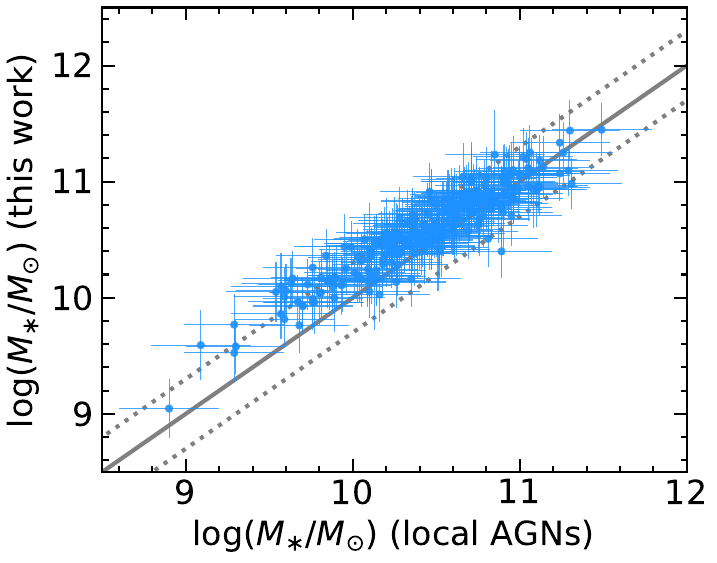}
\caption{Results of applying our stellar mass estimation technique to photometry of the the mostly host-dominated local AGN sample of \citet{Reines2015}. Our stellar masses ($y$-axis) are in very good agreement with those of  \citet{Reines2015}, estimated from mass-to-light ratios after subtracting the AGN contribution from the photometry ($x$-axis). The dotted gray lines of $\pm 0.3$ dex are shown to guide the eye about the solid gray $y=x$ line. All stellar masses are reliably estimated following our procedure This figure demonstrates that our reliable stellar masses are reasonable when the AGN contribution does not dominate the SED. \label{fig:rv15}}
\end{figure}

\begin{figure}
\centering
\includegraphics[width=0.5\textwidth]{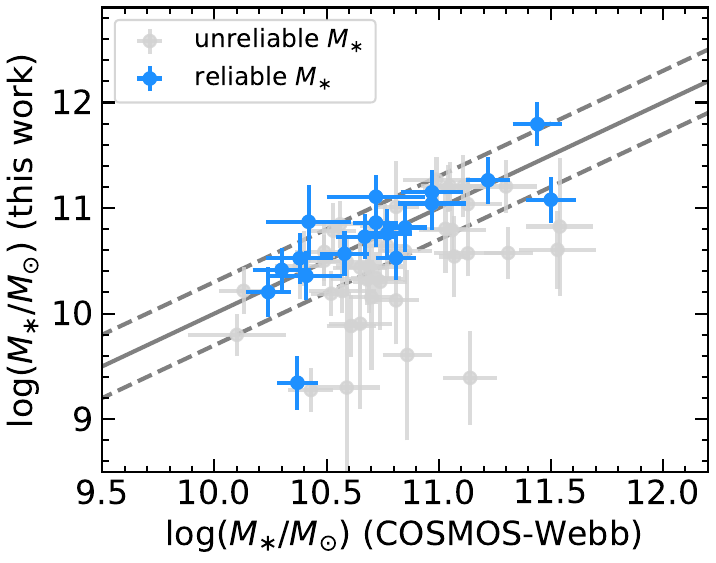}
\caption{Validation of the stellar masses from COSMOS2020 broad-band SED fitting ($x$-axis; this work) to stellar masses from COSMOS-Webb and \emph{HST} archival imaging ($y$-axis; \citealt{Zhuang2023}). The blue points are sources with reliable COSMOS2020-based SED stellar masses, and the gray points have unreliable stellar masses. The stellar masses from COSMOS-Webb and \emph{HST} imaging are based on SED fitting after spatially decomposing the AGN+host emission. Our stellar masses are consistent to $\sim 0.3$ dex (1$\sigma$ scatter, shown as dashed gray lines about the solid gray $y=x$ line), which is comparable to the expected systematic error of $\sim0.3$ dex on SED-based stellar masses \citep{Conroy2013}. This figure demonstrates that our reliable stellar masses are reasonable, even when compared to those measured from AGN+host decomposition with space-based imaging. The good agreement for our sample could partially be due to the larger number of sources with lower AGN luminosities (compared to the host) at a given stellar mass. Our sample of HSC-SSP AGNs includes many more sources outside the COSMOS-Webb and \emph{HST} archival imaging footprint. \label{fig:cosmoswebb}}
\end{figure}

\begin{table*}
\caption{Properties of COSMOS variable AGNs from SED and spectral fitting analysis.   \label{tab:mass}}
\center
\begin{tabular}{cccccccccccccc}
\hline\hline
ID & RA & DEC & i-mag & $z_{{\rm best}}$ & $z_{{\rm ph}}$ & $\log L$ & $\log M_{\rm{BH}}$ & $\log M_{\rm{BH, err}}$ & $\log M_{\ast}$ & $\log M_{\ast, {\rm{err}}}$ & ${\rm{SF}}_{\rm{ex}}$ & $\chi^2_\nu$ & class \\
(1) & (2) & (3) & (4) & (5) & (6) & (7) & (8) & (9) & (10) & (11) & (12) \\
 & deg & deg & AB mag &   &   & $\log$ erg s$^{-1}$ & $\log M_{\odot}$ & $\log M_{\odot}$ & $\log M_{\odot}$ & $\log M_{\odot}$ & \\
\hline
1 & 150.74386 & 2.20245 & 22.71 & 1.58 & 1.58 &  &  &  & 10.64 & 0.12 & 1.1 & 3.0 &  \\
2 & 150.73557 & 2.19957 & 20.36 & 3.5026 & -1.0 & 46.68 & 9.04 & 0.11 & 11.53 & 0.13 & 0.8 & 1.9 & BL \\
3 & 150.73353 & 2.15646 & 20.88 & 0.979 & 1.03 &  &  &  & 11.4 & 0.11 & 2.1 & 0.7 & NeV \\
4 & 150.79702 & 2.13888 & 21.01 & 0.5727 & 0.51 &  &  &  & 10.78 & 0.2 & 3.5 & 1.3 &  \\

\vdots & & & & & \\
491 & 150.03524 & 2.72781 & 21.04 & 0.5093 & 0.52 &  &  &  & 10.49 & 0.06 & 5.4 & 0.7 &  \\
\hline
\end{tabular}
\begin{flushleft}
{\sc Note.} Column (1): Identifier from Table 4 of \citet{Kimura2020}. Column (2): RA. Column (3): Dec. Column (4): $i$ band AB magnitude from \citet{Kimura2020}. Column (5): Best redshift. Column (6): Photometric redshift from COSMOS2020 \textsc{LePHARE}. Column (7): Bolometric luminosity from the spectral continuum. Column (8): Virial black hole mass. Column (9): Inferred \textsc{cigale} stellar mass. Column (10): Excess SF over an AGN dominated model (if $>1.2$, stellar masses are considered reliable). Column (11): Best-fit reduced \textsc{cigale} model $\chi^2$ (recommend $<5$). Column (12): Visual classification of spectrum (see \S\ref{sec:vis}). All uncertainties are $1\sigma$ statistical errors from fitting. This table is published in its entirety in the published version. Only a portion is shown here.
\end{flushleft}
\end{table*}

For galaxies with non-negligible AGN emission, the AGN emission must be constrained or subtracted in order to properly model the star-formation emission in the galaxy and obtain a reliable stellar mass. High resolution optical/NIR imaging combined with source profile fitting can be used to subtract the AGN point source emission from the underlying host galaxy. Alternatively, an AGN template can be fit to the continuum and lines in the spectra and re-scaled to the photometry in order to model the AGN continuum \citep{Reines2015}. When high resolution imaging or spectra is not available or feasible to analyze in large quantities, SED fitting to broad-band catalog photometry can be used with some caveats.

The UV/optical emission from an unobscured AGN accretion disk can be strongly degenerate with stellar emission. When the SED is dominated by an unobscured AGN, the stellar emission component is swamped by the AGN emission and the resulting stellar emission parameters (e.g., stellar mass, star formation history) cannot be reliably constrained (e.g., \citealt{Merloni2010,Ciesla2015}). We employ a model comparison technique to determine whether the stellar emission component, and by extension stellar mass, can be constrained by SED fitting. The model comparison test works as follows. First, we fit the SED using an AGN-dominated model by setting $f_{\rm{AGN}}=0.9999$, where $f_{\rm{AGN}}$ is the AGN fraction computed between observed-frame $0.5-1$ $\mu$m. This wavelength is where the AGN emission is near a minimum, assuring the overall SED is totally dominated by the AGN emission. For technical reasons, the AGN fraction cannot be set to exactly unity in the \textsc{cigale} code \citep{Yang2020}. We have rounded the AGN fraction up to 1 for clarity in the presentation of this paper. Then, we fit the SED using a mixed AGN+stellar emission model by allowing the AGN emission to vary, i.e., varying $f_{\rm{AGN}}$ between 0 and 1 in the observed-frame 0.1$-$0.3 $\mu$m. Note that most of our spectra show significant AGN continuum emission at these wavelengths, and the fact that the sources have been identified from optical band light curves.  Finally, we consider the stellar masses reliable for only those SEDs with significant stellar emission at rest-frame 1.2 $\mu$m that cannot be explained by a totally AGN-dominated model.

To determine whether the stellar emission is constrained, we compute the ratio of the total model fit for the AGN+stellar emission model and the AGN-dominated model. We consider the stellar masses reliable if this ratio is greater than 1.2, as justified in Appendix~\ref{sec:masstests}. This method is a combination of similar approaches used in the literature \citep{Merloni2010,Suh2020,Burke2022des}. Our approach eliminates SEDs that are totally degenerate with AGN-dominated emission at rest-frame 1.2 $\mu$m, where the AGN emission is close to a minimum. We found that simply comparing the reduced $\chi^2$ values was less reliable, because $\chi^2$ values are sensitive to the photometric uncertainties on the data, model parameter choices (i.e., over-fitting stellar emission), and does not always indicate an improved fit in the optical-NIR region where the stellar emission is strongest and from where the stellar masses are derived. Example SEDs from our sample with reliable and unreliable stellar mass estimates are shown in Figure~\ref{fig:seds}. The scatter in the recovered stellar masses is typically $\sim 0.2$ dex. This uncertainty is fully not taken into account in the \textsc{cigale} code, which tends to under-estimate the uncertainties in stellar mass. We add 0.2 dex in quadrature to our \textsc{cigale} uncertainties throughout the figures in this paper.

\subsection{Validation of stellar mass estimation method against the local AGN sample}

\citet{Reines2015} measured the $M_{\rm{BH}}-M_{\ast}$ relation for a sample of $z<0.055$ AGNs selected from SDSS optical spectroscopy. Their stellar masses are estimated using mass-to-light ratios as function of host galaxy color \citep{Zibetti2009}. They estimated the AGN contribution to the integrated photometry by re-scaling a mock AGN spectrum to match the continuum luminosity, then convolving the scaled mock spectrum with the SDSS filter throughput curves. They then subtracted the contribution from AGN emission from the photometry before applying the mass-to-light ratios. Five sources were excluded from their analysis due to being dominated by AGN emission at the $\gtrsim 50\%$ level. 

As a consistency check, we ran our SED fitting procedure on the \citet{Reines2015} sample of local AGNs using Petrosian flux photometry from the NASA Sloan Atlas catalog of GALEX UV and SDSS optical photometry \citep{Blanton2011}, shown in Figure~\ref{fig:rv15}. All of the stellar masses are reliably estimated with our \textsc{cigale} procedure and we find a very good agreement with their results across the entire range of stellar masses. Of course, the photometry of the local AGN sample are comparatively host-dominated compared to our HSC-SSP AGNs. Nevertheless, this demonstrates that our stellar masses are highly reliable when the AGN contribution does not completely dominate the SED. This also ensures that discrepancies between the local $M_{\rm{BH}}-M_{\ast}$ relation and those at higher redshifts are not due to errors in the stellar mass measurements of local AGNs. 

\subsection{Validation of stellar masses using JWST and HST spatially-decomposed and PSF-subtracted stellar masses}

To validate our stellar masses in cases where the AGN contribution to the SED is more substantial, we compare our stellar masses to those estimated by \citet{Zhuang2023}. These authors measured stellar masses for broad-line X-ray selected AGNs with BH masses from \citet{Suh2020} using archival \emph{HST} and \emph{JWST} COSMOS-Webb imaging. Their stellar masses from COSMOS-Webb and \emph{HST} imaging are based on SED fitting with \textsc{cigale} after spatially decomposing the AGN+host emission. Our stellar masses are consistent to $\sim 0.4$ dex (1$\sigma$ scatter, shown as dashed gray lines about the solid gray $y=x$ line) from Figure~\ref{fig:cosmoswebb}, which is comparable to the expected systematic error of $\sim0.3$ dex on SED-based stellar masses \citep{Conroy2013}. The single outlier in the lower left hand corner of the plot has an excess SF model ratio of 1.33 from our SED fitting reliability approach, hence, only a very slight bump near 1.2 $\mu$m from stellar emission. No significant systematic offset is found between our stellar masses and those of \citet{Zhuang2023}. This is encouraging, as it indicates that our COSMOS2020-based stellar masses are not strongly biased due to AGN contamination. We attribute this to our model-comparison technique for selection of reliable stellar masses based on the strength of the stellar emission relative to the AGN emission. Of course, our stellar masses are not as robust as those measured from high resolution \emph{HST} and \emph{JWST} imaging. 

\subsection{Detection limits}

\begin{figure*}
\centering
\includegraphics[width=0.95\textwidth]{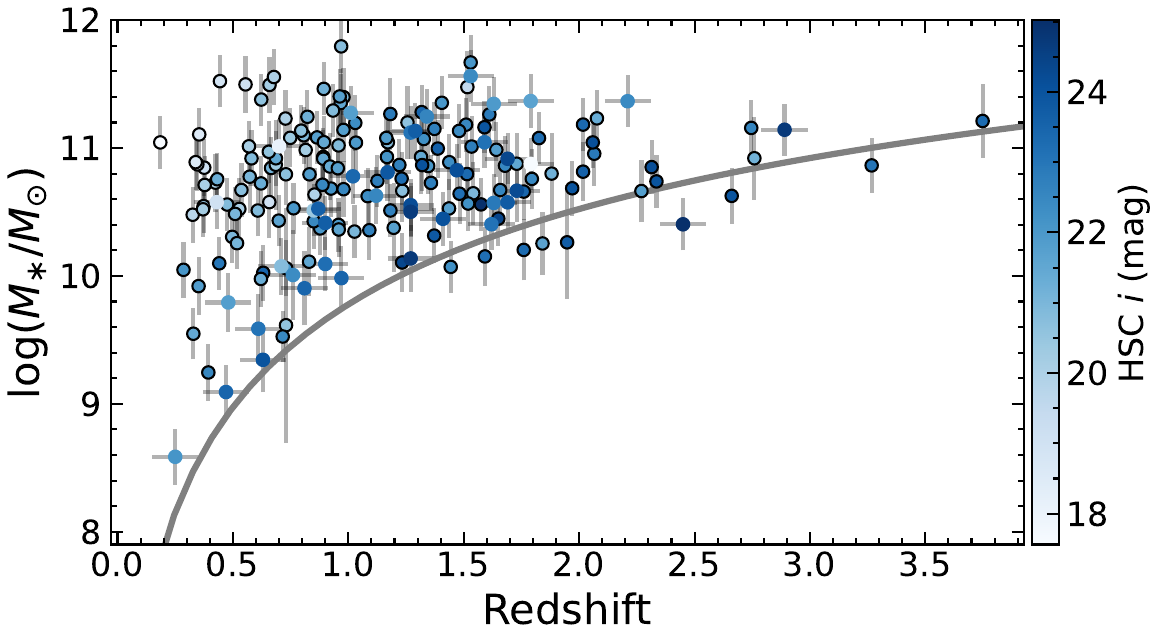}
\caption{Host galaxy stellar mass versus redshift for HSC-SSP variable AGNs with reliable stellar mass estimates from broad-band SED fitting using \textsc{cigale}. Each AGN is shaded by its \emph{HSC} $i$-band apparent magnitude. Symbols with black borders have spectroscopic redshifts from our redshift database. Symbols without black borders have photometric redshifts. The gray curve is the theoretically predicted detection limit. \label{fig:massredshift}}
\end{figure*}

In Figure~\ref{fig:massredshift}, we show our derived stellar mass estimates versus redshift for the HSC-SSP AGNs. The gray curves show the theoretically-predicted BH mass detection limits following \citet{Burke2023}. These BH mass horizon curves are computed assuming a limiting detectable variability amplitude of 0.1 magnitudes and the modified correlations between optical variability amplitude and BH mass (e.g. \citealt{MacLeod2010}) as described in \citet{Burke2023}. The predicted BH mass detection limits are derived assuming a typical amplitude of variability of 0.1 mag and a photometric precision given by \citet{Ivezic2019} with limiting magnitude of $m_5=25.5$.

\section{Spectral Measurements} \label{sec:spec}

\subsection{Description of spectroscopic data}

\begin{figure*}
\centering
\includegraphics[width=0.48\textwidth]{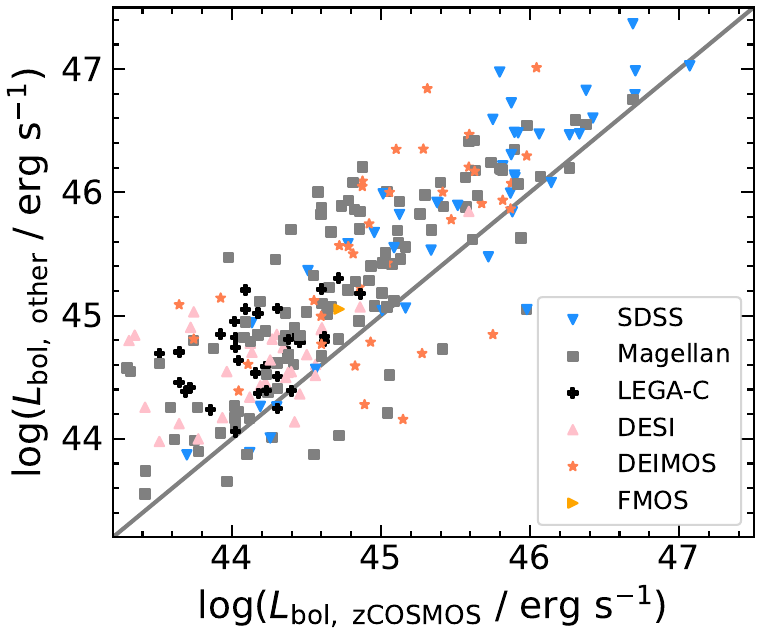}
\includegraphics[width=0.48\textwidth]{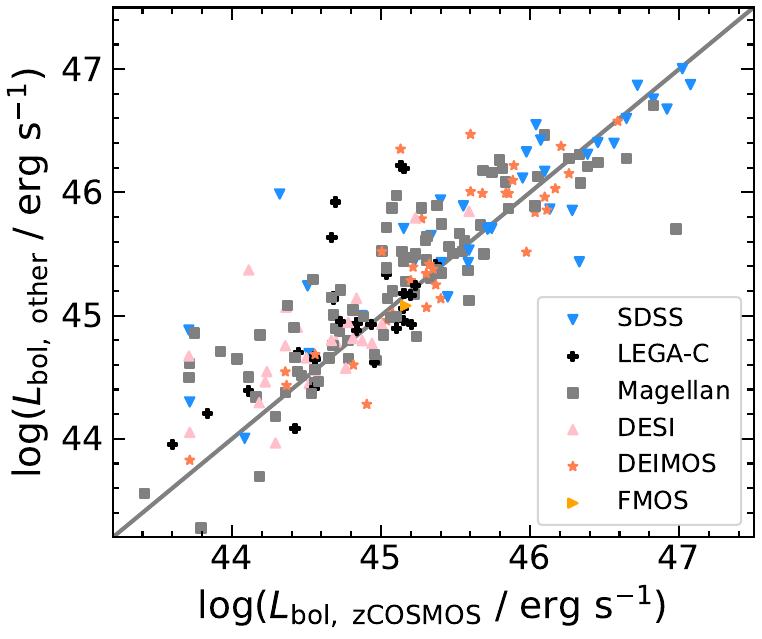}
\caption{Demonstration of absolute flux calibration of our spectra. Each data point is a single HSC SSP AGN with multiple spectra---one from zCOSMOS and at least one from another source. The bolometric luminosity from zCOSMOS spectral continua is plotted against the bolometric luminosity from other spectra (see Table~\ref{tab:redshifts}) before (\emph{left}) and after (\emph{right}) performing absolute flux calibration to the COSMOS2020 photometry. Our resulting flux-calibrated spectra have no systematic offset with zCOSMOS and the fluxes are more consistent as evident by the significantly reduced scatter. The right panel only includes those with successful flux calibration. \label{fig:fluxing}}
\end{figure*}

We downloaded publicly-available spectra for our parent sample of HSC-SSP AGNs from Table~\ref{tab:redshifts}. Our list of downloaded spectra varies slightly from our spectroscopic redshift database, because not all of the spectroscopic data are public in Table~\ref{tab:redshifts}. We obtained spectra from the following programs: zCOSMOS, the \citet{Trump2009} Magellan and IMACS program, VUDS, DEIMOS, C3R2, FMOS-COSMOS, LEGA-C, DESI, and SDSS. The links to download these spectra are given in Appendix~\ref{sec:availibility}. We briefly describe each dataset below.

zCOSMOS was a magnitude-limited $I_{\rm{AB}}<22.5$ sample of about $20,000$ galaxies covering the entire 1.7 deg$^{2}$ COSMOS ACS field plus a deeper 1 deg$^{2}$ targeting galaxies with $I<24$ with about $10,000$ galaxies selected by color-color selection. The data were taken with the VIMOS multi-object spectrograph on the VLT and covers $\sim 5550-9650$ Å with resolutions of $R \sim 600$ (bright) or $R \sim 230$ (deep) \citep{Lilly2007}. 

The \citet{Trump2009} spectroscopic program targeted 677 AGNs in a 2 deg$^{2}$ region of the COSMOS field selected from XMM-Newton with $I_{\rm{AB}}<22$. The data were taken with the Magellan/IMACS and MMT/Hectospec spectrographs. The IMACS data cover $\sim 5600–9200$ Å with a spectral resolution of $\sim 10$ Å or better. The Hectospec data cover $\sim 3800–9200$ Å with a spectral resolution of $\sim 3$ Å \citep{Trump2009}. 

VUDS targeted about 10,000 faint galaxies in the COSMOS, ECDFS and VVDS-02h fields based on color and photometric redshift with $i_{\rm{AB}} < 27$. The data were taken with the VIMOS multi-object spectrograph on the VLT and cover $\sim 3650-9350$ Å with a spectral resolution of $R \sim 230$ in the COSMOS field \citep{LeFevre2013}.  

The DEIMOS COSMOS field program targeted 10,718 sources to limiting magnitudes of $I_{\rm{AB}} = 23.5-25$ with the DEIMOS multi-object spectrograph on the Keck II telescope with a complicated selection function based on many input target catalogs including AGNs and high-redshift galaxies. The observations were taken with two different gratings, covering either $\sim 4800–10000$ Å or $\sim 6700–10500$ Å with spectral resolutions of $R\sim 2000$ or $R \sim 27000$, respectively \citep{Hasinger2018}. 

C3R2 targeting faint galaxies with $i_{\rm{AB}} < 24.5$ with under-explored regions of galaxy color space in the COSMOS, EGS, and VVDS-2h fields. The data were taken with the LIRS, MOSFIRE, and DEIMOS instruments on the Keck I and II telescopes. The DEIMOS data cover $\sim 5000-10000$ Å with a spectral resolution of $R \sim 3000$. The MOSFIRE NIR data have a spectral resolution of $R \sim 3000$. The LRIS data cover $\sim 3200-10000$ Å with a spectral resolution ranging from $R \sim 300$ to $R \sim 5000$. \citep{Masters2017,Masters2019,Stanford2021}.  

FMOS-COSMOS was a NIR survey of 5,484 star-forming galaxies (as of data release 2) in the 1.7 deg$^2$ COSMOS field using the FMOS multi-object spectrograph on the Subaru Telescope. The primary targets were galaxies detected in Herschel far-infrared imaging. The data is intended to cover the H$\alpha$ emission line with a narrow wavelength range of $1.6-1.8$ $\mu$m and a spectral resolution of $R\sim2600$ \citep{Silverman2015,Kashino2019}. 

LEGA-C targeted 4081 $0.6 < z < 1.0$, $K_s$-selected galaxies in the COSMOS ACS field with the VIMOS multi-object spectrograph on the VLT. The data cover $\sim 0.6-0.9$ $\mu$m with a spectral resolution of $R\sim3500$ \citep{vanderWel2016,Straatman2018,vanderWel2021}.

DESI is an ongoing wide-field survey. The spectroscopic targeting sample includes galaxies, quasars, and Milky Way stars. A portion of the early data release includes a targeted campaign of $\sim 4,500$ Lyman break galaxies with $i_{\rm{AB}} < 24.5$ and quasars with $i_{\rm{AB}} < 23.5$ in the COSMOS field. The instrument cover $\sim 3600-9800$ Å with a spectral resolution ranging of $R \sim 2000-5500$ \citep{DESI2023}.  

SDSS is a wide-field spectroscopic survey that overlaps with the COSMOS field. The data cover galaxies and quasars with a variety of selection methods depending on the subset of the SDSS survey. The SDSS spectroscopy cover $\sim 3600-9800$ Å with a spectral resolution of $R\sim 2000$ \citep{Ahn2012,Almeida2023}. 

\subsection{Flux calibration and data cleaning}

In total, we have nearly 1000 spectra of varying quality and wavelength coverage. Below, we describe our procedure for calibrating and cleaning the data. A reasonable estimate of the uncertainties on the data is essential for the least squares minimization and $\chi^2$ estimation. Unfortunately, the flux uncertainties (error spectrum) are not provided for the DEIMOS and zCOSMOS data. In these cases where the error spectrum is not provided, we estimate the uncertainties using the median absolute deviation of the flux spectrum. Data quality (e.g. artifact, spectral gap) masks are used wherever possible. The fitting results of each spectrum is visually inspected. When multiple spectra are available for the same source, both with a valid BH mass estimate, we pick the spectrum with the highest signal-to-noise ratio. We avoided stacking the spectra from different surveys for two reasons. First, we avoided complications resulting from stacking spectra with varying spectral resolution. Second, the lower signal-to-noise spectra tend to have poorer spectrophotometric calibrations that would propagate through the stack.

Accurate flux calibration is important for obtaining virial BH masses, which depends on the continuum or broad-line luminosity. In order to minimize systemic differences between the spectra from different surveys and instruments, we first perform absolute flux calibration for each spectrum. We integrate the spectrum over the available COSMOS2020 bands to generate synthetic photometry. Then, we scale by the error-weighted mean ratio between synthetic photometry and HSC photometry. This procedure is commonly adopted in the literature (e.g. \citealt{Mallery2012}). Figure~\ref{fig:fluxing} demonstrates a significantly improved consistency between the spectral calibration for sources with multiple spectra from different surveys/instruments after performing absolute flux calibration with the COSMOS2020 photometry. Other issues in the data reduction, such as residual instrumental sensitivity, are difficult to correct. In addition, we have not corrected for the effect of variability in the spectra. We estimate a 1$\sigma$ uncertainty of $\sim 10$ percent in the final absolute flux calibration from the scatter in Figure~\ref{fig:fluxing}.

\subsection{Spectral modeling}

We fit the continuum and emission lines in each 1D spectrum using a modified version of the publicly-available \textsc{PyQSOFit} code \citep{Guo2018,Shen2019}, used to measure SDSS quasar properties \citep{Shen2011,Wu2022}. However, unlike the SDSS quasar sample, we found that many of our spectra had a significant underlying host galaxy component. This is not surprising given our much fainter quasar sample, which tend to be more host dominated, compared to the magnitude limit of SDSS quasar spectra of $i<20$. Therefore, we first perform a quasar/host galaxy decomposition using principle component analysis (PCA) with host galaxy templates \citep{Bruzual2003}. After subtracting any significant host galaxy component from the spectrum, the quasar continuum is modeled as a blue power-law plus a 3rd-order polynomial for reddening. Fe~II emission templates \citep{Vestergaard2001} are fitted if including them improves the reduced $\chi^2$ of the continuum fitting by 20 percent. The total model is a linear combination of the continuum and single or multiple Gaussians for the emission lines. Since uncertainties in the continuum model may induce subtle effects on measurements for weak emission lines, we first perform a global fit to the emission-line free region to better quantify the continuum. We then fit multiple Gaussian models to the continuum-subtracted spectra around the H$\beta$ and Mg~II emission line complex regions locally.

Variability is efficient at identifying low luminosity AGNs. Generally speaking, the host galaxy continuum component is significant in the spectra HSC-SSP variable AGNs. It is difficult to correctly decompose the underlying AGN continuum from the host galaxy given the lower AGN luminosities and lower continuum signal-to-noise ratios of the spectra owing to their higher redshifts and faintness. For this reason, we chose to use the broad-line luminosities instead of AGN continuum luminosities to estimate BH masses.

\subsection{Visual Spectral Classification}
\label{sec:vis}

In addition to our automated broad-line detection and fitting approach, we have visually inspected each spectrum to identify AGN signatures and present initial classifications. We classify sources with at least one broad line feature (``broad line''), sources without strong broad-line features but with Ne~V emission line indicating a probable AGN (``Ne~V''), and sources without either AGN feature (``host-dominated''), or low $S/N$ spectrum (``noisy''). Due to varying quality of the spectrophotometric calibrations, we do not attempt to identify AGN continuum features for non broad-line/Ne~V emitting sources. We additionally identify J095835.9+015156 (ID=280) as a broad absorption line (``BAL'') quasar. 252 of our sources have AGN features (broad line or Ne~V line), leaving 128 without obvious AGN-like features in their spectra (host-dominated or noisy). The classifications are presented in Table~\ref{tab:mass}. We caution that the absence of AGN spectral features that we have identified does not necessarily imply the absence of an AGN. For example, the spectrum could be in a host-dominated state due to variability, or the spectrum simply does not cover any of the broad emission lines or the Ne~V line given the redshift. A note on the nature of SN 1000+0216 (ID=451) is presented in Appendix~\ref{sec:451}.

\subsection{Black Hole Masses}

Following \citet{Shen2011}, we estimate the BH masses from broad-emission lines (e.g., \citealt{Greene2005}) using the single-epoch virial method. This method assumes that the broad-line region (BLR) is virialized and uses the continuum or broad-line luminosity and broad-line FWHM as a proxy for the BLR radius and virial velocity respectively. We use the broad-line rather than continuum luminosities to estimate BH masses. The continuum luminosities are not well-constrained for some of our sources given the spectral quality (Appendix~\ref{sec:conti}). Under these assumptions, the BH mass can be estimated by:
\begin{equation}
\begin{split}
    \log{\left(\frac{M_{\rm{BH}} }{M_{\odot}} \right)} = a + b &\log{ \left( \frac{L_{\rm{br}}}{10^{44}\ \rm{ erg\ s}^{-1}} \right) } \\&+ 2 \log{ \left( \frac{\rm{FWHM}_{\rm{br}}}{\rm{ km\ s}^{-1}} \right) }
\end{split}
\label{eq:BHmass}
\end{equation}
where $L_{\rm{br}}$ and FWHM$_{\rm{br}}$ are the broad-line luminosity and full-width-at-half-maximum (FWHM) with an intrinsic scatter of $\sim0.4$ dex in BH mass. The coefficients $a$ and $b$ are empirically calibrated against local AGNs with BH masses measured from reverberation mapping. We adopt the calibrations \citep{Vestergaard2006} from in \citet{Shen2011} derived by \citet{Shaw2012}:
\begin{equation}
    (a, b) = (1.63, 0.49), \quad \rm H\beta
\end{equation}
\vspace{-6mm}
\begin{equation}
    (a, b) = (1.70, 0.63), \quad \rm Mg\ II
\end{equation}
\vspace{-6mm}
\begin{equation}
    (a, b) = (1.52, 0.46), \quad \rm C\ IV.
\end{equation}
Following \citet{Shen2011}, depending on the wavelength coverage, redshift, and line $S/N$, we adopt a fiducial or preferred ``best'' BH mass following the ordering above. We only consider BH masses when the broad line component is detected over the residual with a $S/N > 2$ as defined in \citet{Burke2023blazar}. When more than one spectrum exists with a valid BH mass for a single source, we adopt the spectrum with the highest median per-pixel $S/N$. Although an extinction curve is fitted prior to measuring the line luminosities, the C~IV and Mg~II masses are expected to be more prone to intrinsic reddening than the H$\beta$ masses \citep{Shen2019}. Although these systematics may be partially folded into the virial coefficients (see \citealt{Shen2012}).

\subsection{X-ray Properties}

\begin{figure}
\centering
\includegraphics[width=0.5\textwidth]{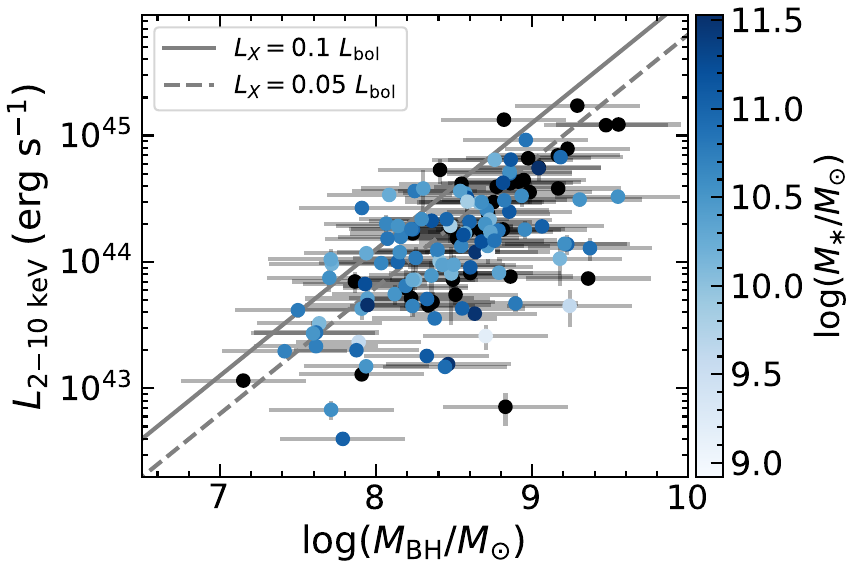}
\caption{2$-$10 keV X-ray luminosities against our measured broad-line black hole masses. The points are colored by their stellar mass when deemed reliable. We assume an uncertainty of $0.4$ dex on the black hole masses. \label{fig:xray}}
\end{figure}

\citet{Kimura2020} show that $\sim 90$ percent of their sources are detected in the X-ray. For non X-ray detected sources, their X-ray stacking analysis favors an AGN emission origin for the X-ray stacked sample. We have re-done the matching and plot the 2$-$10 keV X-ray fluxes from the Chandra-COSMOS Legacy Survey catalog \citep{Civano2016}. We convert the fluxes to luminosities (uncorrected for absorption) using:
\begin{equation}
L_{2-10\ {\rm keV}} = 4 \pi d^2\ (1+z)^{\Gamma-2} f_{2-10\ {\rm keV}},
\end{equation}
where $f_{2-10\ {\rm keV}}$ is the flux given in the Chandra-COSMOS Legacy Survey catalog. We take $\Gamma=1.8$, which is typical of low-luminosity AGNs (e.g., \citealt{Ho2009}).

We show the X-ray luminosities against our broad-line black hole masses in Figure~\ref{fig:xray}. \citet{Kimura2020} demonstrated that the X-ray luminosities are too high to be explained by X-ray binary populations. Assuming a typical bolometric correction of $L_{\rm{bol}}/L_{2-10\ \rm{{keV}}} = 10$ \citep{Duras2020}, the sources have a typical (median) Eddington ratio of $\sim 0.05$, slightly lower than the median Eddington ratio of $\sim 0.1$ of the sample from \citet{Suh2020}. Our estimated Eddington ratios and black hole masses are broadly reasonable given the correlation between these parameters and optical variability amplitude (e.g. \citealt{MacLeod2010}) and the typical variability amplitude of the HSC-SSP AGNs of $\sim 0.1$ magitudes. 

\section{$M_{\rm{BH}}-M_{\ast}$ relation} \label{sec:relation}

\begin{figure*}
\centering
\includegraphics[width=0.95\textwidth]{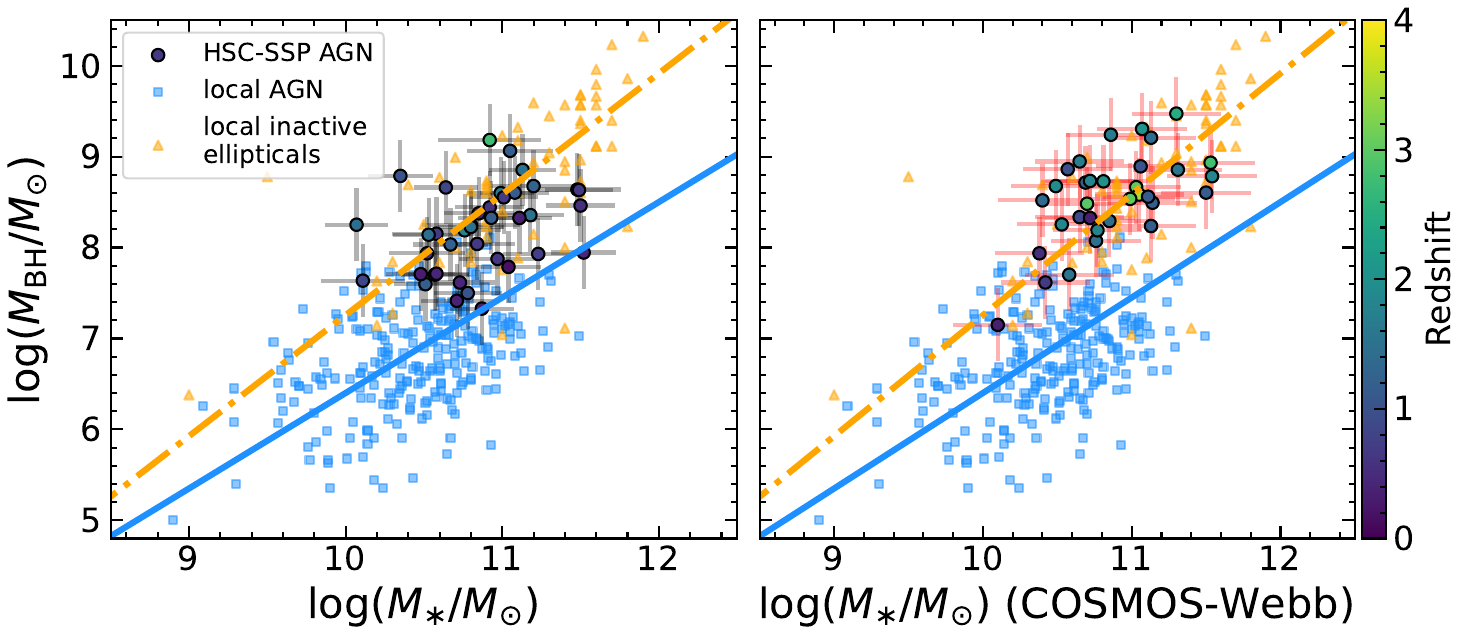}
\caption{$M_{\rm{BH}}-M_{\ast}$ relation for the HSC-SSP AGNs with reliable BH mass estimates from this work, and reliable stellar masses from this work (\emph{left}) or from AGN+host decomposition from \emph{HST} and COSMOS-Webb archival imaging from \citet{Zhuang2023} (\emph{right}). Points are shaded by redshift. We assume uncertainties of $\sim 0.4$ dex on the BH masses and $\sim0.4$ dex on the COSMOS-Webb stellar masses. The local $M_{\rm{BH}}-M_{\ast}$ relation and data for AGNs is plotted as the solid blue line and square symbols \citep{Reines2015} and for inactive elliptical galaxies as the dash-dotted orange line and orange triangle symbols \citep{Greene2020}. 
\label{fig:relation}}
\end{figure*}

\begin{figure*}
\centering
\includegraphics[width=0.95\textwidth]{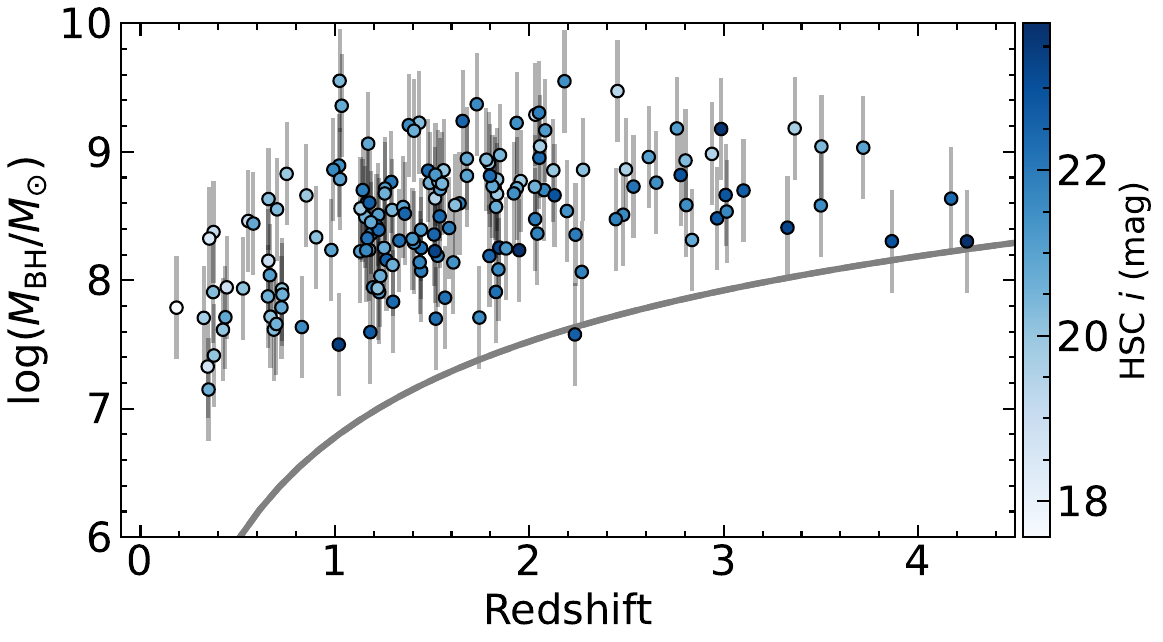}
\caption{Black hole mass versus redshift for HSC-SSP variable AGNs with reliable broad-line BH mass estimates from broad-band SED fitting using \textsc{cigale}. Each AGN is shaded by its \emph{HSC} $i$-band apparent magnitude. The gray curve is the theoretically predicted detection limit, which is deeper than the BH mass detection limit. \label{fig:bhmassredshift}}
\end{figure*}

We show the $M_{\rm{BH}}-M_{\ast}$ relation for our sample of variability-selected AGNs with reliable stellar masses from COSMOS2020 photometry and virial black hole masses measured from our spectroscopic database in Figure~\ref{fig:relation}. We also show the $M_{\rm{BH}}-M_{\ast}$ relation using the \emph{HST} and \emph{JWST} stellar masses from \citet{Zhuang2023} and virial black hole masses measured from our spectroscopic database. In either case, the $M_{\rm{BH}}-M_{\ast}$ relation for our sample of $z\sim0.5-4$ (median redshift of $\sim 1.5$) AGNs is more consistent with the relation for local inactive elliptical galaxies \citep{Greene2020} than local ($z<0.055$) AGNs \citep{Reines2015}. However, the variability-selected AGNs show a possible trend toward the local AGN relation at lower redshifts. In contrast, the \citet{Zhuang2023} sources are highly consistent with the relation for local inactive elliptical galaxies. This difference could be due to having more $z\lesssim1$ and lower luminosity AGNs (with relatively more host-dominated SEDs) in our figure panel.

At face value, this result could indicate an evolution in the $M_{\rm{BH}}-M_{\ast}$ relation for AGNs with redshift. Such an evolution could be connected to the host star-formation or black hole accretion activity (the densities rates of both peak near $z \sim 2$) \citep{Zhuang2023nat}. On the other hand, \citet{Bongiorno2012} studied the star-formation rates and AGN activity in a sample of AGNs selected from X-ray and optical spectroscopy up to $z\sim3$ in the COSMOS field, finding no strong evidence for a connection between the AGN activity and star-formation processes in their host galaxies. However, \citet{Hickox2014} point out that the weak correlations between observed AGN properties and host SFR or stellar mass could be explained by the much shorter $\sim$ Myr timescales of AGN activity compared to the $\sim 100$ Myr timescales of star formation in galaxies. These weak trends (see also \citealt{Lutz2010,Bonfield2011}) could also partly be due to large scatter in the stellar mass and SFR estimates (see \S\ref{sec:suh}).

\subsection{Selection Bias}

Measurements of the $M_{\rm{BH}}-M_{\ast}$ relation are strongly influenced by selection biases at low and high redshift \citep{Lauer2007,Shen2010,Shankar2019} which we discuss below. The probability of a source being included in our $M_{\rm{BH}}-M_{\ast}$ relation sample is given by the probability that (a) the source is variable and detected, (b) the source has a spectrum and the broad-line is detected in the spectrum, (c) the probability that the stellar mass measurement is reliable. Our spectroscopic sample contains spectra from a variety of survey and quasar targeting programs with different targeting criteria, flux limits, and spectral sensitivities. This makes it difficult quantify the effect of the selection biases on our result. \citet{Lauer2007} and \citet{Shen2010} point out several selection biases that can result in a false evolution of the $M_{\rm{BH}}-M_{\ast}$ relation. Inevitably, we have selected a sample of sources based on AGN activity, which biases our sample toward more luminous and larger black hole masses as redshift increases. Given the spectral sensitivity limit of $\sim 10^8 M_{\odot}$ and lack of obvious dependence on the stellar mass reliability with stellar mass (Figure~\ref{fig:xray}), the resulting $M_{\rm{BH}}-M_{\ast}$ relation seems unlikely to be dominated by selection biases (b) and (c). Given other authors have found a similar relation for X-ray and spectroscopically-selected AGNs at similar redshifts for higher AGN luminosities \citep{Merloni2010,Zhuang2023}, it seems unlikely that the variable population is probing a significantly different parameter space of the Type 1 AGN population. A model of the detectable AGN population at these redshifts may be able to quantify effects of the selection biases on our particular sample (e.g., \citealt{Pacucci2023}), but requires knowing the true intrinsic scatter of the $M_{\rm{BH}}-M_{\ast}$ relation. Modeling these selection biases is well beyond the scope of this work.

\section{Discussion} \label{sec:discussion}

\begin{figure*}
\centering
\includegraphics[width=0.99\textwidth]{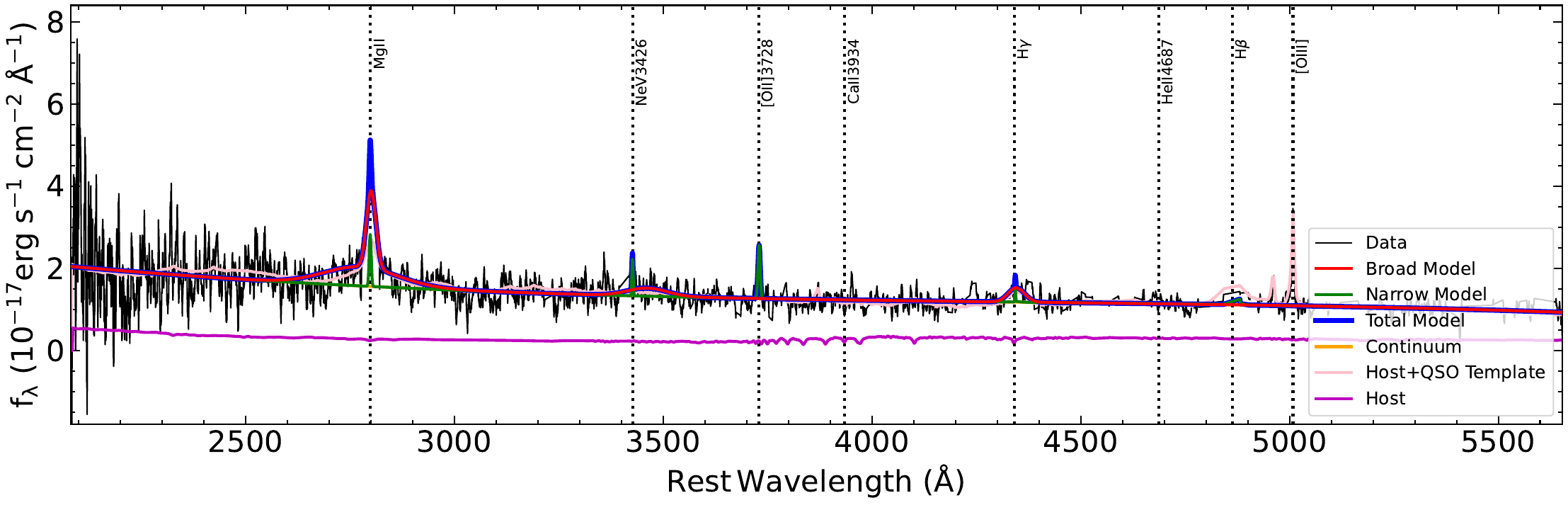}
\caption{Spectral fitting results for \citet{Kimura2020} ID 290 ($M_{\ast} \sim 10^{10} M_{\odot}$). The DESI spectral data (black) is plotted with the model fitting components: polynomial+power-law continuum (orange), total line model flux (blue), broad line model flux (red), narrow line model flux (green) fitted host+QSO template (pink) and fitted host template (magenta). In this case, the PCA-derived host flux is small. The resulting broad-line black hole mass is $M_{\rm{BH}} \sim 10^8 M_{\odot}$ at $z=0.7300$. \label{fig:spec}}
\end{figure*}

\subsection{IMBHs in low stellar mass galaxies at similar redshifts}

Intriguingly, our results in combination with estimates at higher redshift suggest redshift evolution in the $M_{\rm{BH}}-M_{\ast}$ relation. For example, the recently proposed model of \citet{Pacucci2024}, motivated by higher redshift over-massive black holes seen in \emph{JWST} data (e.g., \citealt{Harikane2023,Maiolino2023,Kocevski2023,Ubler2023,Kokorev2023,Bogdan2023,Natarajan2023,Goulding2023}), predict that at lower redshifts $z \sim 0.5 - 3$ black holes are expected to be $3 - 10$ times more massive than the local AGN relation, similar to what is seen in our sample. In the \citep{Pacucci2024} model, it is argued the $M_{\rm{BH}}/M_{\ast}$ ratio sets the average star formation efficiency in the galaxy, until the galaxy is no longer able to efficiently form stars, bringing it into agreement with the local $M_{\rm{BH}}-M_{\ast}$ relation. Earlier models of BH-galaxy coevolution proposed by \citet{Wyithe2003,Caplar2018} predict similar levels of redshift evolution in the $M_{\rm{BH}}-M_{\ast}$ relation attributed to self-regulated feedback before quenching at a critical value of $M_{\rm{BH}}/M_{\ast}$. While all these models attempt to modulate the SFRs to account for stellar assembly in BH host galaxies, they do not address the key issue of the relationship between the BH accretion rate and the SFR, which likely holds the key to co-evolution. The detailed dependence of this ratio $\dot{M}_{\rm{acc}}/{\dot{M}_\ast}$ on feedback; the environment and gas content of galaxies remains to be understood.

In order to build volume and given that the density and accretion rates of black holes peak near $z \sim 2$, several studies have identified IMBH candidates in dwarf galaxies beyond $z=0$. The higher than expected black hole masses than the local AGN relation, whether due to selection effects or intrinsic evolution in the $M_{\rm{BH}}-M_{\ast}$ relation, unfortunately casts doubt on the IMBH nature of previously-identified $L_{2-10\ \rm{{keV}}} \sim 10^{43-44}$ erg s$^{-1}$ X-ray AGNs at similar redshifts \citep{Mezcua2018,Mezcua2019,Zou2023}. Simply extrapolating the local inactive $M_{\rm{BH}}-M_{\ast}$ relation from Figure~\ref{fig:relation}, we expect an AGN with stellar mass of $M_{\ast} \sim 10^{10} M_{\odot}$ to have a black hole mass of $M_{\rm{BH}} \approx 10^{7.2} M_{\odot}$ rather than $M_{\rm{BH}} \approx 10^{6.4} M_{\odot}$ at similar redshifts.

We have visually inspected the spectra for dwarf galaxies in our sample with $M_{\ast} < 10^{10} M_{\odot}$ with reliable stellar mass estimates from SED fitting. One source, \citet{Kimura2020} ID 290 ($M_{\ast} = 10^{9.6\pm0.9} M_{\odot}$), has a sufficiently-detected broad Mg~II line with a black hole mass estimate of $M_{\rm{BH}} \sim 10^{7.89\pm0.05} M_{\odot}$ (statistical fitting error) at $z=0.73$ (Figure~\ref{fig:spec}). Five do not have a spectrum, and the remainder are either too host-dominated or noisy to obtain a broad-line black hole mass estimate. This highlights the need for higher $S/N$ spectroscopy for detecting broad lines in these relatively more host-dominated and fainter variability-selected AGNs in lower mass galaxies.


\subsection{Biases in AGN Stellar Mass Estimation and the $M_{\rm{BH}}-M_{\ast}$ Relation} \label{sec:suh}

\begin{figure*}
\centering
\includegraphics[width=0.48\textwidth]{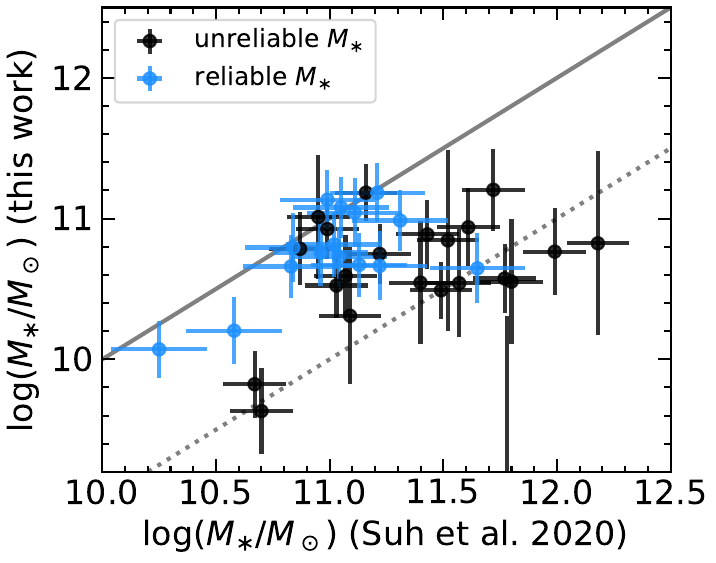}
\includegraphics[width=0.48\textwidth]{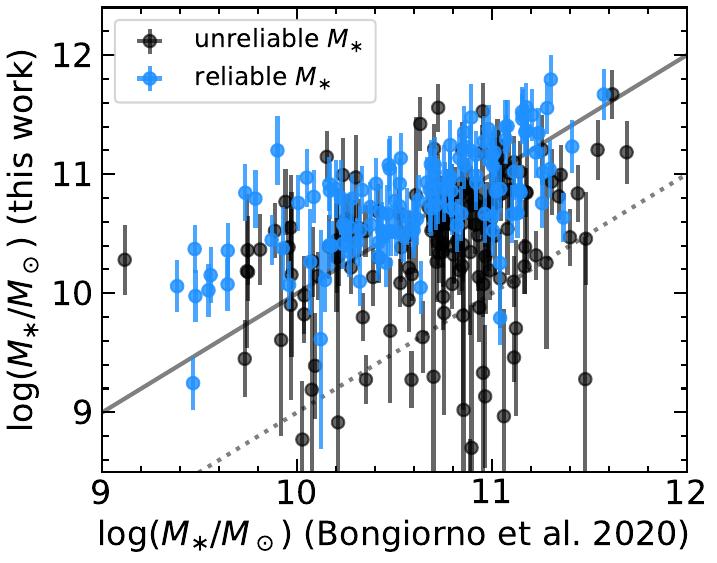}
\caption{Comparison of stellar masses estimated by \citet{Suh2020} (\emph{left}) and \citet{Bongiorno2012} (\emph{right}) and our stellar masses when our stellar mass estimates are considered reliable (blue symbols) or unreliable (black symbols). The solid gray line is the line of $y=x$, and the dotted line falls 1 dex below the line of $y=x$. \label{fig:suh}}
\end{figure*}

\cite{Suh2020} found a sample of $z\sim0.5-2.5$ X-ray selected AGN to be consistent with the local AGN relation of \citet{Reines2015}, contrary to our findings and the findings of several other authors \citep{Merloni2010,Mezcua2023,Li2021,Zhang2023,Zhuang2023,Stone2023,Tanaka2024}. \citet{Zhuang2023} have used black hole masses taken from \cite{Suh2020}, and their black hole masses are consistent with ours given the systematic uncertainties in virial mass estimates and the spectral calibration. In order to scrutinize the possible origin of this discrepancy, we compare our stellar masses to those from \cite{Suh2020} in the left panel of Figure~\ref{fig:suh}. We find that the \cite{Suh2020} stellar masses are over-estimated on average compared to our stellar masses and those estimated from COSMOS-Webb AGN-host decomposition by \citet{Zhuang2023}. In some cases, the \citet{Suh2020} stellar masses are too large by up to an order of magnitude. The sources with unreliable stellar masses due to being swamped by AGN emission are more likely to be over-estimated by \citet{Suh2017,Suh2019,Suh2020}. We attribute this discrepancy to inadequate separation of star-formation and AGN-dominated SEDs. There is also a slight tendency for even the reliable stellar masses to be over-estimated. It is possible this originates from their requirement that the UV part of the SED be AGN dominated, which can under-count the contribution from young stars even when the UV AGN fraction is low. The on-average over-estimated stellar masses by \citet{Suh2020} would explain why their sources appear to fall on the local AGN $M_{\rm{BH}}-M_{\ast}$ relation, in contrast with our results. We conclude that the \citet{Suh2020} stellar masses are significantly over-estimated on average. This highlights the extreme caution that must be taken when estimating stellar masses for AGNs without decomposing the AGN+host emission.

\citet{Bongiorno2012} studied the host galaxy properties if both Type 1 and 2 AGNs in the COSMOS field selected from optical spectroscopic and X-ray selection. They derive SED fitting-based stellar masses and SFRs using a similar approach to differentiate between AGN and star-formation dominated SEDs in the NIR. We compare our stellar masses to those from \citet{Bongiorno2012} in the right panel of Figure~\ref{fig:suh}. We only consider sources with constrained stellar mass measurements according to \citet{Bongiorno2012}. First, it is evident that a large fraction of the \citet{Bongiorno2012} stellar mass measurements are considered unreliable according to our analysis, having a low ratio between the total model fit for our AGN$+$stellar emission SED model and the AGN-dominated model. Second, it appears that the \citet{Bongiorno2012} stellar masses are systematically lower than our stellar masses below our $M_{\ast} \sim 10^{10.8} M_{\odot}$ for those stellar masses that we consider to be reliable. To reassure that this is not an issue with our stellar masses, we confirm that a substantial fraction (17/96 with $\Delta M_{\ast} > 0.5$ dex) of the \citet{Bongiorno2012} stellar masses are also under-estimated compared to those from COSMOS-Webb AGN-host decomposition \citep{Zhuang2023}. \citet{Bongiorno2012} have used the AGN template from \citet{Richards2006}, while the \textsc{cigale} AGN models are more flexible while retaining energy balance. \citet{Bongiorno2012} note that their SED fitting results tend to over-estimate the AGN component, which would in-turn under-estimate the stellar emission component. This could likely be the source of under-estimated stellar masses. 

Recently, \cite{Hoshi2024} studied the $M_{\rm{BH}}-M_{\ast}$ relation for the same parent sample \citep{Kimura2020}. Their results are broadly consistent with ours, with their $M_{\rm{BH}}/M_{\ast}$ ratios being more consistent with the inactive early-type relation. The much larger scatter in their $M_{\rm{BH}}-M_{\ast}$ relation compared to our results, and especially the COSMOS-Webb $M_{\rm{BH}}-M_{\ast}$ relation, could be due to less accurate stellar masses (less robust accounting for the AGN component and sources where no reliable stellar mass is feasible) or less accurate BH masses (lack of absolute spectral flux calibration and possible unreliable continuum measurements for some sources) compared to our work.

\subsection{Forecasts for Rubin Observatory}

The expected single-epoch limiting magnitude of $r\sim24.5$ for LSST is more shallow than the HSC-SSP limiting magnitude of $r\sim25.5$. The single-epoch liming magnitude per exposure depends mostly the exposure time, sky brightness, seeing, and airmass. The single-epoch liming magnitude scales logarithmically with the effective ``visit'' exposure time as $\propto 1.25\ \log_{10}(t_{\rm{vis}})$ \citep{Ivezic2019}. For example, co-adding an additional 20 visits in the deep drilling fields (in e.g., groups of a few nights) would increase the single-epoch imaging depth by about 1.6 magnitudes, potentially matching or out-performing HSC-SSP in depth. We recommend taking this approach early-on in the survey, which will significantly expand the discovery space to identify faint variables in early Rubin data. After several years of operation, these light curves may be sufficiently long to allow for estimation of the BH mass from the variability timescale \citep{Burke2021}. This would potentially circumvent the need for spectroscopic follow-up to estimate broad-line BH masses.

\section{Conclusions}  \label{sec:conclusion}

Using the sample of variable AGNs selected from the HSC-SSP COSMOS field \citep{Kimura2020}, we have obtained improved photometric redshifts and multi-wavelength photometry from the COSMOS2020 catalog and used SED fitting to estimate their stellar masses. We have devised an approach for determining the reliability of the stellar mass estimates using the stellar emission strength at 1.2 $\mu$m, where the AGN emission is expected to be at a minimum. After constructing a database of publicly-available spectra from the literature, we measured their virial black hole masses from the detected broad emission lines when $S/N$ permitted. We compared our stellar masses to those from those estimated from AGN-host decomposition using \emph{HST} and \emph{JWST} imaging \citep{Zhuang2023} and measured the $M_{\rm{BH}}-M_{\ast}$ relation for our variable AGNs at $0.5 \lesssim z \lesssim 3$. Our results concur with previous findings using other AGN samples of more massive black holes at a given stellar mass than the local AGN relation would suggest. These results suggest that AGNs selected from optical variability are not vastly different from samples of AGNs selected from broad lines at similar redshifts \textbf{at fixed luminosity}.

Using these results as a proxy for LSST Rubin capability, we have demonstrated that black holes with $M_{\rm{BH}} \sim 10^8 M_{\odot}$ are detectable out to at least $z\sim4$ in $M_{*} \sim 10^{11} M_{\odot}$ host galaxies using optical variability. Future work combining Rubin-selected AGNs and Roman host galaxy imaging will vastly increase the sample size of sources with reliable stellar masses estimate from AGN$+$host decomposition. Follow-up analysis of this sample will apply \texttt{Scarlet} source deblending \citep{Melchior2018} to high-resolution time-resolved HSC COSMOS imaging in order to extract the SEDs of the variable AGN and their host galaxies, as well as the host galaxy morphologies, providing further details on host galaxy properties for AGN-dominated cases (Ward et al., in prep). High resolution imaging from JWST could be incorporated into this analysis framework to further improve the AGN and host decomposition, especially at high redshifts.

We are currently performing a similar analysis with variable AGNs from the Dark Energy Survey deep fields \citep{Burke2022des} and an ongoing repeat imaging survey of DES and Rubin deep drilling fields with the Dark Energy Camera (Zhuang et al. in prep). This catalog is not as deep as HSC-SSP, with a single-epoch photometric precision of $g \sim 24.5$, but is a larger area of $\sim 4.6$ deg$^2$, which will allow us to fill in the space at $z<1.5$. This regime is interesting because it corresponds to the epoch when the globally averaged SFR has declined revealing that the overall gas supply available in galaxies for both star formation and accretion have dwindled and the merger rate of galaxies - additional mechanism for injecting gas into galactic nuclei - has also declined. Besides, this epoch lies conveniently in between the local and intermediate-redshift regimes.

We have not investigated narrow emission line ratio diagnostics in this paper. Given the varying wavelength coverage and redshifts of the spectra, the majority of the spectra do not cover both the H$\beta$ and H$\alpha$ spectral complexes. The majority of the sources have AGN features in their spectra (either a broad line detection or Ne V emission line), but a significant fraction are either too noisy or host-dominated to detect any obvious AGN features. Some of the latter could be false positives (i.e., non-AGN galaxy or transient interlopers). This work highlights the challenges of obtaining sufficient spectroscopy to investigate low luminosity AGNs at the redshifts that LSST Rubin will unveil. However, we are optimistic on the capability of optical variability to identify a relatively lower luminosity samples of Type 1 AGNs at low and intermediate redshifts.

\section*{Acknowledgements}

We thank the anonymous referee for a careful review and for providing comments which improved this work. This research has made use of the SIMBAD database, operated at CDS, Strasbourg, France. This research made use of \textsc{astroquery} \citep{Ginsburg2019}. We are grateful to Ming-Yang Zhuang, Guang Yang, Kedar Phadke, Qian Yang, Meg Urry, Marla Geha, and Yue Shen for useful discussions. CJB is supported by an NSF Astronomy and Astrophysics Postdoctoral Fellowship under award AST-2303803. This material is based upon work supported by the National Science Foundation under Award No. 2303803. This research award is partially funded by a generous gift of Charles Simonyi to the NSF Division of Astronomical Sciences. The award is made in recognition of significant contributions to Rubin Observatory’s Legacy Survey of Space and Time. 
Y.L. and X.L. acknowledge support from NSF grant AST-2308077. P.N. acknowledges support from the Gordon and Betty Moore Foundation and the John Templeton Foundation that fund the Black Hole Initiative (BHI) at Harvard University where she serves as one of the PIs.



\appendix

\section{DATA AVAILABILITY} \label{sec:availibility}

The HSC PDR3 redshift catalog is available at: \\ \url{https://hsc-release.mtk.nao.ac.jp/doc/index.php/catalog-of-spectroscopic-redshifts__pdr3/}.

\noindent The COSMOS2020 catalogs are available at: \url{https://cosmos2020.calet.org}.

\noindent The publicly-available spectra were collected from the following web pages:
\vspace{-2mm}
\begin{enumerate}
    \item zCOSMOS, VUDS, DEIMOS, Magellan: \url{https://irsa.ipac.caltech.edu/data/COSMOS/spectra/}
    \item C3R2: \url{https://koa.ipac.caltech.edu/Datasets/C3R2/}
    \item DESI: \url{https://data.desi.lbl.gov/public/edr/}
    \item FMOS-COSMOS: \url{https://member.ipmu.jp/fmos-cosmos/FMOS-COSMOS.html}
    \item LEGA-C: \url{https://users.ugent.be/~avdrwel/research.html}
    \item SDSS: \url{https://dr18.sdss.org/optical/plate/search}
\end{enumerate}

\section{STELLAR MASS RECOVERY TESTS} \label{sec:masstests}

\begin{figure*}
\centering
\includegraphics[width=0.49\textwidth]{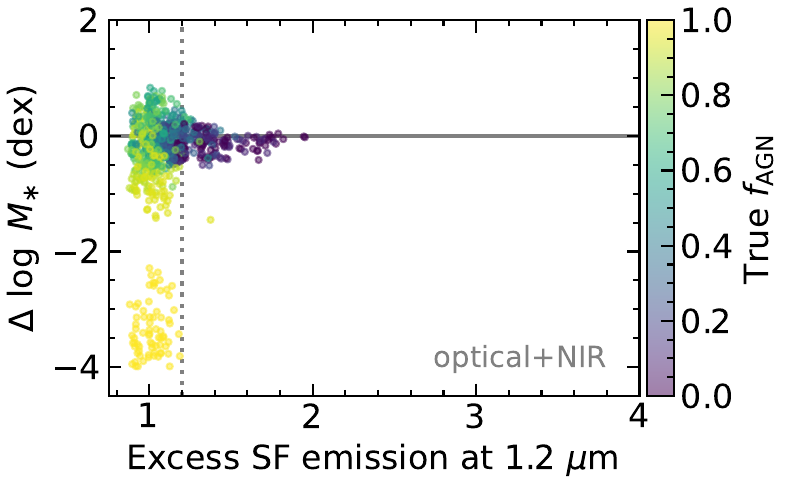}
\includegraphics[width=0.49\textwidth]{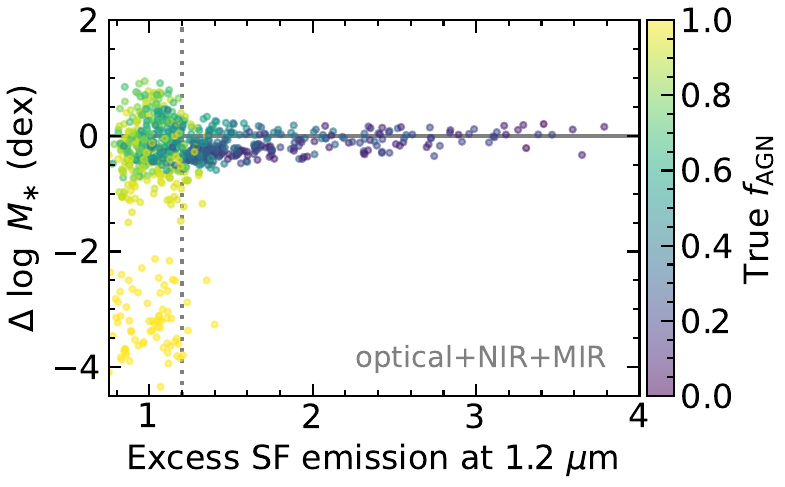}
\includegraphics[width=0.49\textwidth]{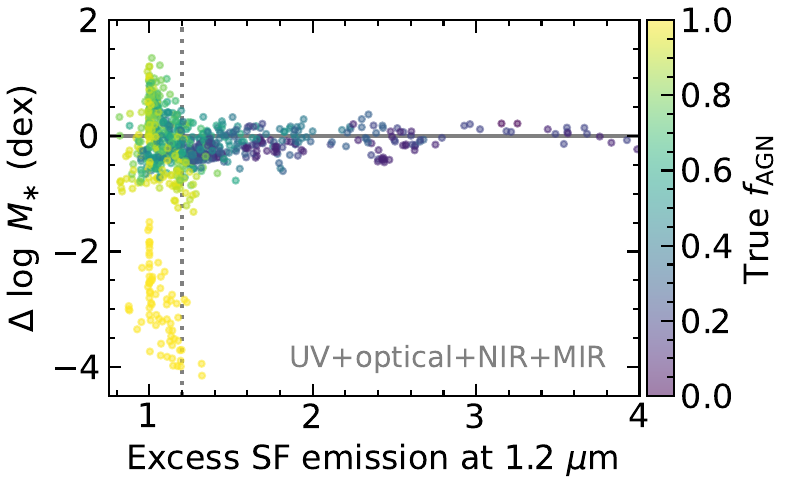}
\includegraphics[width=0.49\textwidth]{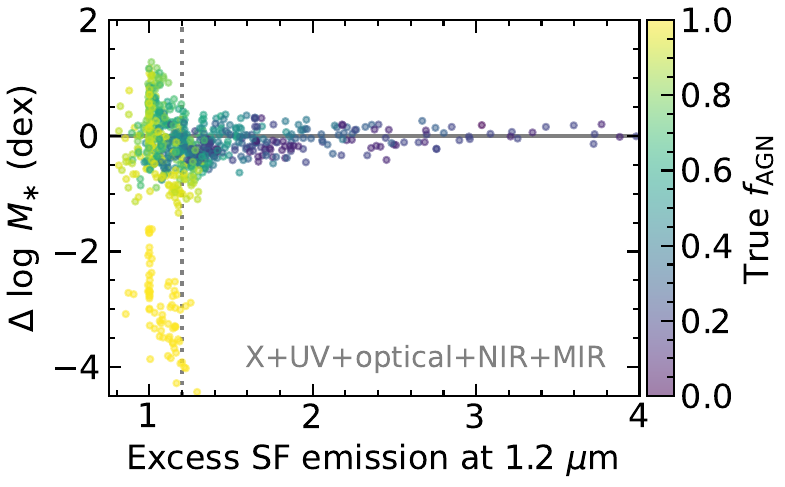}
\caption{Recoverability of stellar mass in AGNs using mock SEDs with varying multiwavelength coverage generate by \textsc{cigale}. The difference between the log true stellar mass and log recovered stellar mass (in dex) is shown versus the ratio of the AGN continuum emission of an AGN-dominated SED and the total continuum emission of an mixed star-formation $+$ AGN SED calculated near $1.2$ $\mu$m. Points are colored by their true AGN fraction calculated between $0.5-1$ $\mu$m. Sources with a very high AGN fraction ($f_{\rm AGN } \gtrsim 0.8$) are generally well-fitted by an AGN-dominated model which makes inferring the stellar mass difficult. Sources with true $f_{\rm AGN } \sim 1$ are very prone to catastrophically-underestimated stellar masses. In this case, the AGN emission swamps the star-formation emission, and the stellar mass is not well constrained. The inclusion of additional wavelength coverage helps to distinguish between the two different models, constraining the AGN emission, and leading to a more robust stellar mass estimate. \label{fig:Mstarrecovery}}
\end{figure*}

Degeneracies between AGN and stellar emission can lead to highly unreliable stellar mass estimates from SED fitting, depending on the strength of the AGN emission compared to the underlying stellar emission (i.e., the AGN fraction $f_{\rm AGN }$). Although \textsc{cigale} can infer the AGN fraction, it has been shown that quantity cannot be reliably inferred except in AGN-dominated sources ($f_{\rm AGN } \gtrsim 0.8$). For example, a young stellar population can be fitted to a UV/optical AGN continuum emission, leading to catastrophically-underestimated stellar masses. In order to overcome this degeneracy, we separate our sample into sources with ``reliable'' and ``unreliable'' stellar mass estimates using a model comparison test. Our approach is summarized in \S\ref{sec:Mstarrecovery}, and we describe it in detail below:
\begin{enumerate}
    \item Fit the observed SED with a totally AGN-dominated model by setting $f_{\rm{AGN}} = 0.9999$, computed over the wavelength range $0.5-1$ $\mu$m. 
    \item Fit the observed SED with an AGN$+$SF model with $f_{\rm{AGN}}$ as a free parameter varying between $0-1$, computed at observed-frame $0.1-0.3$ $\mu$m. This model assures that the UV component of the SED is AGN-dominated, but allows the redder optical and near-infrared part of the SED to be AGN or SF dominated. 
    \item Compute the ratio of the the best-fit total continuum emission from the AGN$+$SF model over the best-fit AGN continuum emission from the AGN-dominated model at rest-frame 1.2 $\mu$m (where the AGN emission is at a minimum).
    \item Recovered stellar masses from the AGN$+$SF model are considered reliable if its best-fit reduced $\chi^2 <5$ and the excess stellar emission from (3) is greater than 1.2, as justified using mock tests.
\end{enumerate}

In order to test this procedure, we use the ``savefluxes'' mode of \textsc{cigale} to generate mock photometry at varying AGN fractions. We allow the stellar and AGN parameters as in Table~\ref{tab:cigale}. We then run the mock photometry through the same \textsc{cigale} fitting procedure and stellar-mass reliability tests as our real data. The recovered stellar masses given the true mock photometry is shown in Figure~\ref{fig:Mstarrecovery}. The value of 1.2 is chosen to eliminate $>95\%$ of the catastrophic failures and does not depend strongly on the wavelength coverage (or redshift) of the SED. Although increased wavelength coverage does help increase the number of sources that meet this reliable stellar mass criteria.

The obvious limitation of this test is that it does not include systematic uncertainties in the recovered stellar masses beyond the parameter choices in the mock SEDs we have generated. Nevertheless, our results are broadly consistent with detailed tests of stellar mass estimates from SED fitting of AGNs using independently-calculated photometry from detailed star formation histories \citep{Ciesla2015}. We have also used stellar masses of our AGNs from spatially-decomposed imaging in the literature as an independent check on our stellar masses in Figure~\ref{fig:cosmoswebb}. 

We use the ``Bayesian-like'' parameter estimates and uncertainties provided by \textsc{cigale} throughout this paper. These parameters are estimated by weighting each solution by the $\exp(\chi^2/2)$ likelihood \citep{Boquien2019}. The parameters and uncertainties are then estimated from the likelihood-weighted mean and standard deviation. Artificially small $\chi^2$ values from under-estimated photometric uncertainties can result in under-estimated uncertainties on the recovered parameters, such as stellar mass. In severe cases, the best-fitted SED and its associated best-fit stellar mass may be biased. We caution against using these small stellar mass uncertainties estimated by \textsc{cigale}, and opt to assume the typical systematic uncertainty of $\sim 0.3$ dex in the figures. We checked that despite the under-estimated photometric errors in the NASA Sloan Atlas and many reduced $\chi^2$ values below 1, the \citet{Reines2015} sources have very reliable likelihood-weighted stellar masses (rms dispersion of $\sim 0.16$ dex) and are generally more reliable than the best-fit stellar mass (rms dispersion of $\sim 0.21$ dex) when compared to the \citet{Reines2015} estimates. For this reason, we adopt the ``Bayesian-like'' quantities.

\section{RELIABILITY OF SPECTRAL CONTINUUM MEASUREMENTS} \label{sec:conti}

\begin{figure}
\centering
\includegraphics[width=0.5\textwidth]{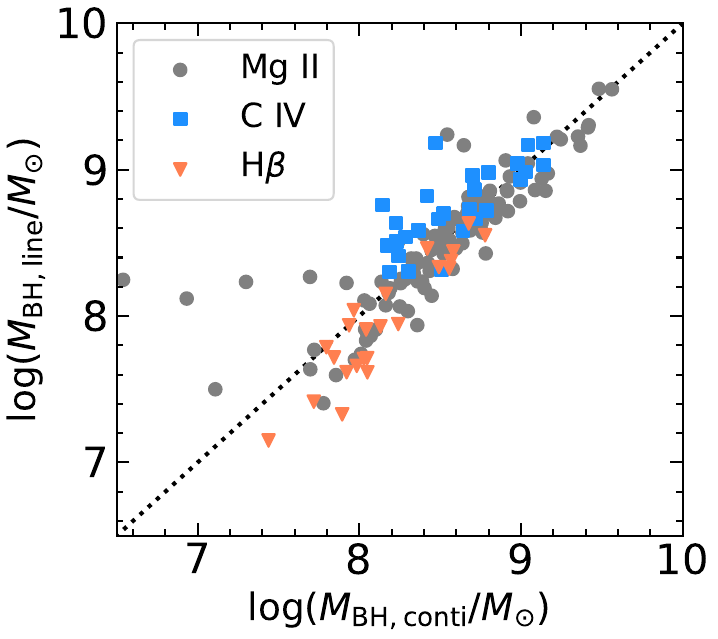}
\caption{Black hole masses estimated from the broad-line luminosity versus ($y$ axis) AGN continuum luminosity ($x$ axis). The black hole masses from the AGN continuum luminosity are susceptible to a few catastrophically under-estimated masses when the AGN continuum is not well constrained. \label{fig:conti}}
\end{figure}

AGNs with bolometric luminosities $L_{\rm{bol}} \lesssim 10^{45}$ erg s$^{-1}$ tend to have significant contribution from the host galaxy \citep{Shen2011,Kimura2020}. Reliably constraining the quasar continuum luminosity is essential to obtaining a virial black hole mass estimate using the prescriptions that use the continuum luminosity as a proxy for the BLR luminosity \citep{Shen2011}. PCA decomposition can constrain the quasar continuum for sources below $L_{\rm{bol}} \sim 10^{45}$ erg s$^{-1}$ for high quality spectra. However, our spectral quality (calibration and $S/N$) vary considerably depending on the instrument and spectral reduction. In order to test whether the quasar continuum luminosities are well-constrained, we computed the virial black hole masses using both the broad-line luminosity (Equation~\ref{eq:BHmass}) or continuum luminosity approaches using the equation:
\begin{equation}
\begin{split}
    \log{\left(\frac{M_{\rm{BH}} }{M_{\odot}} \right)} = a + b \log{ \left( \frac{\lambda L_\lambda}{10^{44}\ \rm{ erg\ s}^{-1}} \right) } + 2 \log{ \left( \frac{\rm{FWHM}_{\rm{br}}}{\rm{ km\ s}^{-1}} \right) }
\end{split}
\end{equation}
where $\lambda L_\lambda$ and FWHM$_{\rm{br}}$ are the continuum luminosity and broad-line full-width-at-half-maximum (FWHM) with an intrinsic scatter of $\sim0.4$ dex in BH mass.  We adopt the calibrations \citep{Vestergaard2006} used in \citet{Shen2011}:
\begin{equation}
    (a, b) = (0.910, 0.50), \quad \rm H\beta
\end{equation}
\vspace{-6mm}
\begin{equation}
    (a, b) = (0.740, 0.62), \quad \rm Mg\ II
\end{equation}
\vspace{-6mm}
\begin{equation}
    (a, b) = (0.660, 0.53), \quad \rm C\ IV.
\end{equation}

If the quasar continuum luminosities are well constrained, we expect the two approaches to yield consistent results within systematic uncertainties between the two prescriptions. We found that the black hole masses from the AGN continuum luminosity are susceptible to a few catastrophically under-estimated masses when the AGN continuum is not well constrained, as shown in Figure~\ref{fig:conti}. For consistency and simplicity, we adopt black hole masses estimated from the broad line luminosity from Equation~\ref{eq:BHmass}.

\section{SN 1000+0216}
\label{sec:451}

One source, SN 1000+0216 (ID=451), is reported as a superluminous supernova (SLSN) by \citet{Cooke2012} based on variability from years 2005$-$2008. The $z=3.8993$ source does not have obvious SN or AGN features in the spectrum presented in \citet{Cooke2012}. It is possible the spectrum was taken in a host-dominated state, i.e., when the AGN luminosity happened to be at a minimum. Since the source is still variable in 2014$-$2017 HSC-SSP light curves, the quasar interpretation may be more likely. A detailed study of this source and further scrutiny of its light curve is required to reach a firm conclusion.


\bibliography{aastex631}{}

\begin{thebibliography}{}
\expandafter\ifx\csname natexlab\endcsname\relax\def\natexlab#1{#1}\fi
\providecommand{\url}[1]{\href{#1}{#1}}
\providecommand{\dodoi}[1]{doi:~\href{http://doi.org/#1}{\nolinkurl{#1}}}
\providecommand{\doeprint}[1]{\href{http://ascl.net/#1}{\nolinkurl{http://ascl.net/#1}}}
\providecommand{\doarXiv}[1]{\href{https://arxiv.org/abs/#1}{\nolinkurl{https://arxiv.org/abs/#1}}}

\bibitem[{{Ahn} {et~al.}(2012){Ahn}, {Alexandroff}, {Allende Prieto},
  {Anderson}, {Anderton}, {Andrews}, {Aubourg}, {Bailey}, {Balbinot}, {Barnes},
  {Bautista}, {Beers}, {Beifiori}, {Berlind}, {Bhardwaj}, {Bizyaev}, {Blake},
  {Blanton}, {Blomqvist}, {Bochanski}, {Bolton}, {Borde}, {Bovy}, {Brandt},
  {Brinkmann}, {Brown}, {Brownstein}, {Bundy}, {Busca}, {Carithers}, {Carnero},
  {Carr}, {Casetti-Dinescu}, {Chen}, {Chiappini}, {Comparat}, {Connolly},
  {Crepp}, {Cristiani}, {Croft}, {Cuesta}, {da Costa}, {Davenport}, {Dawson},
  {de Putter}, {De Lee}, {Delubac}, {Dhital}, {Ealet}, {Ebelke}, {Edmondson},
  {Eisenstein}, {Escoffier}, {Esposito}, {Evans}, {Fan}, {Femen{\'\i}a
  Castell{\'a}}, {Fern{\'a}ndez Alvar}, {Ferreira}, {Filiz Ak}, {Finley},
  {Fleming}, {Font-Ribera}, {Frinchaboy}, {Garc{\'\i}a-Hern{\'a}ndez},
  {Garc{\'\i}a P{\'e}rez}, {Ge}, {G{\'e}nova-Santos}, {Gillespie}, {Girardi},
  {Gonz{\'a}lez Hern{\'a}ndez}, {Grebel}, {Gunn}, {Guo}, {Haggard}, {Hamilton},
  {Harris}, {Hawley}, {Hearty}, {Ho}, {Hogg}, {Holtzman}, {Honscheid},
  {Huehnerhoff}, {Ivans}, {Ivezi{\'c}}, {Jacobson}, {Jiang}, {Johansson},
  {Johnson}, {Kauffmann}, {Kirkby}, {Kirkpatrick}, {Klaene}, {Knapp}, {Kneib},
  {Le Goff}, {Leauthaud}, {Lee}, {Lee}, {Long}, {Loomis}, {Lucatello},
  {Lundgren}, {Lupton}, {Ma}, {Ma}, {MacDonald}, {Mack}, {Mahadevan}, {Maia},
  {Majewski}, {Makler}, {Malanushenko}, {Malanushenko}, {Manchado},
  {Mandelbaum}, {Manera}, {Maraston}, {Margala}, {Martell}, {McBride},
  {McGreer}, {McMahon}, {M{\'e}nard}, {Meszaros}, {Miralda-Escud{\'e}},
  {Montero-Dorta}, {Montesano}, {Morrison}, {Muna}, {Munn}, {Murayama},
  {Myers}, {Neto}, {Nguyen}, {Nichol}, {Nidever}, {Noterdaeme}, {Nuza},
  {Ogando}, {Olmstead}, {Oravetz}, {Owen}, {Padmanabhan},
  {Palanque-Delabrouille}, {Pan}, {Parejko}, {Parihar}, {P{\^a}ris},
  {Pattarakijwanich}, {Pepper}, {Percival}, {P{\'e}rez-Fournon},
  {P{\'e}rez-R{\`a}fols}, {Petitjean}, {Pforr}, {Pieri}, {Pinsonneault}, {Porto
  de Mello}, {Prada}, {Price-Whelan}, {Raddick}, {Rebolo}, {Rich}, {Richards},
  {Robin}, {Rocha-Pinto}, {Rockosi}, {Roe}, {Ross}, {Ross}, {Rossi},
  {Rubi{\~n}o-Martin}, {Samushia}, {Sanchez Almeida}, {S{\'a}nchez},
  {Santiago}, {Sayres}, {Schlegel}, {Schlesinger}, {Schmidt}, {Schneider},
  {Schultheis}, {Schwope}, {Sc{\'o}ccola}, {Seljak}, {Sheldon}, {Shen}, {Shu},
  {Simmerer}, {Simmons}, {Skibba}, {Skrutskie}, {Slosar}, {Sobreira}, {Sobeck},
  {Stassun}, {Steele}, {Steinmetz}, {Strauss}, {Streblyanska}, {Suzuki},
  {Swanson}, {Tal}, {Thakar}, {Thomas}, {Thompson}, {Tinker}, {Tojeiro},
  {Tremonti}, {Vargas Maga{\~n}a}, {Verde}, {Viel}, {Vikas}, {Vogt}, {Wake},
  {Wang}, {Weaver}, {Weinberg}, {Weiner}, {West}, {White}, {Wilson},
  {Wisniewski}, {Wood-Vasey}, {Yanny}, {Y{\`e}che}, {York}, {Zamora},
  {Zasowski}, {Zehavi}, {Zhao}, {Zheng}, {Zhu}, \& {Zinn}}]{Ahn2012}
{Ahn}, C.~P., {Alexandroff}, R., {Allende Prieto}, C., {et~al.} 2012, \apjs,
  203, 21, \dodoi{10.1088/0067-0049/203/2/21}

\bibitem[{{Ahumada} {et~al.}(2020){Ahumada}, {Allende Prieto}, {Almeida},
  {Anders}, {Anderson}, {Andrews}, {Anguiano}, {Arcodia}, {Armengaud},
  {Aubert}, {Avila}, {Avila-Reese}, {Badenes}, {Balland}, {Barger},
  {Barrera-Ballesteros}, {Basu}, {Bautista}, {Beaton}, {Beers}, {Benavides},
  {Bender}, {Bernardi}, {Bershady}, {Beutler}, {Bidin}, {Bird}, {Bizyaev},
  {Blanc}, {Blanton}, {Boquien}, {Borissova}, {Bovy}, {Brandt}, {Brinkmann},
  {Brownstein}, {Bundy}, {Bureau}, {Burgasser}, {Burtin}, {Cano-D{\'\i}az},
  {Capasso}, {Cappellari}, {Carrera}, {Chabanier}, {Chaplin}, {Chapman},
  {Cherinka}, {Chiappini}, {Doohyun Choi}, {Chojnowski}, {Chung}, {Clerc},
  {Coffey}, {Comerford}, {Comparat}, {da Costa}, {Cousinou}, {Covey}, {Crane},
  {Cunha}, {Ilha}, {Dai}, {Damsted}, {Darling}, {Davidson}, {Davies}, {Dawson},
  {De}, {de la Macorra}, {De Lee}, {Queiroz}, {Deconto Machado}, {de la Torre},
  {Dell'Agli}, {du Mas des Bourboux}, {Diamond-Stanic}, {Dillon}, {Donor},
  {Drory}, {Duckworth}, {Dwelly}, {Ebelke}, {Eftekharzadeh}, {Davis Eigenbrot},
  {Elsworth}, {Eracleous}, {Erfanianfar}, {Escoffier}, {Fan}, {Farr},
  {Fern{\'a}ndez-Trincado}, {Feuillet}, {Finoguenov}, {Fofie},
  {Fraser-McKelvie}, {Frinchaboy}, {Fromenteau}, {Fu}, {Galbany}, {Garcia},
  {Garc{\'\i}a-Hern{\'a}ndez}, {Garma Oehmichen}, {Ge}, {Geimba Maia},
  {Geisler}, {Gelfand}, {Goddy}, {Gonzalez-Perez}, {Grabowski}, {Green},
  {Grier}, {Guo}, {Guy}, {Harding}, {Hasselquist}, {Hawken}, {Hayes}, {Hearty},
  {Hekker}, {Hogg}, {Holtzman}, {Horta}, {Hou}, {Hsieh}, {Huber}, {Hunt}, {Ider
  Chitham}, {Imig}, {Jaber}, {Jimenez Angel}, {Johnson}, {Jones},
  {J{\"o}nsson}, {Jullo}, {Kim}, {Kinemuchi}, {Kirkpatrick}, {Kite}, {Klaene},
  {Kneib}, {Kollmeier}, {Kong}, {Kounkel}, {Krishnarao}, {Lacerna}, {Lan},
  {Lane}, {Law}, {Le Goff}, {Leung}, {Lewis}, {Li}, {Lian}, {Lin}, {Long},
  {Longa-Pe{\~n}a}, {Lundgren}, {Lyke}, {Mackereth}, {MacLeod}, {Majewski},
  {Manchado}, {Maraston}, {Martini}, {Masseron}, {Masters}, {Mathur},
  {McDermid}, {Merloni}, {Merrifield}, {M{\'e}sz{\'a}ros}, {Miglio}, {Minniti},
  {Minsley}, {Miyaji}, {Mohammad}, {Mosser}, {Mueller}, {Muna},
  {Mu{\~n}oz-Guti{\'e}rrez}, {Myers}, {Nadathur}, {Nair}, {Nandra}, {Correa do
  Nascimento}, {Nevin}, {Newman}, {Nidever}, {Nitschelm}, {Noterdaeme},
  {O'Connell}, {Olmstead}, {Oravetz}, {Oravetz}, {Osorio}, {Pace}, {Padilla},
  {Palanque-Delabrouille}, {Palicio}, {Pan}, {Pan}, {Parker}, {Paviot},
  {Peirani}, {Ram{\'r}ez}, {Penny}, {Percival}, {Perez-Fournon},
  {P{\'e}rez-R{\`a}fols}, {Petitjean}, {Pieri}, {Pinsonneault}, {Poovelil},
  {Povick}, {Prakash}, {Price-Whelan}, {Raddick}, {Raichoor}, {Ray}, {Rembold},
  {Rezaie}, {Riffel}, {Riffel}, {Rix}, {Robin}, {Roman-Lopes},
  {Rom{\'a}n-Z{\'u}{\~n}iga}, {Rose}, {Ross}, {Rossi}, {Rowlands}, {Rubin},
  {Salvato}, {S{\'a}nchez}, {S{\'a}nchez-Menguiano}, {S{\'a}nchez-Gallego},
  {Sayres}, {Schaefer}, {Schiavon}, {Schimoia}, {Schlafly}, {Schlegel},
  {Schneider}, {Schultheis}, {Schwope}, {Seo}, {Serenelli}, {Shafieloo},
  {Shamsi}, {Shao}, {Shen}, {Shetrone}, {Shirley}, {Silva Aguirre}, {Simon},
  {Skrutskie}, {Slosar}, {Smethurst}, {Sobeck}, {Sodi}, {Souto}, {Stark},
  {Stassun}, {Steinmetz}, {Stello}, {Stermer}, {Storchi-Bergmann},
  {Streblyanska}, {Stringfellow}, {Stutz}, {Su{\'a}rez}, {Sun},
  {Taghizadeh-Popp}, {Talbot}, {Tayar}, {Thakar}, {Theriault}, {Thomas},
  {Thomas}, {Tinker}, {Tojeiro}, {Toledo}, {Tremonti}, {Troup}, {Tuttle},
  {Unda-Sanzana}, {Valentini}, {Vargas-Gonz{\'a}lez}, {Vargas-Maga{\~n}a},
  {V{\'a}zquez-Mata}, {Vivek}, {Wake}, {Wang}, {Weaver}, {Weijmans}, {Wild},
  {Wilson}, {Wilson}, {Wolthuis}, {Wood-Vasey}, {Yan}, {Yang}, {Y{\`e}che},
  {Zamora}, {Zarrouk}, {Zasowski}, {Zhang}, {Zhao}, {Zhao}, {Zheng}, {Zheng},
  {Zhu}, \& {Zou}}]{Ahumada2020}
{Ahumada}, R., {Allende Prieto}, C., {Almeida}, A., {et~al.} 2020, \apjs, 249,
  3, \dodoi{10.3847/1538-4365/ab929e}

\bibitem[{{Aihara} {et~al.}(2022){Aihara}, {AlSayyad}, {Ando}, {Armstrong},
  {Bosch}, {Egami}, {Furusawa}, {Furusawa}, {Harasawa}, {Harikane}, {Hsieh},
  {Ikeda}, {Ito}, {Iwata}, {Kodama}, {Koike}, {Kokubo}, {Komiyama}, {Li},
  {Liang}, {Lin}, {Lupton}, {Lust}, {MacArthur}, {Mawatari}, {Mineo},
  {Miyatake}, {Miyazaki}, {More}, {Morishima}, {Murayama}, {Nakajima},
  {Nakata}, {Nishizawa}, {Oguri}, {Okabe}, {Okura}, {Ono}, {Osato}, {Ouchi},
  {Pan}, {Plazas Malag{\'o}n}, {Price}, {Reed}, {Rykoff}, {Shibuya},
  {Simunovic}, {Strauss}, {Sugimori}, {Suto}, {Suzuki}, {Takada}, {Takagi},
  {Takata}, {Takita}, {Tanaka}, {Tang}, {Taranu}, {Terai}, {Toba}, {Turner},
  {Uchiyama}, {Vijarnwannaluk}, {Waters}, {Yamada}, {Yamamoto}, \&
  {Yamashita}}]{Aihara2022}
{Aihara}, H., {AlSayyad}, Y., {Ando}, M., {et~al.} 2022, \pasj, 74, 247,
  \dodoi{10.1093/pasj/psab122}

\bibitem[{{Alam} {et~al.}(2015){Alam}, {Albareti}, {Allende Prieto}, {Anders},
  {Anderson}, {Anderton}, {Andrews}, {Armengaud}, {Aubourg}, {Bailey}, {Basu},
  {Bautista}, {Beaton}, {Beers}, {Bender}, {Berlind}, {Beutler}, {Bhardwaj},
  {Bird}, {Bizyaev}, {Blake}, {Blanton}, {Blomqvist}, {Bochanski}, {Bolton},
  {Bovy}, {Shelden Bradley}, {Brandt}, {Brauer}, {Brinkmann}, {Brown},
  {Brownstein}, {Burden}, {Burtin}, {Busca}, {Cai}, {Capozzi}, {Carnero
  Rosell}, {Carr}, {Carrera}, {Chambers}, {Chaplin}, {Chen}, {Chiappini},
  {Chojnowski}, {Chuang}, {Clerc}, {Comparat}, {Covey}, {Croft}, {Cuesta},
  {Cunha}, {da Costa}, {Da Rio}, {Davenport}, {Dawson}, {De Lee}, {Delubac},
  {Deshpande}, {Dhital}, {Dutra-Ferreira}, {Dwelly}, {Ealet}, {Ebelke},
  {Edmondson}, {Eisenstein}, {Ellsworth}, {Elsworth}, {Epstein}, {Eracleous},
  {Escoffier}, {Esposito}, {Evans}, {Fan}, {Fern{\'a}ndez-Alvar}, {Feuillet},
  {Filiz Ak}, {Finley}, {Finoguenov}, {Flaherty}, {Fleming}, {Font-Ribera},
  {Foster}, {Frinchaboy}, {Galbraith-Frew}, {Garc{\'\i}a},
  {Garc{\'\i}a-Hern{\'a}ndez}, {Garc{\'\i}a P{\'e}rez}, {Gaulme}, {Ge},
  {G{\'e}nova-Santos}, {Georgakakis}, {Ghezzi}, {Gillespie}, {Girardi},
  {Goddard}, {Gontcho}, {Gonz{\'a}lez Hern{\'a}ndez}, {Grebel}, {Green},
  {Grieb}, {Grieves}, {Gunn}, {Guo}, {Harding}, {Hasselquist}, {Hawley},
  {Hayden}, {Hearty}, {Hekker}, {Ho}, {Hogg}, {Holley-Bockelmann}, {Holtzman},
  {Honscheid}, {Huber}, {Huehnerhoff}, {Ivans}, {Jiang}, {Johnson},
  {Kinemuchi}, {Kirkby}, {Kitaura}, {Klaene}, {Knapp}, {Kneib}, {Koenig},
  {Lam}, {Lan}, {Lang}, {Laurent}, {Le Goff}, {Leauthaud}, {Lee}, {Lee},
  {Licquia}, {Liu}, {Long}, {L{\'o}pez-Corredoira}, {Lorenzo-Oliveira},
  {Lucatello}, {Lundgren}, {Lupton}, {Mack}, {Mahadevan}, {Maia}, {Majewski},
  {Malanushenko}, {Malanushenko}, {Manchado}, {Manera}, {Mao}, {Maraston},
  {Marchwinski}, {Margala}, {Martell}, {Martig}, {Masters}, {Mathur},
  {McBride}, {McGehee}, {McGreer}, {McMahon}, {M{\'e}nard}, {Menzel},
  {Merloni}, {M{\'e}sz{\'a}ros}, {Miller}, {Miralda-Escud{\'e}}, {Miyatake},
  {Montero-Dorta}, {More}, {Morganson}, {Morice-Atkinson}, {Morrison},
  {Mosser}, {Muna}, {Myers}, {Nandra}, {Newman}, {Neyrinck}, {Nguyen},
  {Nichol}, {Nidever}, {Noterdaeme}, {Nuza}, {O'Connell}, {O'Connell},
  {O'Connell}, {Ogando}, {Olmstead}, {Oravetz}, {Oravetz}, {Osumi}, {Owen},
  {Padgett}, {Padmanabhan}, {Paegert}, {Palanque-Delabrouille}, {Pan},
  {Parejko}, {P{\^a}ris}, {Park}, {Pattarakijwanich}, {Pellejero-Ibanez},
  {Pepper}, {Percival}, {P{\'e}rez-Fournon}, {P{\'e}rez-R{\`a}fols},
  {Petitjean}, {Pieri}, {Pinsonneault}, {Porto de Mello}, {Prada}, {Prakash},
  {Price-Whelan}, {Protopapas}, {Raddick}, {Rahman}, {Reid}, {Rich}, {Rix},
  {Robin}, {Rockosi}, {Rodrigues}, {Rodr{\'\i}guez-Torres}, {Roe}, {Ross},
  {Ross}, {Rossi}, {Ruan}, {Rubi{\~n}o-Mart{\'\i}n}, {Rykoff},
  {Salazar-Albornoz}, {Salvato}, {Samushia}, {S{\'a}nchez}, {Santiago},
  {Sayres}, {Schiavon}, {Schlegel}, {Schmidt}, {Schneider}, {Schultheis},
  {Schwope}, {Sc{\'o}ccola}, {Scott}, {Sellgren}, {Seo}, {Serenelli}, {Shane},
  {Shen}, {Shetrone}, {Shu}, {Silva Aguirre}, {Sivarani}, {Skrutskie},
  {Slosar}, {Smith}, {Sobreira}, {Souto}, {Stassun}, {Steinmetz}, {Stello},
  {Strauss}, {Streblyanska}, {Suzuki}, {Swanson}, {Tan}, {Tayar}, {Terrien},
  {Thakar}, {Thomas}, {Thomas}, {Thompson}, {Tinker}, {Tojeiro}, {Troup},
  {Vargas-Maga{\~n}a}, {Vazquez}, {Verde}, {Viel}, {Vogt}, {Wake}, {Wang},
  {Weaver}, {Weinberg}, {Weiner}, {White}, {Wilson}, {Wisniewski},
  {Wood-Vasey}, {Ye`che}, {York}, {Zakamska}, {Zamora}, {Zasowski}, {Zehavi},
  {Zhao}, {Zheng}, {Zhou}, {Zhou}, {Zou}, \& {Zhu}}]{Alam2015}
{Alam}, S., {Albareti}, F.~D., {Allende Prieto}, C., {et~al.} 2015, \apjs, 219,
  12, \dodoi{10.1088/0067-0049/219/1/12}

\bibitem[{{Allevato} {et~al.}(2012){Allevato}, {Finoguenov}, {Hasinger},
  {Miyaji}, {Cappelluti}, {Salvato}, {Zamorani}, {Gilli}, {George}, {Tanaka},
  {Brusa}, {Silverman}, {Civano}, {Elvis}, \& {Shankar}}]{Allevato2012}
{Allevato}, V., {Finoguenov}, A., {Hasinger}, G., {et~al.} 2012, \apj, 758, 47,
  \dodoi{10.1088/0004-637X/758/1/47}

\bibitem[{{Almeida} {et~al.}(2023){Almeida}, {Anderson},
  {Argudo-Fern{\'a}ndez}, {Badenes}, {Barger}, {Barrera-Ballesteros}, {Bender},
  {Benitez}, {Besser}, {Bird}, {Bizyaev}, {Blanton}, {Bochanski}, {Bovy},
  {Brandt}, {Brownstein}, {Buchner}, {Bulbul}, {Burchett}, {Cano D{\'\i}az},
  {Carlberg}, {Casey}, {Chandra}, {Cherinka}, {Chiappini}, {Coker}, {Comparat},
  {Conroy}, {Contardo}, {Cortes}, {Covey}, {Crane}, {Cunha}, {Dabbieri},
  {Davidson}, {Davis}, {de Andrade Queiroz}, {De Lee}, {M{\'e}ndez Delgado},
  {Demasi}, {Di Mille}, {Donor}, {Dow}, {Dwelly}, {Eracleous}, {Eriksen},
  {Fan}, {Farr}, {Frederick}, {Fries}, {Frinchaboy}, {G{\"a}nsicke}, {Ge},
  {Gonz{\'a}lez {\'A}vila}, {Grabowski}, {Grier}, {Guiglion}, {Gupta}, {Hall},
  {Hawkins}, {Hayes}, {Hermes}, {Hern{\'a}ndez-Garc{\'\i}a}, {Hogg},
  {Holtzman}, {Ibarra-Medel}, {Ji}, {Jofre}, {Johnson}, {Jones}, {Kinemuchi},
  {Kluge}, {Koekemoer}, {Kollmeier}, {Kounkel}, {Krishnarao}, {Krumpe},
  {Lacerna}, {Lago}, {Laporte}, {Liu}, {Liu}, {Liu}, {Lopes}, {Macktoobian},
  {Majewski}, {Malanushenko}, {Maoz}, {Masseron}, {Masters}, {Matijevic},
  {McBride}, {Medan}, {Merloni}, {Morrison}, {Myers}, {M{\'e}sz{\'a}ros},
  {Negrete}, {Nidever}, {Nitschelm}, {Oravetz}, {Oravetz}, {Pan}, {Peng},
  {Pinsonneault}, {Pogge}, {Qiu}, {Ramirez}, {Rix}, {Fern{\'a}ndez Rosso},
  {Runnoe}, {Salvato}, {Sanchez}, {Santana}, {Saydjari}, {Sayres},
  {Schlaufman}, {Schneider}, {Schwope}, {Serna}, {Shen}, {Sobeck}, {Song},
  {Souto}, {Spoo}, {Stassun}, {Steinmetz}, {Straumit}, {Stringfellow},
  {S{\'a}nchez-Gallego}, {Taghizadeh-Popp}, {Tayar}, {Thakar}, {Tissera},
  {Tkachenko}, {Hernandez Toledo}, {Trakhtenbrot}, {Fern{\'a}ndez-Trincado},
  {Troup}, {Trump}, {Tuttle}, {Ulloa}, {Vazquez-Mata}, {Vera Alfaro},
  {Villanova}, {Wachter}, {Weijmans}, {Wheeler}, {Wilson}, {Wojno}, {Wolf},
  {Xue}, {Ybarra}, {Zari}, \& {Zasowski}}]{Almeida2023}
{Almeida}, A., {Anderson}, S.~F., {Argudo-Fern{\'a}ndez}, M., {et~al.} 2023,
  \apjs, 267, 44, \dodoi{10.3847/1538-4365/acda98}

\bibitem[{{Arnouts} {et~al.}(1999){Arnouts}, {Cristiani}, {Moscardini},
  {Matarrese}, {Lucchin}, {Fontana}, \& {Giallongo}}]{Arnouts1999}
{Arnouts}, S., {Cristiani}, S., {Moscardini}, L., {et~al.} 1999, \mnras, 310,
  540, \dodoi{10.1046/j.1365-8711.1999.02978.x}

\bibitem[{{Baldassare} {et~al.}(2018){Baldassare}, {Geha}, \&
  {Greene}}]{Baldassare2018}
{Baldassare}, V.~F., {Geha}, M., \& {Greene}, J. 2018, \apj, 868, 152,
  \dodoi{10.3847/1538-4357/aae6cf}

\bibitem[{{Baldassare} {et~al.}(2020){Baldassare}, {Geha}, \&
  {Greene}}]{Baldassare2020}
---. 2020, \apj, 896, 10, \dodoi{10.3847/1538-4357/ab8936}

\bibitem[{{Balogh} {et~al.}(2014){Balogh}, {McGee}, {Mok}, {Wilman},
  {Finoguenov}, {Bower}, {Mulchaey}, {Parker}, \& {Tanaka}}]{Balogh2014}
{Balogh}, M.~L., {McGee}, S.~L., {Mok}, A., {et~al.} 2014, \mnras, 443, 2679,
  \dodoi{10.1093/mnras/stu1332}

\bibitem[{{Blanton} {et~al.}(2011){Blanton}, {Kazin}, {Muna}, {Weaver}, \&
  {Price-Whelan}}]{Blanton2011}
{Blanton}, M.~R., {Kazin}, E., {Muna}, D., {Weaver}, B.~A., \& {Price-Whelan},
  A. 2011, \aj, 142, 31, \dodoi{10.1088/0004-6256/142/1/31}

\bibitem[{{Bogd{\'a}n} {et~al.}(2023){Bogd{\'a}n}, {Goulding}, {Natarajan},
  {Kov{\'a}cs}, {Tremblay}, {Chadayammuri}, {Volonteri}, {Kraft}, {Forman},
  {Jones}, {Churazov}, \& {Zhuravleva}}]{Bogdan2023}
{Bogd{\'a}n}, {\'A}., {Goulding}, A.~D., {Natarajan}, P., {et~al.} 2023, Nature
  Astronomy, \dodoi{10.1038/s41550-023-02111-9}

\bibitem[{{Bonfield} {et~al.}(2011){Bonfield}, {Jarvis}, {Hardcastle},
  {Cooray}, {Hatziminaoglou}, {Ivison}, {Page}, {Stevens}, {de Zotti}, {Auld},
  {Baes}, {Buttiglione}, {Cava}, {Dariush}, {Dunlop}, {Dunne}, {Dye}, {Eales},
  {Fritz}, {Hopwood}, {Ibar}, {Maddox}, {Micha{\l}owski}, {Pascale}, {Pohlen},
  {Rigby}, {Rodighiero}, {Serjeant}, {Smith}, {Temi}, \& {van der
  Werf}}]{Bonfield2011}
{Bonfield}, D.~G., {Jarvis}, M.~J., {Hardcastle}, M.~J., {et~al.} 2011, \mnras,
  416, 13, \dodoi{10.1111/j.1365-2966.2011.18826.x}

\bibitem[{{Bongiorno} {et~al.}(2012){Bongiorno}, {Merloni}, {Brusa},
  {Magnelli}, {Salvato}, {Mignoli}, {Zamorani}, {Fiore}, {Rosario}, {Mainieri},
  {Hao}, {Comastri}, {Vignali}, {Balestra}, {Bardelli}, {Berta}, {Civano},
  {Kampczyk}, {Le Floc'h}, {Lusso}, {Lutz}, {Pozzetti}, {Pozzi}, {Riguccini},
  {Shankar}, \& {Silverman}}]{Bongiorno2012}
{Bongiorno}, A., {Merloni}, A., {Brusa}, M., {et~al.} 2012, \mnras, 427, 3103,
  \dodoi{10.1111/j.1365-2966.2012.22089.x}

\bibitem[{{Boquien} {et~al.}(2019){Boquien}, {Burgarella}, {Roehlly}, {Buat},
  {Ciesla}, {Corre}, {Inoue}, \& {Salas}}]{Boquien2019}
{Boquien}, M., {Burgarella}, D., {Roehlly}, Y., {et~al.} 2019, \aap, 622, A103,
  \dodoi{10.1051/0004-6361/201834156}

\bibitem[{{Boutsia} {et~al.}(2018){Boutsia}, {Grazian}, {Giallongo}, {Fiore},
  \& {Civano}}]{Boutsia2018}
{Boutsia}, K., {Grazian}, A., {Giallongo}, E., {Fiore}, F., \& {Civano}, F.
  2018, \apj, 869, 20, \dodoi{10.3847/1538-4357/aae6c7}

\bibitem[{{Brammer} {et~al.}(2008){Brammer}, {van Dokkum}, \&
  {Coppi}}]{Brammer2008}
{Brammer}, G.~B., {van Dokkum}, P.~G., \& {Coppi}, P. 2008, \apj, 686, 1503,
  \dodoi{10.1086/591786}

\bibitem[{{Brusa} {et~al.}(2009){Brusa}, {Comastri}, {Gilli}, {Hasinger},
  {Iwasawa}, {Mainieri}, {Mignoli}, {Salvato}, {Zamorani}, {Bongiorno},
  {Cappelluti}, {Civano}, {Fiore}, {Merloni}, {Silverman}, {Trump}, {Vignali},
  {Capak}, {Elvis}, {Ilbert}, {Impey}, \& {Lilly}}]{Brusa2009}
{Brusa}, M., {Comastri}, A., {Gilli}, R., {et~al.} 2009, \apj, 693, 8,
  \dodoi{10.1088/0004-637X/693/1/8}

\bibitem[{{Bruzual} \& {Charlot}(2003)}]{Bruzual2003}
{Bruzual}, G., \& {Charlot}, S. 2003, \mnras, 344, 1000,
  \dodoi{10.1046/j.1365-8711.2003.06897.x}

\bibitem[{{Burgarella} {et~al.}(2005){Burgarella}, {Buat}, \&
  {Iglesias-P{\'a}ramo}}]{Burgarella2005}
{Burgarella}, D., {Buat}, V., \& {Iglesias-P{\'a}ramo}, J. 2005, \mnras, 360,
  1413, \dodoi{10.1111/j.1365-2966.2005.09131.x}

\bibitem[{{Burke} {et~al.}(2024){Burke}, {Liu}, \& {Shen}}]{Burke2023blazar}
{Burke}, C.~J., {Liu}, X., \& {Shen}, Y. 2024, \mnras, 527, 5356,
  \dodoi{10.1093/mnras/stad3592}

\bibitem[{{Burke} {et~al.}(2020){Burke}, {Shen}, {Chen}, {Scaringi},
  {Faucher-Giguere}, {Liu}, \& {Yang}}]{Burke2020tess}
{Burke}, C.~J., {Shen}, Y., {Chen}, Y.-C., {et~al.} 2020, \apj, 899, 136,
  \dodoi{10.3847/1538-4357/aba3ce}

\bibitem[{{Burke} {et~al.}(2023){Burke}, {Shen}, {Liu}, {Natarajan}, {Caplar},
  {Bellovary}, \& {Wang}}]{Burke2023}
{Burke}, C.~J., {Shen}, Y., {Liu}, X., {et~al.} 2023, \mnras, 518, 1880,
  \dodoi{10.1093/mnras/stac2478}

\bibitem[{{Burke} {et~al.}(2021){Burke}, {Shen}, {Blaes}, {Gammie}, {Horne},
  {Jiang}, {Liu}, {McHardy}, {Morgan}, {Scaringi}, \& {Yang}}]{Burke2021}
{Burke}, C.~J., {Shen}, Y., {Blaes}, O., {et~al.} 2021, Science, 373, 789,
  \dodoi{10.1126/science.abg9933}

\bibitem[{{Burke} {et~al.}(2022){Burke}, {Liu}, {Shen}, {Phadke}, {Yang},
  {Hartley}, {Harrison}, {Palmese}, {Guo}, {Zhang}, {Kron}, {Turner}, {Giles},
  {Lidman}, {Chen}, {Gruendl}, {Choi}, {Amon}, {Sheldon}, {Aguena}, {Allam},
  {Andrade-Oliveira}, {Bacon}, {Bertin}, {Brooks}, {Rosell}, {Kind},
  {Carretero}, {Conselice}, {Costanzi}, {da Costa}, {Pereira}, {Davis}, {De
  Vicente}, {Desai}, {Diehl}, {Everett}, {Ferrero}, {Flaugher},
  {Garc{\'\i}a-Bellido}, {Gaztanaga}, {Gruen}, {Gschwend}, {Gutierrez},
  {Hinton}, {Hollowood}, {Honscheid}, {Hoyle}, {James}, {Kuehn}, {Maia},
  {Marshall}, {Menanteau}, {Miquel}, {Morgan}, {Paz-Chinch{\'o}n}, {Pieres},
  {Malag{\'o}n}, {Reil}, {Romer}, {Sanchez}, {Schubnell}, {Serrano},
  {Sevilla-Noarbe}, {Smith}, {Suchyta}, {Tarle}, {Thomas}, {To}, {Varga},
  {Wilkinson}, \& {DES Collaboration}}]{Burke2022des}
{Burke}, C.~J., {Liu}, X., {Shen}, Y., {et~al.} 2022, \mnras, 516, 2736,
  \dodoi{10.1093/mnras/stac2262}

\bibitem[{{Calzetti} {et~al.}(2000){Calzetti}, {Armus}, {Bohlin}, {Kinney},
  {Koornneef}, \& {Storchi-Bergmann}}]{Calzetti2000}
{Calzetti}, D., {Armus}, L., {Bohlin}, R.~C., {et~al.} 2000, \apj, 533, 682,
  \dodoi{10.1086/308692}

\bibitem[{{Caplar} {et~al.}(2018){Caplar}, {Lilly}, \&
  {Trakhtenbrot}}]{Caplar2018}
{Caplar}, N., {Lilly}, S.~J., \& {Trakhtenbrot}, B. 2018, \apj, 867, 148,
  \dodoi{10.3847/1538-4357/aae691}

\bibitem[{{Chabrier}(2003)}]{Chabrier2003}
{Chabrier}, G. 2003, \apjl, 586, L133, \dodoi{10.1086/374879}

\bibitem[{{Ciesla} {et~al.}(2015){Ciesla}, {Charmandaris}, {Georgakakis},
  {Bernhard}, {Mitchell}, {Buat}, {Elbaz}, {LeFloc'h}, {Lacey}, {Magdis}, \&
  {Xilouris}}]{Ciesla2015}
{Ciesla}, L., {Charmandaris}, V., {Georgakakis}, A., {et~al.} 2015, \aap, 576,
  A10, \dodoi{10.1051/0004-6361/201425252}

\bibitem[{{Civano} {et~al.}(2016){Civano}, {Marchesi}, {Comastri}, {Urry},
  {Elvis}, {Cappelluti}, {Puccetti}, {Brusa}, {Zamorani}, {Hasinger},
  {Aldcroft}, {Alexander}, {Allevato}, {Brunner}, {Capak}, {Finoguenov},
  {Fiore}, {Fruscione}, {Gilli}, {Glotfelty}, {Griffiths}, {Hao}, {Harrison},
  {Jahnke}, {Kartaltepe}, {Karim}, {LaMassa}, {Lanzuisi}, {Miyaji}, {Ranalli},
  {Salvato}, {Sargent}, {Scoville}, {Schawinski}, {Schinnerer}, {Silverman},
  {Smolcic}, {Stern}, {Toft}, {Trakhtenbrot}, {Treister}, \&
  {Vignali}}]{Civano2016}
{Civano}, F., {Marchesi}, S., {Comastri}, A., {et~al.} 2016, \apj, 819, 62,
  \dodoi{10.3847/0004-637X/819/1/62}

\bibitem[{{Coil} {et~al.}(2011){Coil}, {Blanton}, {Burles}, {Cool},
  {Eisenstein}, {Moustakas}, {Wong}, {Zhu}, {Aird}, {Bernstein}, {Bolton}, \&
  {Hogg}}]{Coil2011}
{Coil}, A.~L., {Blanton}, M.~R., {Burles}, S.~M., {et~al.} 2011, \apj, 741, 8,
  \dodoi{10.1088/0004-637X/741/1/8}

\bibitem[{{Conroy}(2013)}]{Conroy2013}
{Conroy}, C. 2013, \araa, 51, 393, \dodoi{10.1146/annurev-astro-082812-141017}

\bibitem[{{Cooke} {et~al.}(2012){Cooke}, {Sullivan}, {Gal-Yam}, {Barton},
  {Carlberg}, {Ryan-Weber}, {Horst}, {Omori}, \& {D{\'\i}az}}]{Cooke2012}
{Cooke}, J., {Sullivan}, M., {Gal-Yam}, A., {et~al.} 2012, \nat, 491, 228,
  \dodoi{10.1038/nature11521}

\bibitem[{{Cool} {et~al.}(2013){Cool}, {Moustakas}, {Blanton}, {Burles},
  {Coil}, {Eisenstein}, {Wong}, {Zhu}, {Aird}, {Bernstein}, {Bolton}, {Hogg},
  \& {Mendez}}]{Cool2013}
{Cool}, R.~J., {Moustakas}, J., {Blanton}, M.~R., {et~al.} 2013, \apj, 767,
  118, \dodoi{10.1088/0004-637X/767/2/118}

\bibitem[{{Dale} {et~al.}(2014){Dale}, {Helou}, {Magdis}, {Armus},
  {D{\'\i}az-Santos}, \& {Shi}}]{Dale2014}
{Dale}, D.~A., {Helou}, G., {Magdis}, G.~E., {et~al.} 2014, \apj, 784, 83,
  \dodoi{10.1088/0004-637X/784/1/83}

\bibitem[{{Damjanov} {et~al.}(2018){Damjanov}, {Zahid}, {Geller}, {Fabricant},
  \& {Hwang}}]{Damjanov2018}
{Damjanov}, I., {Zahid}, H.~J., {Geller}, M.~J., {Fabricant}, D.~G., \&
  {Hwang}, H.~S. 2018, \apjs, 234, 21, \dodoi{10.3847/1538-4365/aaa01c}

\bibitem[{{DESI Collaboration} {et~al.}(2023){DESI Collaboration}, {Adame},
  {Aguilar}, {Ahlen}, {Alam}, {Aldering}, {Alexander}, {Alfarsy}, {Allende
  Prieto}, {Alvarez}, {Alves}, {Anand}, {Andrade-Oliveira}, {Armengaud},
  {Asorey}, {Avila}, {Aviles}, {Bailey}, {Balaguera-Antol{\'\i}nez},
  {Ballester}, {Baltay}, {Bault}, {Bautista}, {Behera}, {Beltran}, {BenZvi},
  {Beraldo e Silva}, {Bermejo-Climent}, {Berti}, {Besuner}, {Beutler},
  {Bianchi}, {Blake}, {Blum}, {Bolton}, {Brieden}, {Brodzeller}, {Brooks},
  {Brown}, {Buckley-Geer}, {Burtin}, {Cabayol-Garcia}, {Cai}, {Canning},
  {Cardiel-Sas}, {Carnero Rosell}, {Castander}, {Cervantes-Cota}, {Chabanier},
  {Chaussidon}, {Chaves-Montero}, {Chen}, {Chuang}, {Claybaugh}, {Cole},
  {Cooper}, {Cuceu}, {Davis}, {Dawson}, {de Belsunce}, {de la Cruz}, {de la
  Macorra}, {de Mattia}, {Demina}, {Demirbozan}, {DeRose}, {Dey}, {Dey},
  {Dhungana}, {Ding}, {Ding}, {Doel}, {Doshi}, {Douglass}, {Edge},
  {Eftekharzadeh}, {Eisenstein}, {Elliott}, {Escoffier}, {Fagrelius}, {Fan},
  {Fanning}, {Fawcett}, {Ferraro}, {Ereza}, {Flaugher}, {Font-Ribera},
  {Forero-S{\'a}nchez}, {Forero-Romero}, {Frenk}, {G{\"a}nsicke},
  {Garc{\'\i}a}, {Garc{\'\i}a-Bellido}, {Garcia-Quintero}, {Garrison},
  {Gil-Mar{\'\i}n}, {Golden-Marx}, {Gontcho}, {Gonzalez-Morales},
  {Gonzalez-Perez}, {Gordon}, {Graur}, {Green}, {Gruen}, {Guy}, {Hadzhiyska},
  {Hahn}, {Han}, {Hanif}, {Herrera-Alcantar}, {Honscheid}, {Hou}, {Howlett},
  {Huterer}, {Ir{\v{s}}i{\v{c}}}, {Ishak}, {Jacques}, {Jana}, {Jiang},
  {Jimenez}, {Jing}, {Joudaki}, {Jullo}, {Juneau}, {Kizhuprakkat},
  {Kara{\c{c}}ayl{\i}}, {Karim}, {Kehoe}, {Kent}, {Khederlarian}, {Kim},
  {Kirkby}, {Kisner}, {Kitaura}, {Kneib}, {Koposov}, {Kov{\'a}cs}, {Kremin},
  {Krolewski}, {L'Huillier}, {Lambert}, {Lamman}, {Lan}, {Landriau}, {Lang},
  {Lange}, {Lasker}, {Le Guillou}, {Leauthaud}, {Levi}, {Li}, {Linder},
  {Lyons}, {Magneville}, {Manera}, {Manser}, {Margala}, {Martini}, {McDonald},
  {Medina}, {Medina-Varela}, {Meisner}, {Mena-Fern{\'a}ndez}, {Meneses-Rizo},
  {Mezcua}, {Miquel}, {Montero-Camacho}, {Moon}, {Moore}, {Moustakas},
  {Mueller}, {Mundet}, {Mu{\~n}oz-Guti{\'e}rrez}, {Myers}, {Nadathur},
  {Napolitano}, {Neveux}, {Newman}, {Nie}, {Nikutta}, {Niz}, {Norberg},
  {Noriega}, {Paillas}, {Palanque-Delabrouille}, {Palmese}, {Zhiwei},
  {Parkinson}, {Penmetsa}, {Percival}, {P{\'e}rez-Fern{\'a}ndez},
  {P{\'e}rez-R{\`a}fols}, {Pieri}, {Poppett}, {Porredon}, {Pothier}, {Prada},
  {Pucha}, {Raichoor}, {Ram{\'\i}rez-P{\'e}rez}, {Ramirez-Solano},
  {Rashkovetskyi}, {Ravoux}, {Rocher}, {Rockosi}, {Ross}, {Rossi}, {Ruggeri},
  {Ruhlmann-Kleider}, {Sabiu}, {Said}, {Saintonge}, {Samushia}, {Sanchez},
  {Saulder}, {Schaan}, {Schlafly}, {Schlegel}, {Scholte}, {Schubnell}, {Seo},
  {Shafieloo}, {Sharples}, {Sheu}, {Silber}, {Sinigaglia}, {Siudek}, {Slepian},
  {Smith}, {Sprayberry}, {Stephey}, {Su{\'a}rez-P{\'e}rez}, {Sun}, {Tan},
  {Tarl{\'e}}, {Tojeiro}, {Ure{\~n}a-L{\'o}pez}, {Vaisakh}, {Valcin}, {Valdes},
  {Valluri}, {Vargas-Maga{\~n}a}, {Variu}, {Verde}, {Walther}, {Wang}, {Wang},
  {Weaver}, {Weaverdyck}, {Wechsler}, {White}, {Xie}, {Yang}, {Y{\`e}che},
  {Yu}, {Yuan}, {Zhang}, {Zhang}, {Zhao}, {Zheng}, {Zhou}, {Zhou}, {Zou},
  {Zou}, \& {Zu}}]{DESI2023}
{DESI Collaboration}, {Adame}, A.~G., {Aguilar}, J., {et~al.} 2023, arXiv
  e-prints, arXiv:2306.06308, \dodoi{10.48550/arXiv.2306.06308}

\bibitem[{{Ding} {et~al.}(2020){Ding}, {Silverman}, {Treu}, {Schulze},
  {Schramm}, {Birrer}, {Park}, {Jahnke}, {Bennert}, {Kartaltepe}, {Koekemoer},
  {Malkan}, \& {Sanders}}]{Ding2020}
{Ding}, X., {Silverman}, J., {Treu}, T., {et~al.} 2020, \apj, 888, 37,
  \dodoi{10.3847/1538-4357/ab5b90}

\bibitem[{{Draine} \& {Li}(2007)}]{Draine2007}
{Draine}, B.~T., \& {Li}, A. 2007, \apj, 657, 810, \dodoi{10.1086/511055}

\bibitem[{{Draine} {et~al.}(2014){Draine}, {Aniano}, {Krause}, {Groves},
  {Sandstrom}, {Braun}, {Leroy}, {Klaas}, {Linz}, {Rix}, {Schinnerer},
  {Schmiedeke}, \& {Walter}}]{Draine2014}
{Draine}, B.~T., {Aniano}, G., {Krause}, O., {et~al.} 2014, \apj, 780, 172,
  \dodoi{10.1088/0004-637X/780/2/172}

\bibitem[{{Duras} {et~al.}(2020){Duras}, {Bongiorno}, {Ricci}, {Piconcelli},
  {Shankar}, {Lusso}, {Bianchi}, {Fiore}, {Maiolino}, {Marconi}, {Onori},
  {Sani}, {Schneider}, {Vignali}, \& {La Franca}}]{Duras2020}
{Duras}, F., {Bongiorno}, A., {Ricci}, F., {et~al.} 2020, \aap, 636, A73,
  \dodoi{10.1051/0004-6361/201936817}

\bibitem[{{Ginsburg} {et~al.}(2019){Ginsburg}, {Sip{\H o}cz}, {Brasseur},
  {Cowperthwaite}, {Craig}, {Deil}, {Guillochon}, {Guzman}, {Liedtke}, {Lian
  Lim}, {Lockhart}, {Mommert}, {Morris}, {Norman}, {Parikh}, {Persson},
  {Robitaille}, {Segovia}, {Singer}, {Tollerud}, {de Val-Borro}, {Valtchanov},
  {Woillez}, {The Astroquery collaboration}, \& {a subset of the astropy
  collaboration}}]{Ginsburg2019}
{Ginsburg}, A., {Sip{\H o}cz}, B.~M., {Brasseur}, C.~E., {et~al.} 2019, \aj,
  157, 98, \dodoi{10.3847/1538-3881/aafc33}

\bibitem[{{Goulding} {et~al.}(2023){Goulding}, {Greene}, {Setton}, {Labbe},
  {Bezanson}, {Miller}, {Atek}, {Bogd{\'a}n}, {Brammer}, {Chemerynska},
  {Cutler}, {Dayal}, {Fudamoto}, {Fujimoto}, {Furtak}, {Kokorev}, {Khullar},
  {Leja}, {Marchesini}, {Natarajan}, {Nelson}, {Oesch}, {Pan}, {Papovich},
  {Price}, {van Dokkum}, {Wang}, {Weaver}, {Whitaker}, \&
  {Zitrin}}]{Goulding2023}
{Goulding}, A.~D., {Greene}, J.~E., {Setton}, D.~J., {et~al.} 2023, \apjl, 955,
  L24, \dodoi{10.3847/2041-8213/acf7c5}

\bibitem[{{Greene} \& {Ho}(2005)}]{Greene2005}
{Greene}, J.~E., \& {Ho}, L.~C. 2005, \apj, 630, 122, \dodoi{10.1086/431897}

\bibitem[{{Greene} {et~al.}(2020){Greene}, {Strader}, \& {Ho}}]{Greene2020}
{Greene}, J.~E., {Strader}, J., \& {Ho}, L.~C. 2020, \araa, 58, 257,
  \dodoi{10.1146/annurev-astro-032620-021835}

\bibitem[{{Guo} {et~al.}(2018){Guo}, {Shen}, \& {Wang}}]{Guo2018}
{Guo}, H., {Shen}, Y., \& {Wang}, S. 2018, {PyQSOFit: Python code to fit the
  spectrum of quasars}, Astrophysics Source Code Library, record ascl:1809.008.
\newblock \doeprint{1809.008}

\bibitem[{{Guo} {et~al.}(2020){Guo}, {Burke}, {Liu}, {Phadke}, {Zhang}, {Chen},
  {Gruendl}, {Lidman}, {Shen}, {Morganson}, {Aguena}, {Allam}, {Avila},
  {Bertin}, {Brooks}, {Rosell}, {Carollo}, {Kind}, {Costanzi}, {da Costa}, {De
  Vicente}, {Desai}, {Doel}, {Eifler}, {Everett}, {Garc{\'\i}a-Bellido},
  {Gaztanaga}, {Gerdes}, {Gruen}, {Gschwend}, {Gutierrez}, {Hinton},
  {Hollowood}, {Honscheid}, {James}, {Kuehn}, {Lima}, {Maia}, {Menanteau},
  {Miquel}, {M{\"o}ller}, {Ogando}, {Palmese}, {Paz-Chinch{\'o}n}, {Plazas},
  {Romer}, {Roodman}, {Sanchez}, {Scarpine}, {Schubnell}, {Serrano}, {Smith},
  {Soares-Santos}, {Sommer}, {Suchyta}, {Swanson}, {Tarle}, {Tucker}, {Varga},
  \& {(DES Collaboration)}}]{Guo2020}
{Guo}, H., {Burke}, C.~J., {Liu}, X., {et~al.} 2020, \mnras, 496, 3636,
  \dodoi{10.1093/mnras/staa1803}

\bibitem[{{Haehnelt} {et~al.}(1998){Haehnelt}, {Natarajan}, \&
  {Rees}}]{Haehnelt1998}
{Haehnelt}, M.~G., {Natarajan}, P., \& {Rees}, M.~J. 1998, \mnras, 300, 817,
  \dodoi{10.1046/j.1365-8711.1998.01951.x}

\bibitem[{{Halevi} {et~al.}(2019){Halevi}, {Goulding}, {Greene}, {Coupon},
  {Golob}, {Gwyn}, {Johnson}, {Moutard}, {Sawicki}, {Suh}, \&
  {Toba}}]{Halevi2019}
{Halevi}, G., {Goulding}, A., {Greene}, J., {et~al.} 2019, \apjl, 885, L3,
  \dodoi{10.3847/2041-8213/ab4b4f}

\bibitem[{{Harikane} {et~al.}(2023){Harikane}, {Zhang}, {Nakajima}, {Ouchi},
  {Isobe}, {Ono}, {Hatano}, {Xu}, \& {Umeda}}]{Harikane2023}
{Harikane}, Y., {Zhang}, Y., {Nakajima}, K., {et~al.} 2023, \apj, 959, 39,
  \dodoi{10.3847/1538-4357/ad029e}

\bibitem[{{Harrison} {et~al.}(2016){Harrison}, {Alexander}, {Mullaney},
  {Stott}, {Swinbank}, {Arumugam}, {Bauer}, {Bower}, {Bunker}, \&
  {Sharples}}]{Harrison2016}
{Harrison}, C.~M., {Alexander}, D.~M., {Mullaney}, J.~R., {et~al.} 2016,
  \mnras, 456, 1195, \dodoi{10.1093/mnras/stv2727}

\bibitem[{{Harrison} {et~al.}(2017){Harrison}, {Johnson}, {Swinbank}, {Stott},
  {Bower}, {Smail}, {Tiley}, {Bunker}, {Cirasuolo}, {Sobral}, {Sharples},
  {Best}, {Bureau}, {Jarvis}, \& {Magdis}}]{Harrison2017}
{Harrison}, C.~M., {Johnson}, H.~L., {Swinbank}, A.~M., {et~al.} 2017, \mnras,
  467, 1965, \dodoi{10.1093/mnras/stx217}

\bibitem[{{Hasinger} {et~al.}(2018){Hasinger}, {Capak}, {Salvato}, {Barger},
  {Cowie}, {Faisst}, {Hemmati}, {Kakazu}, {Kartaltepe}, {Masters}, {Mobasher},
  {Nayyeri}, {Sanders}, {Scoville}, {Suh}, {Steinhardt}, \&
  {Yang}}]{Hasinger2018}
{Hasinger}, G., {Capak}, P., {Salvato}, M., {et~al.} 2018, \apj, 858, 77,
  \dodoi{10.3847/1538-4357/aabacf}

\bibitem[{{Hickox} {et~al.}(2014){Hickox}, {Mullaney}, {Alexander}, {Chen},
  {Civano}, {Goulding}, \& {Hainline}}]{Hickox2014}
{Hickox}, R.~C., {Mullaney}, J.~R., {Alexander}, D.~M., {et~al.} 2014, \apj,
  782, 9, \dodoi{10.1088/0004-637X/782/1/9}

\bibitem[{{Ho}(2009)}]{Ho2009}
{Ho}, L.~C. 2009, \apj, 699, 626, \dodoi{10.1088/0004-637X/699/1/626}

\bibitem[{{Hoshi} {et~al.}(2024){Hoshi}, {Yamada}, {Kokubo}, {Matsuoka}, \&
  {Nagao}}]{Hoshi2024}
{Hoshi}, A., {Yamada}, T., {Kokubo}, M., {Matsuoka}, Y., \& {Nagao}, T. 2024,
  arXiv e-prints, arXiv:2404.13561, \dodoi{10.48550/arXiv.2404.13561}

\bibitem[{{Ilbert} {et~al.}(2006){Ilbert}, {Arnouts}, {McCracken},
  {Bolzonella}, {Bertin}, {Le F{\`e}vre}, {Mellier}, {Zamorani}, {Pell{\`o}},
  {Iovino}, {Tresse}, {Le Brun}, {Bottini}, {Garilli}, {Maccagni}, {Picat},
  {Scaramella}, {Scodeggio}, {Vettolani}, {Zanichelli}, {Adami}, {Bardelli},
  {Cappi}, {Charlot}, {Ciliegi}, {Contini}, {Cucciati}, {Foucaud}, {Franzetti},
  {Gavignaud}, {Guzzo}, {Marano}, {Marinoni}, {Mazure}, {Meneux}, {Merighi},
  {Paltani}, {Pollo}, {Pozzetti}, {Radovich}, {Zucca}, {Bondi}, {Bongiorno},
  {Busarello}, {de La Torre}, {Gregorini}, {Lamareille}, {Mathez}, {Merluzzi},
  {Ripepi}, {Rizzo}, \& {Vergani}}]{Ilbert2006}
{Ilbert}, O., {Arnouts}, S., {McCracken}, H.~J., {et~al.} 2006, \aap, 457, 841,
  \dodoi{10.1051/0004-6361:20065138}

\bibitem[{{Inoue}(2011)}]{Inoue2011}
{Inoue}, A.~K. 2011, \mnras, 415, 2920,
  \dodoi{10.1111/j.1365-2966.2011.18906.x}

\bibitem[{{Ivezi{\'c}} {et~al.}(2019){Ivezi{\'c}}, {Kahn}, {Tyson}, {Abel},
  {Acosta}, {Allsman}, {Alonso}, {AlSayyad}, {Anderson}, {Andrew}, {Angel},
  {Angeli}, {Ansari}, {Antilogus}, {Araujo}, {Armstrong}, {Arndt}, {Astier},
  {Aubourg}, {Auza}, {Axelrod}, {Bard}, {Barr}, {Barrau}, {Bartlett}, {Bauer},
  {Bauman}, {Baumont}, {Bechtol}, {Bechtol}, {Becker}, {Becla}, {Beldica},
  {Bellavia}, {Bianco}, {Biswas}, {Blanc}, {Blazek}, {Blandford}, {Bloom},
  {Bogart}, {Bond}, {Booth}, {Borgland}, {Borne}, {Bosch}, {Boutigny},
  {Brackett}, {Bradshaw}, {Brandt}, {Brown}, {Bullock}, {Burchat}, {Burke},
  {Cagnoli}, {Calabrese}, {Callahan}, {Callen}, {Carlin}, {Carlson},
  {Chandrasekharan}, {Charles-Emerson}, {Chesley}, {Cheu}, {Chiang}, {Chiang},
  {Chirino}, {Chow}, {Ciardi}, {Claver}, {Cohen-Tanugi}, {Cockrum}, {Coles},
  {Connolly}, {Cook}, {Cooray}, {Covey}, {Cribbs}, {Cui}, {Cutri}, {Daly},
  {Daniel}, {Daruich}, {Daubard}, {Daues}, {Dawson}, {Delgado}, {Dellapenna},
  {de Peyster}, {de Val-Borro}, {Digel}, {Doherty}, {Dubois},
  {Dubois-Felsmann}, {Durech}, {Economou}, {Eifler}, {Eracleous}, {Emmons},
  {Fausti Neto}, {Ferguson}, {Figueroa}, {Fisher-Levine}, {Focke}, {Foss},
  {Frank}, {Freemon}, {Gangler}, {Gawiser}, {Geary}, {Gee}, {Geha}, {Gessner},
  {Gibson}, {Gilmore}, {Glanzman}, {Glick}, {Goldina}, {Goldstein}, {Goodenow},
  {Graham}, {Gressler}, {Gris}, {Guy}, {Guyonnet}, {Haller}, {Harris},
  {Hascall}, {Haupt}, {Hernandez}, {Herrmann}, {Hileman}, {Hoblitt}, {Hodgson},
  {Hogan}, {Howard}, {Huang}, {Huffer}, {Ingraham}, {Innes}, {Jacoby}, {Jain},
  {Jammes}, {Jee}, {Jenness}, {Jernigan}, {Jevremovi{\'c}}, {Johns}, {Johnson},
  {Johnson}, {Jones}, {Juramy-Gilles}, {Juri{\'c}}, {Kalirai}, {Kallivayalil},
  {Kalmbach}, {Kantor}, {Karst}, {Kasliwal}, {Kelly}, {Kessler}, {Kinnison},
  {Kirkby}, {Knox}, {Kotov}, {Krabbendam}, {Krughoff}, {Kub{\'a}nek},
  {Kuczewski}, {Kulkarni}, {Ku}, {Kurita}, {Lage}, {Lambert}, {Lange},
  {Langton}, {Le Guillou}, {Levine}, {Liang}, {Lim}, {Lintott}, {Long},
  {Lopez}, {Lotz}, {Lupton}, {Lust}, {MacArthur}, {Mahabal}, {Mandelbaum},
  {Markiewicz}, {Marsh}, {Marshall}, {Marshall}, {May}, {McKercher}, {McQueen},
  {Meyers}, {Migliore}, {Miller}, {Mills}, {Miraval}, {Moeyens}, {Moolekamp},
  {Monet}, {Moniez}, {Monkewitz}, {Montgomery}, {Morrison}, {Mueller},
  {Muller}, {Mu{\~n}oz Arancibia}, {Neill}, {Newbry}, {Nief}, {Nomerotski},
  {Nordby}, {O'Connor}, {Oliver}, {Olivier}, {Olsen}, {O'Mullane}, {Ortiz},
  {Osier}, {Owen}, {Pain}, {Palecek}, {Parejko}, {Parsons}, {Pease},
  {Peterson}, {Peterson}, {Petravick}, {Libby Petrick}, {Petry},
  {Pierfederici}, {Pietrowicz}, {Pike}, {Pinto}, {Plante}, {Plate}, {Plutchak},
  {Price}, {Prouza}, {Radeka}, {Rajagopal}, {Rasmussen}, {Regnault}, {Reil},
  {Reiss}, {Reuter}, {Ridgway}, {Riot}, {Ritz}, {Robinson}, {Roby}, {Roodman},
  {Rosing}, {Roucelle}, {Rumore}, {Russo}, {Saha}, {Sassolas}, {Schalk},
  {Schellart}, {Schindler}, {Schmidt}, {Schneider}, {Schneider}, {Schoening},
  {Schumacher}, {Schwamb}, {Sebag}, {Selvy}, {Sembroski}, {Seppala}, {Serio},
  {Serrano}, {Shaw}, {Shipsey}, {Sick}, {Silvestri}, {Slater}, {Smith},
  {Smith}, {Sobhani}, {Soldahl}, {Storrie-Lombardi}, {Stover}, {Strauss},
  {Street}, {Stubbs}, {Sullivan}, {Sweeney}, {Swinbank}, {Szalay}, {Takacs},
  {Tether}, {Thaler}, {Thayer}, {Thomas}, {Thornton}, {Thukral}, {Tice},
  {Trilling}, {Turri}, {Van Berg}, {Vanden Berk}, {Vetter}, {Virieux},
  {Vucina}, {Wahl}, {Walkowicz}, {Walsh}, {Walter}, {Wang}, {Wang}, {Warner},
  {Wiecha}, {Willman}, {Winters}, {Wittman}, {Wolff}, {Wood-Vasey}, {Wu},
  {Xin}, {Yoachim}, \& {Zhan}}]{Ivezic2019}
{Ivezi{\'c}}, {\v{Z}}., {Kahn}, S.~M., {Tyson}, J.~A., {et~al.} 2019, \apj,
  873, 111, \dodoi{10.3847/1538-4357/ab042c}

\bibitem[{{Izumi} {et~al.}(2019){Izumi}, {Onoue}, {Matsuoka}, {Nagao},
  {Strauss}, {Imanishi}, {Kashikawa}, {Fujimoto}, {Kohno}, {Toba}, {Umehata},
  {Goto}, {Ueda}, {Shirakata}, {Silverman}, {Greene}, {Harikane}, {Hashimoto},
  {Ikarashi}, {Iono}, {Iwasawa}, {Lee}, {Minezaki}, {Nakanishi}, {Tamura},
  {Tang}, \& {Taniguchi}}]{Izumi2019}
{Izumi}, T., {Onoue}, M., {Matsuoka}, Y., {et~al.} 2019, \pasj, 71, 111,
  \dodoi{10.1093/pasj/psz096}

\bibitem[{{Izumi} {et~al.}(2021){Izumi}, {Matsuoka}, {Fujimoto}, {Onoue},
  {Strauss}, {Umehata}, {Imanishi}, {Kohno}, {Kawaguchi}, {Kawamuro}, {Baba},
  {Nagao}, {Toba}, {Inayoshi}, {Silverman}, {Inoue}, {Ikarashi}, {Iwasawa},
  {Kashikawa}, {Hashimoto}, {Nakanishi}, {Ueda}, {Schramm}, {Lee}, \&
  {Suh}}]{Izumi2021}
{Izumi}, T., {Matsuoka}, Y., {Fujimoto}, S., {et~al.} 2021, \apj, 914, 36,
  \dodoi{10.3847/1538-4357/abf6dc}

\bibitem[{{Jamal} {et~al.}(2018){Jamal}, {Le Brun}, {Le F{\`e}vre}, {Vibert},
  {Schmitt}, {Surace}, {Copin}, {Garilli}, {Moresco}, \&
  {Pozzetti}}]{Jamal2018}
{Jamal}, S., {Le Brun}, V., {Le F{\`e}vre}, O., {et~al.} 2018, \aap, 611, A53,
  \dodoi{10.1051/0004-6361/201731305}

\bibitem[{{Kartaltepe} {et~al.}(2015){Kartaltepe}, {Sanders}, {Silverman},
  {Kashino}, {Chu}, {Zahid}, {Hasinger}, {Kewley}, {Matsuoka}, {Nagao},
  {Riguccini}, {Salvato}, {Schawinski}, {Taniguchi}, {Treister}, {Capak},
  {Daddi}, \& {Ohta}}]{Kartaltepe2015}
{Kartaltepe}, J.~S., {Sanders}, D.~B., {Silverman}, J.~D., {et~al.} 2015,
  \apjl, 806, L35, \dodoi{10.1088/2041-8205/806/2/L35}

\bibitem[{{Kashino} {et~al.}(2019){Kashino}, {Silverman}, {Sanders},
  {Kartaltepe}, {Daddi}, {Renzini}, {Rodighiero}, {Puglisi}, {Valentino},
  {Juneau}, {Arimoto}, {Nagao}, {Ilbert}, {Le F{\`e}vre}, \&
  {Koekemoer}}]{Kashino2019}
{Kashino}, D., {Silverman}, J.~D., {Sanders}, D., {et~al.} 2019, \apjs, 241,
  10, \dodoi{10.3847/1538-4365/ab06c4}

\bibitem[{{Kessler} {et~al.}(2015){Kessler}, {Marriner}, {Childress},
  {Covarrubias}, {D'Andrea}, {Finley}, {Fischer}, {Foley}, {Goldstein},
  {Gupta}, {Kuehn}, {Marcha}, {Nichol}, {Papadopoulos}, {Sako}, {Scolnic},
  {Smith}, {Sullivan}, {Wester}, {Yuan}, {Abbott}, {Abdalla}, {Allam},
  {Benoit-L{\'e}vy}, {Bernstein}, {Bertin}, {Brooks}, {Carnero Rosell},
  {Carrasco Kind}, {Castander}, {Crocce}, {da Costa}, {Desai}, {Diehl},
  {Eifler}, {Fausti Neto}, {Flaugher}, {Frieman}, {Gerdes}, {Gruen}, {Gruendl},
  {Honscheid}, {James}, {Kuropatkin}, {Li}, {Maia}, {Marshall}, {Martini},
  {Miller}, {Miquel}, {Nord}, {Ogando}, {Plazas}, {Reil}, {Romer}, {Roodman},
  {Sanchez}, {Sevilla-Noarbe}, {Smith}, {Soares-Santos}, {Sobreira}, {Tarle},
  {Thaler}, {Thomas}, {Tucker}, {Walker}, \& {DES Collaboration}}]{Kessler2015}
{Kessler}, R., {Marriner}, J., {Childress}, M., {et~al.} 2015, \aj, 150, 172,
  \dodoi{10.1088/0004-6256/150/6/172}

\bibitem[{{Kimura} {et~al.}(2020){Kimura}, {Yamada}, {Kokubo}, {Yasuda},
  {Morokuma}, {Nagao}, \& {Matsuoka}}]{Kimura2020}
{Kimura}, Y., {Yamada}, T., {Kokubo}, M., {et~al.} 2020, \apj, 894, 24,
  \dodoi{10.3847/1538-4357/ab83f3}

\bibitem[{{Knobel} {et~al.}(2012){Knobel}, {Lilly}, {Iovino}, {Kova{\v{c}}},
  {Bschorr}, {Presotto}, {Oesch}, {Kampczyk}, {Carollo}, {Contini}, {Kneib},
  {Le Fevre}, {Mainieri}, {Renzini}, {Scodeggio}, {Zamorani}, {Bardelli},
  {Bolzonella}, {Bongiorno}, {Caputi}, {Cucciati}, {de la Torre}, {de Ravel},
  {Franzetti}, {Garilli}, {Lamareille}, {Le Borgne}, {Le Brun}, {Maier},
  {Mignoli}, {Pello}, {Peng}, {Perez Montero}, {Silverman}, {Tanaka}, {Tasca},
  {Tresse}, {Vergani}, {Zucca}, {Barnes}, {Bordoloi}, {Cappi}, {Cimatti},
  {Coppa}, {Koekemoer}, {L{\'o}pez-Sanjuan}, {McCracken}, {Moresco}, {Nair},
  {Pozzetti}, \& {Welikala}}]{Knobel2012}
{Knobel}, C., {Lilly}, S.~J., {Iovino}, A., {et~al.} 2012, \apj, 753, 121,
  \dodoi{10.1088/0004-637X/753/2/121}

\bibitem[{{Kocevski} {et~al.}(2023){Kocevski}, {Onoue}, {Inayoshi}, {Trump},
  {Haro}, {Grazian}, {Dickinson}, {Finkelstein}, {Kartaltepe}, {Hirschmann},
  {Aird}, {Holwerda}, {Fujimoto}, {Juneau}, {Amor{\'\i}n}, {Backhaus},
  {Bagley}, {Barro}, {Bell}, {Bisigello}, {Calabr{\`o}}, {Cleri}, {Cooper},
  {Ding}, {Grogin}, {Ho}, {Hutchison}, {Inoue}, {Jiang}, {Jones}, {Koekemoer},
  {Li}, {Li}, {McGrath}, {Molina}, {Papovich}, {P{\'e}rez-Gonz{\'a}lez},
  {Pirzkal}, {Wilkins}, {Yang}, \& {Yung}}]{Kocevski2023}
{Kocevski}, D.~D., {Onoue}, M., {Inayoshi}, K., {et~al.} 2023, \apjl, 954, L4,
  \dodoi{10.3847/2041-8213/ace5a0}

\bibitem[{{Kokorev} {et~al.}(2023){Kokorev}, {Fujimoto}, {Labbe}, {Greene},
  {Bezanson}, {Dayal}, {Nelson}, {Atek}, {Brammer}, {Caputi}, {Chemerynska},
  {Cutler}, {Feldmann}, {Fudamoto}, {Furtak}, {Goulding}, {de Graaff}, {Leja},
  {Marchesini}, {Miller}, {Nanayakkara}, {Oesch}, {Pan}, {Price}, {Setton},
  {Smit}, {Stefanon}, {Wang}, {Weaver}, {Whitaker}, {Williams}, \&
  {Zitrin}}]{Kokorev2023}
{Kokorev}, V., {Fujimoto}, S., {Labbe}, I., {et~al.} 2023, \apjl, 957, L7,
  \dodoi{10.3847/2041-8213/ad037a}

\bibitem[{{Koprowski} {et~al.}(2016){Koprowski}, {Dunlop}, {Micha{\l}owski},
  {Roseboom}, {Geach}, {Cirasuolo}, {Aretxaga}, {Bowler}, {Banerji}, {Bourne},
  {Coppin}, {Chapman}, {Hughes}, {Jenness}, {McLure}, {Symeonidis}, \&
  {Werf}}]{Koprowski2016}
{Koprowski}, M.~P., {Dunlop}, J.~S., {Micha{\l}owski}, M.~J., {et~al.} 2016,
  \mnras, 458, 4321, \dodoi{10.1093/mnras/stw564}

\bibitem[{{Kormendy} \& {Ho}(2013)}]{Kormendy2013}
{Kormendy}, J., \& {Ho}, L.~C. 2013, \araa, 51, 511,
  \dodoi{10.1146/annurev-astro-082708-101811}

\bibitem[{{Lang} {et~al.}(2016){Lang}, {Hogg}, \& {Mykytyn}}]{Lang2016}
{Lang}, D., {Hogg}, D.~W., \& {Mykytyn}, D. 2016, {The Tractor: Probabilistic
  astronomical source detection and measurement}, Astrophysics Source Code
  Library, record ascl:1604.008.
\newblock \doeprint{1604.008}

\bibitem[{{Lauer} {et~al.}(2007){Lauer}, {Tremaine}, {Richstone}, \&
  {Faber}}]{Lauer2007}
{Lauer}, T.~R., {Tremaine}, S., {Richstone}, D., \& {Faber}, S.~M. 2007, \apj,
  670, 249, \dodoi{10.1086/522083}

\bibitem[{{Le F{\`e}vre} {et~al.}(2013){Le F{\`e}vre}, {Cassata}, {Cucciati},
  {Garilli}, {Ilbert}, {Le Brun}, {Maccagni}, {Moreau}, {Scodeggio}, {Tresse},
  {Zamorani}, {Adami}, {Arnouts}, {Bardelli}, {Bolzonella}, {Bondi},
  {Bongiorno}, {Bottini}, {Cappi}, {Charlot}, {Ciliegi}, {Contini}, {de la
  Torre}, {Foucaud}, {Franzetti}, {Gavignaud}, {Guzzo}, {Iovino}, {Lemaux},
  {L{\'o}pez-Sanjuan}, {McCracken}, {Marano}, {Marinoni}, {Mazure}, {Mellier},
  {Merighi}, {Merluzzi}, {Paltani}, {Pell{\`o}}, {Pollo}, {Pozzetti},
  {Scaramella}, {Tasca}, {Vergani}, {Vettolani}, {Zanichelli}, \&
  {Zucca}}]{LeFevre2013}
{Le F{\`e}vre}, O., {Cassata}, P., {Cucciati}, O., {et~al.} 2013, \aap, 559,
  A14, \dodoi{10.1051/0004-6361/201322179}

\bibitem[{{Leitherer} {et~al.}(2002){Leitherer}, {Li}, {Calzetti}, \&
  {Heckman}}]{Leitherer2002}
{Leitherer}, C., {Li}, I.~H., {Calzetti}, D., \& {Heckman}, T.~M. 2002, \apjs,
  140, 303, \dodoi{10.1086/342486}

\bibitem[{{Li} {et~al.}(2024){Li}, {Silverman}, {Shen}, {Volonteri}, {Jahnke},
  {Zhuang}, {Scoggins}, {Ding}, {Harikane}, {Onoue}, \& {Tanaka}}]{Li+2024}
{Li}, J., {Silverman}, J.~D., {Shen}, Y., {et~al.} 2024, arXiv e-prints,
  arXiv:2403.00074, \dodoi{10.48550/arXiv.2403.00074}

\bibitem[{{Li} {et~al.}(2021){Li}, {Shen}, {Ho}, {Brandt}, {Dalla Bont{\`a}},
  {Fonseca Alvarez}, {Grier}, {Hernandez Santisteban}, {Homayouni}, {Horne},
  {Peterson}, {Schneider}, \& {Trump}}]{Li2021}
{Li}, J. I.~H., {Shen}, Y., {Ho}, L.~C., {et~al.} 2021, \apj, 906, 103,
  \dodoi{10.3847/1538-4357/abc8e6}

\bibitem[{{Lilly} {et~al.}(2007){Lilly}, {Le F{\`e}vre}, {Renzini}, {Zamorani},
  {Scodeggio}, {Contini}, {Carollo}, {Hasinger}, {Kneib}, {Iovino}, {Le Brun},
  {Maier}, {Mainieri}, {Mignoli}, {Silverman}, {Tasca}, {Bolzonella},
  {Bongiorno}, {Bottini}, {Capak}, {Caputi}, {Cimatti}, {Cucciati}, {Daddi},
  {Feldmann}, {Franzetti}, {Garilli}, {Guzzo}, {Ilbert}, {Kampczyk}, {Kovac},
  {Lamareille}, {Leauthaud}, {Le Borgne}, {McCracken}, {Marinoni}, {Pello},
  {Ricciardelli}, {Scarlata}, {Vergani}, {Sanders}, {Schinnerer}, {Scoville},
  {Taniguchi}, {Arnouts}, {Aussel}, {Bardelli}, {Brusa}, {Cappi}, {Ciliegi},
  {Finoguenov}, {Foucaud}, {Franceschini}, {Halliday}, {Impey}, {Knobel},
  {Koekemoer}, {Kurk}, {Maccagni}, {Maddox}, {Marano}, {Marconi}, {Meneux},
  {Mobasher}, {Moreau}, {Peacock}, {Porciani}, {Pozzetti}, {Scaramella},
  {Schiminovich}, {Shopbell}, {Smail}, {Thompson}, {Tresse}, {Vettolani},
  {Zanichelli}, \& {Zucca}}]{Lilly2007}
{Lilly}, S.~J., {Le F{\`e}vre}, O., {Renzini}, A., {et~al.} 2007, \apjs, 172,
  70, \dodoi{10.1086/516589}

\bibitem[{{Lilly} {et~al.}(2009){Lilly}, {Le Brun}, {Maier}, {Mainieri},
  {Mignoli}, {Scodeggio}, {Zamorani}, {Carollo}, {Contini}, {Kneib}, {Le
  F{\`e}vre}, {Renzini}, {Bardelli}, {Bolzonella}, {Bongiorno}, {Caputi},
  {Coppa}, {Cucciati}, {de la Torre}, {de Ravel}, {Franzetti}, {Garilli},
  {Iovino}, {Kampczyk}, {Kovac}, {Knobel}, {Lamareille}, {Le Borgne}, {Pello},
  {Peng}, {P{\'e}rez-Montero}, {Ricciardelli}, {Silverman}, {Tanaka}, {Tasca},
  {Tresse}, {Vergani}, {Zucca}, {Ilbert}, {Salvato}, {Oesch}, {Abbas},
  {Bottini}, {Capak}, {Cappi}, {Cassata}, {Cimatti}, {Elvis}, {Fumana},
  {Guzzo}, {Hasinger}, {Koekemoer}, {Leauthaud}, {Maccagni}, {Marinoni},
  {McCracken}, {Memeo}, {Meneux}, {Porciani}, {Pozzetti}, {Sanders},
  {Scaramella}, {Scarlata}, {Scoville}, {Shopbell}, \& {Taniguchi}}]{Lilly2009}
{Lilly}, S.~J., {Le Brun}, V., {Maier}, C., {et~al.} 2009, \apjs, 184, 218,
  \dodoi{10.1088/0067-0049/184/2/218}

\bibitem[{{Lutz} {et~al.}(2010){Lutz}, {Mainieri}, {Rafferty}, {Shao},
  {Hasinger}, {Wei{\ss}}, {Walter}, {Smail}, {Alexander}, {Brandt}, {Chapman},
  {Coppin}, {F{\"o}rster Schreiber}, {Gawiser}, {Genzel}, {Greve}, {Ivison},
  {Koekemoer}, {Kurczynski}, {Menten}, {Nordon}, {Popesso}, {Schinnerer},
  {Silverman}, {Wardlow}, \& {Xue}}]{Lutz2010}
{Lutz}, D., {Mainieri}, V., {Rafferty}, D., {et~al.} 2010, \apj, 712, 1287,
  \dodoi{10.1088/0004-637X/712/2/1287}

\bibitem[{{Lyke} {et~al.}(2020){Lyke}, {Higley}, {McLane}, {Schurhammer},
  {Myers}, {Ross}, {Dawson}, {Chabanier}, {Martini}, {Busca}, {Mas des
  Bourboux}, {Salvato}, {Streblyanska}, {Zarrouk}, {Burtin}, {Anderson},
  {Bautista}, {Bizyaev}, {Brandt}, {Brinkmann}, {Brownstein}, {Comparat},
  {Green}, {de la Macorra}, {Mu{\~n}oz Guti{\'e}rrez}, {Hou}, {Newman},
  {Palanque-Delabrouille}, {P{\^a}ris}, {Percival}, {Petitjean}, {Rich},
  {Rossi}, {Schneider}, {Smith}, {Vivek}, \& {Weaver}}]{Lyke2020}
{Lyke}, B.~W., {Higley}, A.~N., {McLane}, J.~N., {et~al.} 2020, \apjs, 250, 8,
  \dodoi{10.3847/1538-4365/aba623}

\bibitem[{{MacLeod} {et~al.}(2010){MacLeod}, {Ivezi{\'c}}, {Kochanek},
  {Koz{\l}owski}, {Kelly}, {Bullock}, {Kimball}, {Sesar}, {Westman}, {Brooks},
  {Gibson}, {Becker}, \& {de Vries}}]{MacLeod2010}
{MacLeod}, C.~L., {Ivezi{\'c}}, {\v{Z}}., {Kochanek}, C.~S., {et~al.} 2010,
  \apj, 721, 1014, \dodoi{10.1088/0004-637X/721/2/1014}

\bibitem[{{Magorrian} {et~al.}(1998){Magorrian}, {Tremaine}, {Richstone},
  {Bender}, {Bower}, {Dressler}, {Faber}, {Gebhardt}, {Green}, {Grillmair},
  {Kormendy}, \& {Lauer}}]{Magorrian1998}
{Magorrian}, J., {Tremaine}, S., {Richstone}, D., {et~al.} 1998, \aj, 115,
  2285, \dodoi{10.1086/300353}

\bibitem[{{Maiolino} {et~al.}(2023){Maiolino}, {Scholtz}, {Curtis-Lake},
  {Carniani}, {Baker}, {de Graaff}, {Tacchella}, {{\"U}bler}, {D'Eugenio},
  {Witstok}, {Curti}, {Arribas}, {Bunker}, {Charlot}, {Chevallard},
  {Eisenstein}, {Egami}, {Ji}, {Jones}, {Lyu}, {Rawle}, {Robertson},
  {Rujopakarn}, {Perna}, {Sun}, {Venturi}, {Williams}, \&
  {Willott}}]{Maiolino2023}
{Maiolino}, R., {Scholtz}, J., {Curtis-Lake}, E., {et~al.} 2023, arXiv
  e-prints, arXiv:2308.01230, \dodoi{10.48550/arXiv.2308.01230}

\bibitem[{{Mallery} {et~al.}(2012){Mallery}, {Mobasher}, {Capak}, {Kakazu},
  {Masters}, {Ilbert}, {Hemmati}, {Scarlata}, {Salvato}, {McCracken},
  {LeFevre}, \& {Scoville}}]{Mallery2012}
{Mallery}, R.~P., {Mobasher}, B., {Capak}, P., {et~al.} 2012, \apj, 760, 128,
  \dodoi{10.1088/0004-637X/760/2/128}

\bibitem[{{Marchesi} {et~al.}(2016){Marchesi}, {Civano}, {Elvis}, {Salvato},
  {Brusa}, {Comastri}, {Gilli}, {Hasinger}, {Lanzuisi}, {Miyaji}, {Treister},
  {Urry}, {Vignali}, {Zamorani}, {Allevato}, {Cappelluti}, {Cardamone},
  {Finoguenov}, {Griffiths}, {Karim}, {Laigle}, {LaMassa}, {Jahnke}, {Ranalli},
  {Schawinski}, {Schinnerer}, {Silverman}, {Smolcic}, {Suh}, \&
  {Trakhtenbrot}}]{Marchesi2016}
{Marchesi}, S., {Civano}, F., {Elvis}, M., {et~al.} 2016, \apj, 817, 34,
  \dodoi{10.3847/0004-637X/817/1/34}

\bibitem[{{Masters} {et~al.}(2017){Masters}, {Stern}, {Cohen}, {Capak},
  {Rhodes}, {Castander}, \& {Paltani}}]{Masters2017}
{Masters}, D.~C., {Stern}, D.~K., {Cohen}, J.~G., {et~al.} 2017, \apj, 841,
  111, \dodoi{10.3847/1538-4357/aa6f08}

\bibitem[{{Masters} {et~al.}(2019){Masters}, {Stern}, {Cohen}, {Capak},
  {Stanford}, {Hernitschek}, {Galametz}, {Davidzon}, {Rhodes}, {Sanders},
  {Mobasher}, {Castander}, {Pruett}, \& {Fotopoulou}}]{Masters2019}
---. 2019, \apj, 877, 81, \dodoi{10.3847/1538-4357/ab184d}

\bibitem[{{McCracken} {et~al.}(2012){McCracken}, {Milvang-Jensen}, {Dunlop},
  {Franx}, {Fynbo}, {Le F{\`e}vre}, {Holt}, {Caputi}, {Goranova}, {Buitrago},
  {Emerson}, {Freudling}, {Hudelot}, {L{\'o}pez-Sanjuan}, {Magnard}, {Mellier},
  {M{\o}ller}, {Nilsson}, {Sutherland}, {Tasca}, \& {Zabl}}]{McCracken2012}
{McCracken}, H.~J., {Milvang-Jensen}, B., {Dunlop}, J., {et~al.} 2012, \aap,
  544, A156, \dodoi{10.1051/0004-6361/201219507}

\bibitem[{Melchior {et~al.}(2018)Melchior, Moolekamp, Jerdee, Armstrong, Sun,
  Bosch, \& Lupton}]{Melchior2018}
Melchior, P., Moolekamp, F., Jerdee, M., {et~al.} 2018, Astronomy and
  Computing, 24, 129, \dodoi{https://doi.org/10.1016/j.ascom.2018.07.001}

\bibitem[{{Merloni} {et~al.}(2010){Merloni}, {Bongiorno}, {Bolzonella},
  {Brusa}, {Civano}, {Comastri}, {Elvis}, {Fiore}, {Gilli}, {Hao}, {Jahnke},
  {Koekemoer}, {Lusso}, {Mainieri}, {Mignoli}, {Miyaji}, {Renzini}, {Salvato},
  {Silverman}, {Trump}, {Vignali}, {Zamorani}, {Capak}, {Lilly}, {Sanders},
  {Taniguchi}, {Bardelli}, {Carollo}, {Caputi}, {Contini}, {Coppa}, {Cucciati},
  {de la Torre}, {de Ravel}, {Franzetti}, {Garilli}, {Hasinger}, {Impey},
  {Iovino}, {Iwasawa}, {Kampczyk}, {Kneib}, {Knobel}, {Kova{\v{c}}},
  {Lamareille}, {Le Borgne}, {Le Brun}, {Le F{\`e}vre}, {Maier}, {Pello},
  {Peng}, {Perez Montero}, {Ricciardelli}, {Scodeggio}, {Tanaka}, {Tasca},
  {Tresse}, {Vergani}, \& {Zucca}}]{Merloni2010}
{Merloni}, A., {Bongiorno}, A., {Bolzonella}, M., {et~al.} 2010, \apj, 708,
  137, \dodoi{10.1088/0004-637X/708/1/137}

\bibitem[{{Mezcua} {et~al.}(2018){Mezcua}, {Civano}, {Marchesi}, {Suh},
  {Fabbiano}, \& {Volonteri}}]{Mezcua2018}
{Mezcua}, M., {Civano}, F., {Marchesi}, S., {et~al.} 2018, \mnras, 478, 2576,
  \dodoi{10.1093/mnras/sty1163}

\bibitem[{{Mezcua} {et~al.}(2024){Mezcua}, {Pacucci}, {Suh}, {Siudek}, \&
  {Natarajan}}]{Mezcua+2024}
{Mezcua}, M., {Pacucci}, F., {Suh}, H., {Siudek}, M., \& {Natarajan}, P. 2024,
  \apjl, 966, L30, \dodoi{10.3847/2041-8213/ad3c2a}

\bibitem[{{Mezcua} {et~al.}(2023){Mezcua}, {Siudek}, {Suh}, {Valiante},
  {Spinoso}, \& {Bonoli}}]{Mezcua2023}
{Mezcua}, M., {Siudek}, M., {Suh}, H., {et~al.} 2023, \apjl, 943, L5,
  \dodoi{10.3847/2041-8213/acae25}

\bibitem[{{Mezcua} {et~al.}(2019){Mezcua}, {Suh}, \& {Civano}}]{Mezcua2019}
{Mezcua}, M., {Suh}, H., \& {Civano}, F. 2019, \mnras, 488, 685,
  \dodoi{10.1093/mnras/stz1760}

\bibitem[{{Momcheva} {et~al.}(2016){Momcheva}, {Brammer}, {van Dokkum},
  {Skelton}, {Whitaker}, {Nelson}, {Fumagalli}, {Maseda}, {Leja}, {Franx},
  {Rix}, {Bezanson}, {Da Cunha}, {Dickey}, {F{\"o}rster Schreiber},
  {Illingworth}, {Kriek}, {Labb{\'e}}, {Ulf Lange}, {Lundgren}, {Magee},
  {Marchesini}, {Oesch}, {Pacifici}, {Patel}, {Price}, {Tal}, {Wake}, {van der
  Wel}, \& {Wuyts}}]{Momcheva2016}
{Momcheva}, I.~G., {Brammer}, G.~B., {van Dokkum}, P.~G., {et~al.} 2016, \apjs,
  225, 27, \dodoi{10.3847/0067-0049/225/2/27}

\bibitem[{{Moneti} {et~al.}(2022){Moneti}, {McCracken}, {Shuntov}, {Kauffmann},
  {Capak}, {Davidzon}, {Ilbert}, {Scarlata}, {Toft}, {Weaver}, {Chary}, {Cuby},
  {Faisst}, {Masters}, {McPartland}, {Mobasher}, {Sanders}, {Scaramella},
  {Stern}, {Szapudi}, {Teplitz}, {Zalesky}, {Amara}, {Auricchio}, {Bodendorf},
  {Bonino}, {Branchini}, {Brau-Nogue}, {Brescia}, {Brinchmann}, {Capobianco},
  {Carbone}, {Carretero}, {Castander}, {Castellano}, {Cavuoti}, {Cimatti},
  {Cledassou}, {Congedo}, {Conselice}, {Conversi}, {Copin}, {Corcione},
  {Costille}, {Cropper}, {Da Silva}, {Degaudenzi}, {Douspis}, {Dubath},
  {Duncan}, {Dupac}, {Dusini}, {Farrens}, {Ferriol}, {Fosalba}, {Frailis},
  {Franceschi}, {Fumana}, {Garilli}, {Gillis}, {Giocoli}, {Granett}, {Grazian},
  {Grupp}, {Haugan}, {Hoekstra}, {Holmes}, {Hormuth}, {Hudelot}, {Jahnke},
  {Kermiche}, {Kiessling}, {Kilbinger}, {Kitching}, {Kohley}, {K{\"u}mmel},
  {Kunz}, {Kurki-Suonio}, {Ligori}, {Lilje}, {Lloro}, {Maiorano}, {Mansutti},
  {Marggraf}, {Markovic}, {Marulli}, {Massey}, {Maurogordato}, {Meneghetti},
  {Merlin}, {Meylan}, {Moresco}, {Moscardini}, {Munari}, {Niemi}, {Padilla},
  {Paltani}, {Pasian}, {Pedersen}, {Pires}, {Poncet}, {Popa}, {Pozzetti},
  {Raison}, {Rebolo}, {Rhodes}, {Rix}, {Roncarelli}, {Rossetti}, {Saglia},
  {Schneider}, {Secroun}, {Seidel}, {Serrano}, {Sirignano}, {Sirri}, {Stanco},
  {Tallada-Cresp{\'\i}}, {Taylor}, {Tereno}, {Toledo-Moreo}, {Torradeflot},
  {Wang}, {Welikala}, {Weller}, {Zamorani}, {Zoubian}, {Andreon}, {Bardelli},
  {Camera}, {Graci{\'a}-Carpio}, {Medinaceli}, {Mei}, {Polenta}, {Romelli},
  {Sureau}, {Tenti}, {Vassallo}, {Zacchei}, {Zucca}, {Baccigalupi},
  {Balaguera-Antol{\'\i}nez}, {Bernardeau}, {Biviano}, {Bolzonella}, {Bozzo},
  {Burigana}, {Cabanac}, {Cappi}, {Carvalho}, {Casas}, {Castignani},
  {Colodro-Conde}, {Coupon}, {Courtois}, {Di Ferdinando}, {Farina}, {Finelli},
  {Flose-Reimberg}, {Fotopoulou}, {Galeotta}, {Ganga}, {Garcia-Bellido},
  {Gaztanaga}, {Gozaliasl}, {Hook}, {Joachimi}, {Kansal}, {Keihanen},
  {Kirkpatrick}, {Lindholm}, {Mainetti}, {Maino}, {Maoli}, {Martinelli},
  {Martinet}, {Maturi}, {Metcalf}, {Morgante}, {Morisset}, {Nucita},
  {Patrizii}, {Potter}, {Renzi}, {Riccio}, {S{\'a}nchez}, {Sapone}, {Schirmer},
  {Schultheis}, {Scottez}, {Sefusatti}, {Teyssier}, {Tubio}, {Tutusaus},
  {Valiviita}, {Viel}, \& {Hildebrandt}}]{Euclid2022}
{Moneti}, A., {McCracken}, H.~J., {Shuntov}, M., {et~al.} 2022, \aap, 658,
  A126, \dodoi{10.1051/0004-6361/202142361}

\bibitem[{{Moneti} {et~al.}(2023){Moneti}, {McCracken}, {Hudelot}, {Rouberol},
  {Herent}, {Mellier}, {Dunlop}, {Le Fevre}, {Franx}, {Fynbo}, {Bowler},
  {Caputi}, {Kauffmann}, {Milvang-Jensen}, {Gonzalez-Fernandez},
  {Gonzalez-Solares}, {Irwin}, {Lewis}, {Blake}, {Cross}, {Read}, \&
  {Sutorius}}]{Moneti2023}
{Moneti}, A., {McCracken}, H.~J., {Hudelot}, W., {et~al.} 2023, VizieR Online
  Data Catalog, II/373

\bibitem[{{Monzon} {et~al.}(2020){Monzon}, {Prochaska}, {Lee}, \&
  {Chisholm}}]{Monzon2020}
{Monzon}, J.~S., {Prochaska}, J.~X., {Lee}, K.-G., \& {Chisholm}, J. 2020, \aj,
  160, 37, \dodoi{10.3847/1538-3881/ab94c2}

\bibitem[{{Mountrichas} {et~al.}(2021){Mountrichas}, {Buat}, {Georgantopoulos},
  {Yang}, {Masoura}, {Boquien}, \& {Burgarella}}]{Mountrichas2021}
{Mountrichas}, G., {Buat}, V., {Georgantopoulos}, I., {et~al.} 2021, \aap, 653,
  A70, \dodoi{10.1051/0004-6361/202141273}

\bibitem[{{Mukae} {et~al.}(2020){Mukae}, {Ouchi}, {Hill}, {Gebhardt}, {Cooper},
  {Jeong}, {Saito}, {Fabricius}, {Gawiser}, {Ciardullo}, {Farrow}, {Davis},
  {Zeimann}, {Finkelstein}, {Gronwall}, {Liu}, {Zhang}, {Byrohl}, {Ono},
  {Schneider}, {Jarvis}, {Casey}, \& {Mawatari}}]{Mukae2020}
{Mukae}, S., {Ouchi}, M., {Hill}, G.~J., {et~al.} 2020, \apj, 903, 24,
  \dodoi{10.3847/1538-4357/abb81b}

\bibitem[{{Natarajan}(2011)}]{Natarajan2011}
{Natarajan}, P. 2011, arXiv e-prints, arXiv:1105.4902,
  \dodoi{10.48550/arXiv.1105.4902}

\bibitem[{{Natarajan} {et~al.}(2017){Natarajan}, {Pacucci}, {Ferrara},
  {Agarwal}, {Ricarte}, {Zackrisson}, \& {Cappelluti}}]{Natarajan+2017}
{Natarajan}, P., {Pacucci}, F., {Ferrara}, A., {et~al.} 2017, \apj, 838, 117,
  \dodoi{10.3847/1538-4357/aa6330}

\bibitem[{{Natarajan} {et~al.}(2024){Natarajan}, {Pacucci}, {Ricarte},
  {Bogd{\'a}n}, {Goulding}, \& {Cappelluti}}]{Natarajan2023}
{Natarajan}, P., {Pacucci}, F., {Ricarte}, A., {et~al.} 2024, \apjl, 960, L1,
  \dodoi{10.3847/2041-8213/ad0e76}

\bibitem[{{Noll} {et~al.}(2009){Noll}, {Burgarella}, {Giovannoli}, {Buat},
  {Marcillac}, \& {Mu{\~n}oz-Mateos}}]{Noll2009}
{Noll}, S., {Burgarella}, D., {Giovannoli}, E., {et~al.} 2009, \aap, 507, 1793,
  \dodoi{10.1051/0004-6361/200912497}

\bibitem[{{Ono} {et~al.}(2018){Ono}, {Ouchi}, {Harikane}, {Toshikawa}, {Rauch},
  {Yuma}, {Sawicki}, {Shibuya}, {Shimasaku}, {Oguri}, {Willott}, {Akhlaghi},
  {Akiyama}, {Coupon}, {Kashikawa}, {Komiyama}, {Konno}, {Lin}, {Matsuoka},
  {Miyazaki}, {Nagao}, {Nakajima}, {Silverman}, {Tanaka}, {Taniguchi}, \&
  {Wang}}]{Ono2018}
{Ono}, Y., {Ouchi}, M., {Harikane}, Y., {et~al.} 2018, \pasj, 70, S10,
  \dodoi{10.1093/pasj/psx103}

\bibitem[{{Onodera} {et~al.}(2016){Onodera}, {Carollo}, {Lilly}, {Renzini},
  {Arimoto}, {Capak}, {Daddi}, {Scoville}, {Tacchella}, {Tatehora}, \&
  {Zamorani}}]{Onodera2016}
{Onodera}, M., {Carollo}, C.~M., {Lilly}, S., {et~al.} 2016, \apj, 822, 42,
  \dodoi{10.3847/0004-637X/822/1/42}

\bibitem[{{Pacucci} \& {Loeb}(2024)}]{Pacucci2024}
{Pacucci}, F., \& {Loeb}, A. 2024, \apj, 964, 154,
  \dodoi{10.3847/1538-4357/ad3044}

\bibitem[{{Pacucci} {et~al.}(2023){Pacucci}, {Nguyen}, {Carniani}, {Maiolino},
  \& {Fan}}]{Pacucci2023}
{Pacucci}, F., {Nguyen}, B., {Carniani}, S., {Maiolino}, R., \& {Fan}, X. 2023,
  \apjl, 957, L3, \dodoi{10.3847/2041-8213/ad0158}

\bibitem[{{P{\^a}ris} {et~al.}(2014){P{\^a}ris}, {Petitjean}, {Aubourg},
  {Ross}, {Myers}, {Streblyanska}, {Bailey}, {Hall}, {Strauss}, {Anderson},
  {Bizyaev}, {Borde}, {Brinkmann}, {Bovy}, {Brandt}, {Brewington},
  {Brownstein}, {Cook}, {Ebelke}, {Fan}, {Filiz Ak}, {Finley}, {Font-Ribera},
  {Ge}, {Hamann}, {Ho}, {Jiang}, {Kinemuchi}, {Malanushenko}, {Malanushenko},
  {Marchante}, {McGreer}, {McMahon}, {Miralda-Escud{\'e}}, {Muna},
  {Noterdaeme}, {Oravetz}, {Palanque-Delabrouille}, {Pan}, {Perez-Fournon},
  {Pieri}, {Riffel}, {Schlegel}, {Schneider}, {Simmons}, {Viel}, {Weaver},
  {Wood-Vasey}, {Y{\`e}che}, \& {York}}]{Paris2014}
{P{\^a}ris}, I., {Petitjean}, P., {Aubourg}, {\'E}., {et~al.} 2014, \aap, 563,
  A54, \dodoi{10.1051/0004-6361/201322691}

\bibitem[{{P{\^a}ris} {et~al.}(2018){P{\^a}ris}, {Petitjean}, {Aubourg},
  {Myers}, {Streblyanska}, {Lyke}, {Anderson}, {Armengaud}, {Bautista},
  {Blanton}, {Blomqvist}, {Brinkmann}, {Brownstein}, {Brandt}, {Burtin},
  {Dawson}, {de la Torre}, {Georgakakis}, {Gil-Mar{\'\i}n}, {Green}, {Hall},
  {Kneib}, {LaMassa}, {Le Goff}, {MacLeod}, {Mariappan}, {McGreer}, {Merloni},
  {Noterdaeme}, {Palanque-Delabrouille}, {Percival}, {Ross}, {Rossi},
  {Schneider}, {Seo}, {Tojeiro}, {Weaver}, {Weijmans}, {Y{\`e}che}, {Zarrouk},
  \& {Zhao}}]{Paris2018}
---. 2018, \aap, 613, A51, \dodoi{10.1051/0004-6361/201732445}

\bibitem[{{Ramos Padilla} {et~al.}(2022){Ramos Padilla}, {Wang}, {Ma{\l}ek},
  {Efstathiou}, \& {Yang}}]{Padilla2022}
{Ramos Padilla}, A.~F., {Wang}, L., {Ma{\l}ek}, K., {Efstathiou}, A., \&
  {Yang}, G. 2022, \mnras, 510, 687, \dodoi{10.1093/mnras/stab3486}

\bibitem[{{Reines} \& {Volonteri}(2015)}]{Reines2015}
{Reines}, A.~E., \& {Volonteri}, M. 2015, \apj, 813, 82,
  \dodoi{10.1088/0004-637X/813/2/82}

\bibitem[{{Ricarte} \& {Natarajan}(2018)}]{Ricarte2018}
{Ricarte}, A., \& {Natarajan}, P. 2018, \mnras, 481, 3278,
  \dodoi{10.1093/mnras/sty2448}

\bibitem[{{Richards} {et~al.}(2006){Richards}, {Lacy}, {Storrie-Lombardi},
  {Hall}, {Gallagher}, {Hines}, {Fan}, {Papovich}, {Vanden Berk}, {Trammell},
  {Schneider}, {Vestergaard}, {York}, {Jester}, {Anderson}, {Budav{\'a}ri}, \&
  {Szalay}}]{Richards2006}
{Richards}, G.~T., {Lacy}, M., {Storrie-Lombardi}, L.~J., {et~al.} 2006, \apjs,
  166, 470, \dodoi{10.1086/506525}

\bibitem[{{Rosani} {et~al.}(2020){Rosani}, {Caminha}, {Caputi}, \&
  {Deshmukh}}]{Rosani2020}
{Rosani}, G., {Caminha}, G.~B., {Caputi}, K.~I., \& {Deshmukh}, S. 2020, \aap,
  633, A159, \dodoi{10.1051/0004-6361/201935782}

\bibitem[{{Schulze} {et~al.}(2018){Schulze}, {Silverman}, {Kashino}, {Akiyama},
  {Schramm}, {Sanders}, {Kartaltepe}, {Daddi}, {Rodighiero}, {Renzini},
  {Arimoto}, {Nagao}, {Puglisi}, {Trakhtenbrot}, {Civano}, \&
  {Suh}}]{Schulze2018}
{Schulze}, A., {Silverman}, J.~D., {Kashino}, D., {et~al.} 2018, \apjs, 239,
  22, \dodoi{10.3847/1538-4365/aae82f}

\bibitem[{{Scoville} {et~al.}(2007){Scoville}, {Aussel}, {Brusa}, {Capak},
  {Carollo}, {Elvis}, {Giavalisco}, {Guzzo}, {Hasinger}, {Impey}, {Kneib},
  {LeFevre}, {Lilly}, {Mobasher}, {Renzini}, {Rich}, {Sanders}, {Schinnerer},
  {Schminovich}, {Shopbell}, {Taniguchi}, \& {Tyson}}]{Scoville2007}
{Scoville}, N., {Aussel}, H., {Brusa}, M., {et~al.} 2007, \apjs, 172, 1,
  \dodoi{10.1086/516585}

\bibitem[{{Shankar} {et~al.}(2019){Shankar}, {Bernardi}, {Richardson},
  {Marsden}, {Sheth}, {Allevato}, {Graziani}, {Mezcua}, {Ricci}, {Penny}, {La
  Franca}, \& {Pacucci}}]{Shankar2019}
{Shankar}, F., {Bernardi}, M., {Richardson}, K., {et~al.} 2019, \mnras, 485,
  1278, \dodoi{10.1093/mnras/stz376}

\bibitem[{{Shaw} {et~al.}(2012){Shaw}, {Romani}, {Cotter}, {Healey},
  {Michelson}, {Readhead}, {Richards}, {Max-Moerbeck}, {King}, \&
  {Potter}}]{Shaw2012}
{Shaw}, M.~S., {Romani}, R.~W., {Cotter}, G., {et~al.} 2012, \apj, 748, 49,
  \dodoi{10.1088/0004-637X/748/1/49}

\bibitem[{{Shen} \& {Kelly}(2010)}]{Shen2010}
{Shen}, Y., \& {Kelly}, B.~C. 2010, \apj, 713, 41,
  \dodoi{10.1088/0004-637X/713/1/41}

\bibitem[{{Shen} \& {Liu}(2012)}]{Shen2012}
{Shen}, Y., \& {Liu}, X. 2012, \apj, 753, 125,
  \dodoi{10.1088/0004-637X/753/2/125}

\bibitem[{{Shen} {et~al.}(2011){Shen}, {Richards}, {Strauss}, {Hall},
  {Schneider}, {Snedden}, {Bizyaev}, {Brewington}, {Malanushenko},
  {Malanushenko}, {Oravetz}, {Pan}, \& {Simmons}}]{Shen2011}
{Shen}, Y., {Richards}, G.~T., {Strauss}, M.~A., {et~al.} 2011, \apjs, 194, 45,
  \dodoi{10.1088/0067-0049/194/2/45}

\bibitem[{{Shen} {et~al.}(2019){Shen}, {Hall}, {Horne}, {Zhu}, {McGreer},
  {Simm}, {Trump}, {Kinemuchi}, {Brandt}, {Green}, {Grier}, {Guo}, {Ho},
  {Homayouni}, {Jiang}, {I-Hsiu Li}, {Morganson}, {Petitjean}, {Richards},
  {Schneider}, {Starkey}, {Wang}, {Chambers}, {Kaiser}, {Kudritzki}, {Magnier},
  \& {Waters}}]{Shen2019}
{Shen}, Y., {Hall}, P.~B., {Horne}, K., {et~al.} 2019, \apjs, 241, 34,
  \dodoi{10.3847/1538-4365/ab074f}

\bibitem[{{Silverman} {et~al.}(2015){Silverman}, {Kashino}, {Sanders},
  {Kartaltepe}, {Arimoto}, {Renzini}, {Rodighiero}, {Daddi}, {Zahid}, {Nagao},
  {Kewley}, {Lilly}, {Sugiyama}, {Baronchelli}, {Capak}, {Carollo}, {Chu},
  {Hasinger}, {Ilbert}, {Juneau}, {Kajisawa}, {Koekemoer}, {Kovac}, {Le
  F{\`e}vre}, {Masters}, {McCracken}, {Onodera}, {Schulze}, {Scoville},
  {Strazzullo}, \& {Taniguchi}}]{Silverman2015}
{Silverman}, J.~D., {Kashino}, D., {Sanders}, D., {et~al.} 2015, \apjs, 220,
  12, \dodoi{10.1088/0067-0049/220/1/12}

\bibitem[{{Skelton} {et~al.}(2014){Skelton}, {Whitaker}, {Momcheva}, {Brammer},
  {van Dokkum}, {Labb{\'e}}, {Franx}, {van der Wel}, {Bezanson}, {Da Cunha},
  {Fumagalli}, {F{\"o}rster Schreiber}, {Kriek}, {Leja}, {Lundgren}, {Magee},
  {Marchesini}, {Maseda}, {Nelson}, {Oesch}, {Pacifici}, {Patel}, {Price},
  {Rix}, {Tal}, {Wake}, \& {Wuyts}}]{Skelton2014}
{Skelton}, R.~E., {Whitaker}, K.~E., {Momcheva}, I.~G., {et~al.} 2014, \apjs,
  214, 24, \dodoi{10.1088/0067-0049/214/2/24}

\bibitem[{{Stalevski} {et~al.}(2012){Stalevski}, {Fritz}, {Baes}, {Nakos}, \&
  {Popovi{\'c}}}]{Stalevski2012}
{Stalevski}, M., {Fritz}, J., {Baes}, M., {Nakos}, T., \& {Popovi{\'c}},
  L.~{\v{C}}. 2012, \mnras, 420, 2756, \dodoi{10.1111/j.1365-2966.2011.19775.x}

\bibitem[{{Stalevski} {et~al.}(2016){Stalevski}, {Ricci}, {Ueda}, {Lira},
  {Fritz}, \& {Baes}}]{Stalevski2016}
{Stalevski}, M., {Ricci}, C., {Ueda}, Y., {et~al.} 2016, \mnras, 458, 2288,
  \dodoi{10.1093/mnras/stw444}

\bibitem[{{Stanford} {et~al.}(2021){Stanford}, {Masters}, {Darvish}, {Stern},
  {Cohen}, {Capak}, {Hernitschek}, {Davidzon}, {Rhodes}, {Sanders}, {Mobasher},
  {Castander}, {Paltani}, {Aghanim}, {Amara}, {Auricchio}, {Balestra},
  {Bender}, {Bodendorf}, {Bonino}, {Branchini}, {Brinchmann}, {Capobianco},
  {Carbone}, {Carretero}, {Casas}, {Castellano}, {Cavuoti}, {Cimatti},
  {Cledassou}, {Conselice}, {Corcione}, {Costille}, {Cropper}, {Degaudenzi},
  {Douspis}, {Dubath}, {Dusini}, {Fosalba}, {Frailis}, {Franceschi},
  {Franzetti}, {Fumana}, {Garilli}, {Giocoli}, {Grupp}, {Haugan}, {Hoekstra},
  {Holmes}, {Hormuth}, {Hudelot}, {Jahnke}, {Kiessling}, {Kilbinger},
  {Kitching}, {Kubik}, {K{\"u}mmel}, {Kunz}, {Kurki-Suonio}, {Laureijs},
  {Ligori}, {Lilje}, {Lloro}, {Maiorano}, {Marggraf}, {Markovic}, {Massey},
  {Meneghetti}, {Meylan}, {Moscardini}, {Niemi}, {Padilla}, {Pasian},
  {Pedersen}, {Pettorino}, {Pires}, {Poncet}, {Popa}, {Pozzetti}, {Raison},
  {Roncarelli}, {Rossetti}, {Saglia}, {Scaramella}, {Schneider}, {Secroun},
  {Seidel}, {Serrano}, {Sirignano}, {Sirri}, {Taylor}, {Teplitz}, {Tereno},
  {Toledo-Moreo}, {Valentijn}, {Valenziano}, {Verdoes Kleijn}, {Wang},
  {Zamorani}, {Zoubian}, {Brescia}, {Congedo}, {Conversi}, {Copin}, {Kermiche},
  {Kohley}, {Medinaceli}, {Mei}, {Moresco}, {Morin}, {Munari}, {Polenta},
  {Sureau}, {Tallada Cresp{\'\i}}, {Vassallo}, {Zacchei}, {Andreon}, {Aussel},
  {Baccigalupi}, {Balaguera-Antol{\'\i}nez}, {Baldi}, {Bardelli}, {Biviano},
  {Borsato}, {Bozzo}, {Burigana}, {Cabanac}, {Camera}, {Cappi}, {Carvalho},
  {Casas}, {Castignani}, {Colodro-Conde}, {Coupon}, {Courtois}, {Cuby}, {Da
  Silva}, {de la Torre}, {Di Ferdinando}, {Duncan}, {Dupac}, {Fabricius},
  {Farina}, {Farrens}, {Ferreira}, {Finelli}, {Flose-Reimberg}, {Fotopoulou},
  {Galeotta}, {Ganga}, {Gillard}, {Gozaliasl}, {Graci{\'a}-Carpio}, {Keihanen},
  {Kirkpatrick}, {Lindholm}, {Mainetti}, {Maino}, {Martinet}, {Marulli},
  {Maturi}, {Maurogordato}, {Metcalf}, {Nakajima}, {Neissner}, {Nightingale},
  {Nucita}, {Patrizii}, {Potter}, {Renzi}, {Riccio}, {Romelli}, {S{\'a}nchez},
  {Sapone}, {Schirmer}, {Schultheis}, {Scottez}, {Stanco}, {Tenti}, {Teyssier},
  {Torradeflot}, {Valiviita}, {Viel}, {Whittaker}, {Zucca}, \& {Euclid
  Collaboration}}]{Stanford2021}
{Stanford}, S.~A., {Masters}, D., {Darvish}, B., {et~al.} 2021, \apjs, 256, 9,
  \dodoi{10.3847/1538-4365/ac0833}

\bibitem[{{Stone} {et~al.}(2024){Stone}, {Lyu}, {Rieke}, {Alberts}, \&
  {Hainline}}]{Stone2023}
{Stone}, M.~A., {Lyu}, J., {Rieke}, G.~H., {Alberts}, S., \& {Hainline}, K.~N.
  2024, \apj, 964, 90, \dodoi{10.3847/1538-4357/ad2a57}

\bibitem[{{Straatman} {et~al.}(2018){Straatman}, {van der Wel}, {Bezanson},
  {Pacifici}, {Gallazzi}, {Wu}, {Noeske}, {Bari{\v{s}}i{\'c}}, {Bell},
  {Brammer}, {Calhau}, {Chauke}, {Franx}, {van Houdt}, {Labb{\'e}}, {Maseda},
  {Mu{\~n}oz-Mateos}, {Muzzin}, {van de Sande}, {Sobral}, \&
  {Spilker}}]{Straatman2018}
{Straatman}, C. M.~S., {van der Wel}, A., {Bezanson}, R., {et~al.} 2018, \apjs,
  239, 27, \dodoi{10.3847/1538-4365/aae37a}

\bibitem[{{Suh} {et~al.}(2020){Suh}, {Civano}, {Trakhtenbrot}, {Shankar},
  {Hasinger}, {Sanders}, \& {Allevato}}]{Suh2020}
{Suh}, H., {Civano}, F., {Trakhtenbrot}, B., {et~al.} 2020, \apj, 889, 32,
  \dodoi{10.3847/1538-4357/ab5f5f}

\bibitem[{{Suh} {et~al.}(2017){Suh}, {Civano}, {Hasinger}, {Lusso}, {Lanzuisi},
  {Marchesi}, {Trakhtenbrot}, {Allevato}, {Cappelluti}, {Capak}, {Elvis},
  {Griffiths}, {Laigle}, {Lira}, {Riguccini}, {Rosario}, {Salvato},
  {Schawinski}, \& {Vignali}}]{Suh2017}
{Suh}, H., {Civano}, F., {Hasinger}, G., {et~al.} 2017, \apj, 841, 102,
  \dodoi{10.3847/1538-4357/aa725c}

\bibitem[{{Suh} {et~al.}(2019){Suh}, {Civano}, {Hasinger}, {Lusso}, {Marchesi},
  {Schulze}, {Onodera}, {Rosario}, \& {Sanders}}]{Suh2019}
---. 2019, \apj, 872, 168, \dodoi{10.3847/1538-4357/ab01fb}

\bibitem[{{Tanaka} {et~al.}(2024){Tanaka}, {Silverman}, {Ding}, {Jahnke},
  {Trakhtenbrot}, {Lambrides}, {Onoue}, {Taufik Andika}, {Bongiorno}, {Faisst},
  {Gillman}, {Hayward}, {Hirschmann}, {Koekemoer}, {Kokorev}, {Liu}, {Magdis},
  {Renzini}, {Casey}, {Drakos}, {Franco}, {Gozaliasl}, {Kartaltepe}, {Liu},
  {McCracken}, {Rhodes}, {Robertson}, \& {Toft}}]{Tanaka2024}
{Tanaka}, T.~S., {Silverman}, J.~D., {Ding}, X., {et~al.} 2024, arXiv e-prints,
  arXiv:2401.13742, \dodoi{10.48550/arXiv.2401.13742}

\bibitem[{{Trump} {et~al.}(2009){Trump}, {Impey}, {Elvis}, {McCarthy},
  {Huchra}, {Brusa}, {Salvato}, {Capak}, {Cappelluti}, {Civano}, {Comastri},
  {Gabor}, {Hao}, {Hasinger}, {Jahnke}, {Kelly}, {Lilly}, {Schinnerer},
  {Scoville}, \& {Smol{\v{c}}i{\'c}}}]{Trump2009}
{Trump}, J.~R., {Impey}, C.~D., {Elvis}, M., {et~al.} 2009, \apj, 696, 1195,
  \dodoi{10.1088/0004-637X/696/2/1195}

\bibitem[{{{\"U}bler} {et~al.}(2023){{\"U}bler}, {Maiolino}, {Curtis-Lake},
  {P{\'e}rez-Gonz{\'a}lez}, {Curti}, {Perna}, {Arribas}, {Charlot}, {Marshall},
  {D'Eugenio}, {Scholtz}, {Bunker}, {Carniani}, {Ferruit}, {Jakobsen}, {Rix},
  {Rodr{\'\i}guez Del Pino}, {Willott}, {B{\"o}ker}, {Cresci}, {Jones},
  {Kumari}, \& {Rawle}}]{Ubler2023}
{{\"U}bler}, H., {Maiolino}, R., {Curtis-Lake}, E., {et~al.} 2023, arXiv
  e-prints, arXiv:2302.06647, \dodoi{10.48550/arXiv.2302.06647}

\bibitem[{{van der Wel} {et~al.}(2016){van der Wel}, {Noeske}, {Bezanson},
  {Pacifici}, {Gallazzi}, {Franx}, {Mu{\~n}oz-Mateos}, {Bell}, {Brammer},
  {Charlot}, {Chauk{\'e}}, {Labb{\'e}}, {Maseda}, {Muzzin}, {Rix}, {Sobral},
  {van de Sande}, {van Dokkum}, {Wild}, \& {Wolf}}]{vanderWel2016}
{van der Wel}, A., {Noeske}, K., {Bezanson}, R., {et~al.} 2016, \apjs, 223, 29,
  \dodoi{10.3847/0067-0049/223/2/29}

\bibitem[{{van der Wel} {et~al.}(2021){van der Wel}, {Bezanson}, {D'Eugenio},
  {Straatman}, {Franx}, {van Houdt}, {Maseda}, {Gallazzi}, {Wu}, {Pacifici},
  {Barisic}, {Brammer}, {Munoz-Mateos}, {Vervalcke}, {Zibetti}, {Sobral}, {de
  Graaff}, {Calhau}, {Kaushal}, {Muzzin}, {Bell}, \& {van
  Dokkum}}]{vanderWel2021}
{van der Wel}, A., {Bezanson}, R., {D'Eugenio}, F., {et~al.} 2021, \apjs, 256,
  44, \dodoi{10.3847/1538-4365/ac1356}

\bibitem[{{Vanden Berk} {et~al.}(2001){Vanden Berk}, {Richards}, {Bauer},
  {Strauss}, {Schneider}, {Heckman}, {York}, {Hall}, {Fan}, {Knapp},
  {Anderson}, {Annis}, {Bahcall}, {Bernardi}, {Briggs}, {Brinkmann}, {Brunner},
  {Burles}, {Carey}, {Castander}, {Connolly}, {Crocker}, {Csabai}, {Doi},
  {Finkbeiner}, {Friedman}, {Frieman}, {Fukugita}, {Gunn}, {Hennessy},
  {Ivezi{\'c}}, {Kent}, {Kunszt}, {Lamb}, {Leger}, {Long}, {Loveday}, {Lupton},
  {Meiksin}, {Merelli}, {Munn}, {Newberg}, {Newcomb}, {Nichol}, {Owen}, {Pier},
  {Pope}, {Rockosi}, {Schlegel}, {Siegmund}, {Smee}, {Snir}, {Stoughton},
  {Stubbs}, {SubbaRao}, {Szalay}, {Szokoly}, {Tremonti}, {Uomoto}, {Waddell},
  {Yanny}, \& {Zheng}}]{VandenBerk2001}
{Vanden Berk}, D.~E., {Richards}, G.~T., {Bauer}, A., {et~al.} 2001, \aj, 122,
  549, \dodoi{10.1086/321167}

\bibitem[{{Vestergaard} \& {Peterson}(2006)}]{Vestergaard2006}
{Vestergaard}, M., \& {Peterson}, B.~M. 2006, \apj, 641, 689,
  \dodoi{10.1086/500572}

\bibitem[{{Vestergaard} \& {Wilkes}(2001)}]{Vestergaard2001}
{Vestergaard}, M., \& {Wilkes}, B.~J. 2001, \apjs, 134, 1,
  \dodoi{10.1086/320357}

\bibitem[{{Volonteri} \& {Natarajan}(2009)}]{Volonteri2009}
{Volonteri}, M., \& {Natarajan}, P. 2009, \mnras, 400, 1911,
  \dodoi{10.1111/j.1365-2966.2009.15577.x}

\bibitem[{{Ward} {et~al.}(2022){Ward}, {Gezari}, {Nugent}, {Bellm}, {Dekany},
  {Drake}, {Duev}, {Graham}, {Kasliwal}, {Kool}, {Masci}, \&
  {Riddle}}]{Ward2022}
{Ward}, C., {Gezari}, S., {Nugent}, P., {et~al.} 2022, \apj, 936, 104,
  \dodoi{10.3847/1538-4357/ac8666}

\bibitem[{{Weaver} {et~al.}(2022){Weaver}, {Kauffmann}, {Ilbert}, {McCracken},
  {Moneti}, {Toft}, {Brammer}, {Shuntov}, {Davidzon}, {Hsieh}, {Laigle},
  {Anastasiou}, {Jespersen}, {Vinther}, {Capak}, {Casey}, {McPartland},
  {Milvang-Jensen}, {Mobasher}, {Sanders}, {Zalesky}, {Arnouts}, {Aussel},
  {Dunlop}, {Faisst}, {Franx}, {Furtak}, {Fynbo}, {Gould}, {Greve}, {Gwyn},
  {Kartaltepe}, {Kashino}, {Koekemoer}, {Kokorev}, {Le F{\`e}vre}, {Lilly},
  {Masters}, {Magdis}, {Mehta}, {Peng}, {Riechers}, {Salvato}, {Sawicki},
  {Scarlata}, {Scoville}, {Shirley}, {Silverman}, {Sneppen}, {Smolc̆i{\'c}},
  {Steinhardt}, {Stern}, {Tanaka}, {Taniguchi}, {Teplitz}, {Vaccari}, {Wang},
  \& {Zamorani}}]{Weaver2022}
{Weaver}, J.~R., {Kauffmann}, O.~B., {Ilbert}, O., {et~al.} 2022, \apjs, 258,
  11, \dodoi{10.3847/1538-4365/ac3078}

\bibitem[{{Weaver} {et~al.}(2023){Weaver}, {Zalesky}, {Kokorev}, {McPartland},
  {Chartab}, {Gould}, {Shuntov}, {Davidzon}, {Faisst}, {Stickley}, {Capak},
  {Toft}, {Masters}, {Mobasher}, {Sanders}, {Kauffmann}, {McCracken}, {Ilbert},
  {Brammer}, \& {Moneti}}]{Weaver2023}
{Weaver}, J.~R., {Zalesky}, L., {Kokorev}, V., {et~al.} 2023, \apjs, 269, 20,
  \dodoi{10.3847/1538-4365/acf850}

\bibitem[{{Wu} \& {Shen}(2022)}]{Wu2022}
{Wu}, Q., \& {Shen}, Y. 2022, \apjs, 263, 42, \dodoi{10.3847/1538-4365/ac9ead}

\bibitem[{{Wyithe} \& {Loeb}(2003)}]{Wyithe2003}
{Wyithe}, J. S.~B., \& {Loeb}, A. 2003, \apj, 595, 614, \dodoi{10.1086/377475}

\bibitem[{{Yang} {et~al.}(2020){Yang}, {Boquien}, {Buat}, {Burgarella},
  {Ciesla}, {Duras}, {Stalevski}, {Brandt}, \& {Papovich}}]{Yang2020}
{Yang}, G., {Boquien}, M., {Buat}, V., {et~al.} 2020, \mnras, 491, 740,
  \dodoi{10.1093/mnras/stz3001}

\bibitem[{{Yang} {et~al.}(2022){Yang}, {Boquien}, {Brandt}, {Buat},
  {Burgarella}, {Ciesla}, {Lehmer}, {Ma{\l}ek}, {Mountrichas}, {Papovich},
  {Pons}, {Stalevski}, {Theul{\'e}}, \& {Zhu}}]{Yang2022}
{Yang}, G., {Boquien}, M., {Brandt}, W.~N., {et~al.} 2022, \apj, 927, 192,
  \dodoi{10.3847/1538-4357/ac4971}

\bibitem[{{Zamojski} {et~al.}(2007){Zamojski}, {Schiminovich}, {Rich},
  {Mobasher}, {Koekemoer}, {Capak}, {Taniguchi}, {Sasaki}, {McCracken},
  {Mellier}, {Bertin}, {Aussel}, {Sanders}, {Le F{\`e}vre}, {Ilbert},
  {Salvato}, {Thompson}, {Kartaltepe}, {Scoville}, {Barlow}, {Forster},
  {Friedman}, {Martin}, {Morrissey}, {Neff}, {Seibert}, {Small}, {Wyder},
  {Bianchi}, {Donas}, {Heckman}, {Lee}, {Madore}, {Milliard}, {Szalay},
  {Welsh}, \& {Yi}}]{Zamojski2007}
{Zamojski}, M.~A., {Schiminovich}, D., {Rich}, R.~M., {et~al.} 2007, \apjs,
  172, 468, \dodoi{10.1086/516593}

\bibitem[{{Zhang} {et~al.}(2023){Zhang}, {Ouchi}, {Gebhardt}, {Liu},
  {Harikane}, {Cooper}, {Davis}, {Farrow}, {Gawiser}, {Hill}, {Kollatschny},
  {Ono}, {Schneider}, {Finkelstein}, {Gronwall}, {Jogee}, \&
  {Krumpe}}]{Zhang2023}
{Zhang}, Y., {Ouchi}, M., {Gebhardt}, K., {et~al.} 2023, \apj, 948, 103,
  \dodoi{10.3847/1538-4357/acc2c2}

\bibitem[{{Zhong} {et~al.}(2022){Zhong}, {Inoue}, {Yamanaka}, \&
  {Yamada}}]{Zhong2022}
{Zhong}, Y., {Inoue}, A.~K., {Yamanaka}, S., \& {Yamada}, T. 2022, \apj, 925,
  157, \dodoi{10.3847/1538-4357/ac3edb}

\bibitem[{{Zhuang} \& {Ho}(2023)}]{Zhuang2023nat}
{Zhuang}, M.-Y., \& {Ho}, L.~C. 2023, Nature Astronomy, 7, 1376,
  \dodoi{10.1038/s41550-023-02051-4}

\bibitem[{{Zhuang} {et~al.}(2023){Zhuang}, {Li}, \& {Shen}}]{Zhuang2023}
{Zhuang}, M.-Y., {Li}, J., \& {Shen}, Y. 2023, arXiv e-prints,
  arXiv:2309.03266, \dodoi{10.48550/arXiv.2309.03266}

\bibitem[{{Zibetti} {et~al.}(2009){Zibetti}, {Charlot}, \& {Rix}}]{Zibetti2009}
{Zibetti}, S., {Charlot}, S., \& {Rix}, H.-W. 2009, \mnras, 400, 1181,
  \dodoi{10.1111/j.1365-2966.2009.15528.x}

\bibitem[{{Zou} {et~al.}(2022){Zou}, {Brandt}, {Chen}, {Leja}, {Ni}, {Yan},
  {Yang}, {Zhu}, {Luo}, {Nyland}, {Vito}, \& {Xue}}]{Zou2022}
{Zou}, F., {Brandt}, W.~N., {Chen}, C.-T., {et~al.} 2022, \apjs, 262, 15,
  \dodoi{10.3847/1538-4365/ac7bdf}

\bibitem[{{Zou} {et~al.}(2023){Zou}, {Brandt}, {Ni}, {Zhu}, {Alexander},
  {Bauer}, {Chen}, {Luo}, {Sun}, {Vignali}, {Vito}, {Xue}, \& {Yan}}]{Zou2023}
{Zou}, F., {Brandt}, W.~N., {Ni}, Q., {et~al.} 2023, \apj, 950, 136,
  \dodoi{10.3847/1538-4357/acce39}

\end{thebibliography}
\bibliographystyle{aasjournal}



\end{document}